\documentclass{article}

\usepackage{PRIMEarxiv}

\usepackage{amsmath}
\usepackage{multirow}
\usepackage{tikz}
\usepackage{graphics}
\usepackage{hyperref}
\usepackage{caption}
\usepackage{subcaption}
\usepackage{siunitx}
\usepackage[utf8]{inputenc} 
\usepackage[T1]{fontenc}    
\usepackage{hyperref}       
\usepackage{url}            
\usepackage{booktabs}       
\usepackage{amsfonts}       
\usepackage{nicefrac}       
\usepackage{microtype}      
\usepackage{lipsum}
\usepackage{fancyhdr}       
\usepackage{graphicx}       
\graphicspath{{media/}}     

\newcommand{\overlinerm}[1]{$\overline{\textrm{#1}}$}

\title{Custom Memory Design for Logic--in--Memory: Drawbacks and Improvements over Conventional Memories
}

\author{
  Fabrizio Ottati\thanks{Correspondence: fabrizio.ottati@polito.it.
}, Giovanna Turvani, Marco Vacca, Guido Masera \\
    Department of Electronics and Telecommunications \\
    Politecnico di Torino, Torino, Italy
}

\begin{document}
\maketitle

\begin{abstract}
The speed of modern digital systems is severely limited by memory latency (the ``Memory Wall'' problem). Data exchange between Logic and Memory is also responsible for a large part of the system energy consumption. Logic--In--Memory (LiM) represents an attractive solution to this problem. By performing part of the computations directly inside the memory the system speed can be improved while reducing its energy consumption. LiM solutions that offer the major boost in performance are based on the modification of the memory cell. However, what is the cost of such modifications? How do these impact the memory array performance? \\
In this work, this question is addressed by analysing a LiM memory array implementing an algorithm for the maximum/minimum value computation. The memory array is designed at physical level using the FreePDK $\SI{45}{\nano\meter}$ CMOS process, with three memory cell variants, and its performance is compared to SRAM and CAM memories. Results highlight that read and write operations performance is worsened but in--memory operations result to be very efficient: a 55.26\% reduction in the energy--delay product is measured for the AND operation with respect to the SRAM read one; therefore, the LiM approach represents a very promising solution for low--density and high--performance memories.
\end{abstract}

\keywords{Logic--in--Memory (LiM) \and In--Memory Computing (IMC) \and Memory Wall.}

\section{Introduction}
Modern digital architectures are based on the Von Neumann principle: the system is divided into two main units, a central processing one and a memory. The CPU extracts the data from the memory, elaborates them and writes the results back. This structure represents the main performance bottleneck of modern computing systems: in fact, memories are not able to supply data to CPUs at a speed similar to the processing one, limiting the throughput of the whole system; moreover, high-speed data exchange between CPU and memory leads to large power consumption. This problem is commonly referred to as the ``Memory Wall'' problem or the ``Von Neumann bottleneck''. A complex memory hierarchy is employed to partially compensate for this, but it does not completely solve it: the system results to be still limited by the impossibility to have a memory that is large and very fast at the same time.

For these reasons, companies and researchers are searching for a way to overcome the Memory Wall problem: Logic--in--Memory (LIM), also called In-Memory Computing (IMC) \cite{santoro_turvani_graziano_2019}, is a computing paradigm that is being investigated for this purpose. In this model, part of the computation is executed inside the memory. This result is achieved by modifying the memory architecture by adding logic circuits to it. Since part of the computation is performed directly inside the memory, the CPU is not limited by the memory latency when some operations have to be performed. In addition to this, the rate at which data is exchanged between CPU and memory is reduced, resulting in power consumption reduction.

\smallskip

Many approaches to Logic-In-Memory can be found in literature; however, two main approaches can be distinguished. The first one can be classified as Near--Memory Computing (NMC) \cite{blade_EPFL, nmc_ml_0, nmc_ml_1, nmc_ml_2, nmc_ml_3, nmc_ml_4, nmc_ml_5, nmc_ml_6, nmc_ml_7, nmc_ml_8, nmc_gp_0, nmc_gp_1, nmc_gp_2, nmc_gp_3, nmc_gp_4, nmc_gp_5, nmc_gp_6}, since the memory inner array is not modified and logic circuits are added at the periphery of this; the second one can be instead denoted as Logic--in--Memory (LiM)\cite{lim_ml_0, lim_ml_1, lim_ml_2, lim_ml_3, lim_ml_4, lim_gp_0, lim_gp_1, lim_gp_2, coluccio, marco}, since the memory cell is directly modified by adding logic circuits to it. 

In an NMC architecture, logic and arithmetic circuits are added on the memory array periphery, in some cases exploiting 3D structures; therefore, the distance between computational and memory circuits is shortened, resulting in power saving and latency reduction for the data exchange between these. For instance: in \cite{blade_EPFL}, logic and arithmetic circuits are added on the bottom of an SRAM (Static Random Access Memory) array, where the data are transferred from different memory blocks, elaborated and, then, written back to the array; in \cite{nmc_gp_6}, a DRAM (Dynamic Random Access Memory) is modified to perform logic bitwise operations on the bitlines, and the sense amplifiers are configured as programmable logic gates. Near--Memory Computing allows to maximise the memory density, with minimal modifications to the memory array itself, which is the most critical part of memory design; this results in a limited performance improvement with respect to computing systems based on conventional memories. 

In a LiM architecture, the memory cells and periphery are modified by adding logic and arithmetic circuits to them, resulting in true in-memory processing, with the data being elaborated also inside each memory cell. For instance: in \cite{coluccio}, a XOR logic gate is added to each memory cell to implement a Binary Neural Network (BNN) directly in memory; in \cite{lim_gp_1}, an SRAM is modified at the cell level to perform logic operations directly in the cell, which results are then combined by appositely designed sense amplifiers on the periphery of the array. This approach leads to a reduction in memory density since the cell footprint is increased; nevertheless, the resulting performance boost is huge, since all the data stored in memory can be elaborated at once from the inner array. 

\smallskip

Many applications can benefit from the IMC approach, such as machine learning and deep learning algorithms \cite{nmc_ml_0, nmc_ml_1, nmc_ml_2, nmc_ml_3, nmc_ml_4, nmc_ml_5, nmc_ml_6, nmc_ml_7, nmc_ml_8, lim_ml_0, lim_ml_1, lim_ml_2, lim_ml_3, lim_ml_4}, but also general purpose algorithms \cite{nmc_gp_0, nmc_gp_1, nmc_gp_2, nmc_gp_3, nmc_gp_4, nmc_gp_5, nmc_gp_6, lim_gp_0, lim_gp_1, lim_gp_2}. For instance: in \cite{lim_ml_0}, a 6T SRAM cell is modified by adding two transistors and a capacitor to it, in order to perform analog computing on the whole memory, which allows to implement approximated arithmetic operations for machine learning algorithms; in \cite{nmc_gp_5}, logic layers consisting of latches and LUTs are interleaved with memory ones in an SRAM array, in order to perform different kinds of logic operations directly inside the array; in \cite{lim_gp_2}, the pass transistors of the 6T SRAM cell are modified to perform logic operations directly in the cell, which allows the memory to function as an SRAM, a CAM (Content Addressable Memory) or a LiM architecture. In general, every algorithm that works on high parallelism data and performs many element--wise operations in parallel (e.g. neural networks), is likely to receive a performance improvement when IMC solutions are employed.

Another interesting field of application is represented by Neuromorphic Computing \cite{neuro_1, neuro_2} based on Beyond--CMOS technologies, such as memristive ones. This kind of device is well suited for IMC or LiM applications, thanks to their non--volatile characteristics and low cell area footprint. For instance, in \cite{neuro_0} a VRRAM array is produced for a neuromorphic application, by implementing an in-memory XNOR operation for the synaptic weights.


\smallskip

The modification of the memory cell circuit by the addition of computational elements to it, is a risky solution: memories are circuits with a very high level of optimization; hence, even a minor modification can have a large impact on their behaviour and performance; moreover, this approach results in a reduction of the memory density. At the same time, a large boost in the overall system performance can be obtained, since all the stored data can be processed at once. As a consequence, the LiM approach represents an interesting option for low--density and high--performance memories, like caches. It is important to identify the impact that the modification of a memory cell circuit has on standard memory operations (read and write) and on in-memory logic operations, evaluating objectively the advantages and disadvantages of the approach.

\smallskip

The goal of this work is to understand and quantify this impact. As a case study,  an algorithm for the maximum/minimum computation \cite{marco} based on the bitwise logic AND operation is used. The array is designed and characterised at transistor level in Cadence Virtuoso, using FreePDK $\SI{45}{\nano\meter}$ CMOS process. Three different solutions for the memory cell circuit are investigated, that implements the same logic function, then, the array performance is compared to two conventional memories, a 6T SRAM and a NOR CAM, by considering the latency and energy consumption of each memory operation. The results highlight that modifying the memory certainly affects in a non-negligible way the read and write operations performance, but this impact can be greatly reduced by proper design and optimisation of the memory cell; nevertheless, in--memory logic operations result to be very efficient in terms of energy consumption. In fact, a 44\% reduction in the energy--delay product of the AND operation, with respect to the SRAM read one, is observed. The results obtained suggest that LiM architectures represent a very good alternative for the implementation of algorithm accelerators which can be used as secondary memories, where the execution rate of read and write operations is lower than the in--memory logic operations one.


\smallskip

The paper outline is the following:

\begin{itemize}
\item in \autoref{sec:Reference_architectures}, the design of conventional memories (SRAM and CAM) implementations to act as performance references is discussed.
\item in \autoref{sec:LiM_array}, the design of the LiM array and the three memory cells types is analyzed.
\item in \autoref{sec:Arrays_characterization} the testbench for the characterisation of the memory arrays produced is presented.
\item in \autoref{sec:Simulation_framework}, the simulation framework adopted is discussed.
\item in \autoref{sec:Results}, the obtained results are presented and analysed.
 \item in \autoref{sec:Conclusions}, some considerations about the results and the architecture are provided.
\end{itemize}

The main contributions of this paper are the following:

\begin{itemize}
    \item a LiM array, implementing a specific algorithm \cite{marco} as a case study, is designed at physical level usign the FreePDK $\SI{45}{\nano\meter}$ CMOS process and characterised through extensive SPICE simulations.
    \item three variants of the LiM cell are designed and characterised. 
    \item the LiM array performance are compared to conventional memories ones; in particular, a SRAM and a CAM arrays are designed and simulated using the same parameters of the LiM array.
    \item to characterise the design for large memory sizes, a circuital model that allows to strongly reduce the circuit netlist size is proposed and adopted to shorten as much as possible the simulation time of large arrays. 
    \item to speed--up the design of custom memory arrays such as LiM ones, a scripting approach is proposed and adopted.
\end{itemize}

\section{Reference architectures}
\label{sec:Reference_architectures}

In order to properly characterise the LiM architecture design, two standard memory arrays, SRAM and CAM, are produced in Cadence Virtuoso to be used as reference circuits: the SRAM array is chosen since it provides a lower ground for the memory cell circuit complexity that can be used as a reference by the other memory architectures; the CAM array, instead, is chosen since it is an example of Logic--In--Memory architecture (each memory cell performs an XNOR operation) widely used nowadays. The cell topologies chosen for these memory architectures are the 6T SRAM and the NOR CAM \cite{cam_literature}.

\begin{figure}[t!p]
  \begin{subfigure}[b]{0.5\linewidth}
  \centering
    \scalebox{0.8}{\begin{pgfpicture}{0cm}{0cm}{212pt}{182pt}
\pgfsetxvec{\pgfpoint{1pt}{0pt}}
\pgfsetyvec{\pgfpoint{0pt}{1pt}}
\pgfsetroundjoin 
\pgfsetroundcap
\pgftranslateto{\pgfxy(0,182)}
\begin{pgfmagnify}{1}{-1}
\definecolor{layer0}{rgb}{0.0,0.0,0.0}
\definecolor{layer1}{rgb}{0.0,0.0,0.5}
\definecolor{layer2}{rgb}{1.0,0.0,0.0}
\definecolor{layer3}{rgb}{0.0,0.5,0.5}
\definecolor{layer4}{rgb}{1.0,0.78,0.0}
\definecolor{layer5}{rgb}{0.5,1.0,0.0}
\definecolor{layer6}{rgb}{0.0,1.0,1.0}
\definecolor{layer7}{rgb}{0.0,0.5,0.0}
\definecolor{layer8}{rgb}{0.6,0.8,0.2}
\definecolor{layer9}{rgb}{1.0,0.08,0.58}
\definecolor{layer10}{rgb}{0.71,0.61,0.05}
\definecolor{layer11}{rgb}{0.0,0.5,1.0}
\definecolor{layer12}{rgb}{0.88,0.88,0.88}
\definecolor{layer13}{rgb}{0.64,0.64,0.64}
\definecolor{layer14}{rgb}{0.37,0.37,0.37}
\definecolor{layer15}{rgb}{0.0,0.0,0.0}
\color{layer0}
\pgfsetlinewidth{1.0pt}
\pgfsetdash{}{0pt}
\pgfline{\pgfxy(56.0,66.0)}{\pgfxy(156.0,66.0)}
\pgfline{\pgfxy(26.0,116.0)}{\pgfxy(16.0,116.0)}
\pgfline{\pgfxy(136.0,92.0)}{\pgfxy(146.0,92.0)}
\pgfline{\pgfxy(136.0,80.0)}{\pgfxy(146.0,80.0)}
\pgfline{\pgfxy(146.0,80.0)}{\pgfxy(146.0,76.0)}
\pgfline{\pgfxy(146.0,92.0)}{\pgfxy(146.0,96.0)}
\pgfline{\pgfxy(126.0,86.0)}{\pgfxy(130.0,86.0)}
\pgfline{\pgfxy(134.0,80.0)}{\pgfxy(134.0,92.0)}
\pgfline{\pgfxy(136.0,78.0)}{\pgfxy(136.0,94.0)}
\pgfellipse[stroke]{\pgfxy(132.0,86.0)}{\pgfxy(2.0,0)}{\pgfxy(0,2.0)}
\pgfline{\pgfxy(76.0,80.0)}{\pgfxy(66.0,80.0)}
\pgfline{\pgfxy(76.0,92.0)}{\pgfxy(66.0,92.0)}
\pgfline{\pgfxy(66.0,92.0)}{\pgfxy(66.0,96.0)}
\pgfline{\pgfxy(66.0,80.0)}{\pgfxy(66.0,76.0)}
\pgfline{\pgfxy(86.0,86.0)}{\pgfxy(82.0,86.0)}
\pgfline{\pgfxy(78.0,92.0)}{\pgfxy(78.0,80.0)}
\pgfline{\pgfxy(76.0,94.0)}{\pgfxy(76.0,78.0)}
\pgfellipse[stroke]{\pgfxy(80.0,86.0)}{\pgfxy(2.0,0)}{\pgfxy(0,2.0)}
\pgfline{\pgfxy(76.0,150.0)}{\pgfxy(66.0,150.0)}
\pgfline{\pgfxy(76.0,162.0)}{\pgfxy(66.0,162.0)}
\pgfline{\pgfxy(66.0,162.0)}{\pgfxy(66.0,166.0)}
\pgfline{\pgfxy(66.0,150.0)}{\pgfxy(66.0,146.0)}
\pgfline{\pgfxy(86.0,156.0)}{\pgfxy(78.0,156.0)}
\pgfline{\pgfxy(78.0,162.0)}{\pgfxy(78.0,150.0)}
\pgfline{\pgfxy(76.0,164.0)}{\pgfxy(76.0,148.0)}
\pgfline{\pgfxy(136.0,162.0)}{\pgfxy(146.0,162.0)}
\pgfline{\pgfxy(136.0,150.0)}{\pgfxy(146.0,150.0)}
\pgfline{\pgfxy(146.0,150.0)}{\pgfxy(146.0,146.0)}
\pgfline{\pgfxy(146.0,162.0)}{\pgfxy(146.0,166.0)}
\pgfline{\pgfxy(126.0,156.0)}{\pgfxy(134.0,156.0)}
\pgfline{\pgfxy(134.0,150.0)}{\pgfxy(134.0,162.0)}
\pgfline{\pgfxy(136.0,148.0)}{\pgfxy(136.0,164.0)}
\pgfline{\pgfxy(170.0,106.0)}{\pgfxy(170.0,116.0)}
\pgfline{\pgfxy(182.0,106.0)}{\pgfxy(182.0,116.0)}
\pgfline{\pgfxy(182.0,116.0)}{\pgfxy(186.0,116.0)}
\pgfline{\pgfxy(170.0,116.0)}{\pgfxy(166.0,116.0)}
\pgfline{\pgfxy(176.0,96.0)}{\pgfxy(176.0,104.0)}
\pgfline{\pgfxy(182.0,104.0)}{\pgfxy(170.0,104.0)}
\pgfline{\pgfxy(184.0,106.0)}{\pgfxy(168.0,106.0)}
\pgfline{\pgfxy(30.0,106.0)}{\pgfxy(30.0,116.0)}
\pgfline{\pgfxy(42.0,106.0)}{\pgfxy(42.0,116.0)}
\pgfline{\pgfxy(42.0,116.0)}{\pgfxy(46.0,116.0)}
\pgfline{\pgfxy(30.0,116.0)}{\pgfxy(26.0,116.0)}
\pgfline{\pgfxy(36.0,96.0)}{\pgfxy(36.0,104.0)}
\pgfline{\pgfxy(42.0,104.0)}{\pgfxy(30.0,104.0)}
\pgfline{\pgfxy(44.0,106.0)}{\pgfxy(28.0,106.0)}
\pgfline{\pgfxy(66.0,76.0)}{\pgfxy(66.0,66.0)}
\pgfline{\pgfxy(146.0,66.0)}{\pgfxy(146.0,76.0)}
\pgfline{\pgfxy(146.0,96.0)}{\pgfxy(146.0,146.0)}
\pgfline{\pgfxy(66.0,146.0)}{\pgfxy(66.0,96.0)}
\pgfline{\pgfxy(86.0,86.0)}{\pgfxy(86.0,156.0)}
\pgfline{\pgfxy(126.0,86.0)}{\pgfxy(126.0,156.0)}
\pgfline{\pgfxy(66.0,106.0)}{\pgfxy(126.0,106.0)}
\pgfline{\pgfxy(86.0,136.0)}{\pgfxy(146.0,136.0)}
\pgfline{\pgfxy(146.0,116.0)}{\pgfxy(166.0,116.0)}
\pgfline{\pgfxy(66.0,116.0)}{\pgfxy(46.0,116.0)}
\pgfmoveto{\pgfxy(60.0,170.0)}
\pgflineto{\pgfxy(72.0,170.0)}
\pgflineto{\pgfxy(66.0,176.0)}
\pgflineto{\pgfxy(66.0,176.0)}
\pgfclosepath 
\pgfqstroke 
\pgfline{\pgfxy(66.0,166.0)}{\pgfxy(66.0,170.0)}
\pgfline{\pgfxy(186.0,116.0)}{\pgfxy(196.0,116.0)}
\pgfline{\pgfxy(16.0,46.0)}{\pgfxy(16.0,156.0)}
\pgfline{\pgfxy(196.0,46.0)}{\pgfxy(196.0,156.0)}
\pgfsetlinewidth{0.33pt}
\pgfcircle[fill]{\pgfxy(126,106)}{2.0pt}\pgfcircle[fill]{\pgfxy(66,106)}{2.0pt}\pgfcircle[fill]{\pgfxy(86,136)}{2.0pt}\pgfcircle[fill]{\pgfxy(146,136)}{2.0pt}\pgfsetlinewidth{1.0pt}
\pgfmoveto{\pgfxy(140.0,170.0)}
\pgflineto{\pgfxy(152.0,170.0)}
\pgflineto{\pgfxy(146.0,176.0)}
\pgflineto{\pgfxy(146.0,176.0)}
\pgfclosepath 
\pgfqstroke 
\pgfline{\pgfxy(146.0,166.0)}{\pgfxy(146.0,170.0)}
\pgfline{\pgfxy(16.0,46.0)}{\pgfxy(196.0,46.0)}
\pgfline{\pgfxy(176.0,96.0)}{\pgfxy(176.0,46.0)}
\pgfline{\pgfxy(36.0,96.0)}{\pgfxy(36.0,46.0)}
\begin{pgfmagnify}{1}{-1}
\pgfputat{\pgfxy(186,-26)}{\pgfbox[left,top]{$\overline{\textrm{BL}}$}}
\end{pgfmagnify}
\begin{pgfmagnify}{1}{-1}
\pgfputat{\pgfxy(96,-26)}{\pgfbox[left,top]{WL}}
\end{pgfmagnify}
\begin{pgfmagnify}{1}{-1}
\pgfputat{\pgfxy(6,-26)}{\pgfbox[left,top]{BL}}
\end{pgfmagnify}
\pgfsetlinewidth{0.33pt}
\pgfcircle[fill]{\pgfxy(16,116)}{2.0pt}\pgfcircle[fill]{\pgfxy(196,116)}{2.0pt}\pgfcircle[fill]{\pgfxy(176,46)}{2.0pt}\pgfcircle[fill]{\pgfxy(36,46)}{2.0pt}\pgfcircle[fill]{\pgfxy(66,116)}{2.0pt}\pgfcircle[fill]{\pgfxy(146,116)}{2.0pt}\pgfsetlinewidth{1.0pt}
\pgfline{\pgfxy(10.0,46.0)}{\pgfxy(16.0,46.0)}
\pgfellipse[stroke]{\pgfxy(8.0,46.0)}{\pgfxy(2.0,0)}{\pgfxy(0,2.0)}
\pgfline{\pgfxy(16.0,152.0)}{\pgfxy(16.0,146.0)}
\pgfellipse[stroke]{\pgfxy(16.0,154.0)}{\pgfxy(2.0,0)}{\pgfxy(0,2.0)}
\pgfline{\pgfxy(196.0,152.0)}{\pgfxy(196.0,146.0)}
\pgfellipse[stroke]{\pgfxy(196.0,154.0)}{\pgfxy(2.0,0)}{\pgfxy(0,2.0)}
\pgfline{\pgfxy(202.0,46.0)}{\pgfxy(196.0,46.0)}
\pgfellipse[stroke]{\pgfxy(204.0,46.0)}{\pgfxy(2.0,0)}{\pgfxy(0,2.0)}
\pgfline{\pgfxy(196.0,40.0)}{\pgfxy(196.0,46.0)}
\pgfellipse[stroke]{\pgfxy(196.0,38.0)}{\pgfxy(2.0,0)}{\pgfxy(0,2.0)}
\pgfline{\pgfxy(16.0,40.0)}{\pgfxy(16.0,46.0)}
\pgfellipse[stroke]{\pgfxy(16.0,38.0)}{\pgfxy(2.0,0)}{\pgfxy(0,2.0)}
\end{pgfmagnify}
\end{pgfpicture}}
    \vspace*{-5mm}
    \caption{}
    \label{fig:SRAM_cell}
  \end{subfigure}
  \begin{subfigure}[b]{0.5\linewidth}
    \centering
    \hspace*{-8mm}
    \scalebox{0.8}{\begin{pgfpicture}{0cm}{0cm}{266pt}{192pt}
\pgfsetxvec{\pgfpoint{1pt}{0pt}}
\pgfsetyvec{\pgfpoint{0pt}{1pt}}
\pgfsetroundjoin 
\pgfsetroundcap
\pgftranslateto{\pgfxy(0,192)}
\begin{pgfmagnify}{1}{-1}
\definecolor{layer0}{rgb}{0.0,0.0,0.0}
\definecolor{layer1}{rgb}{0.0,0.0,0.5}
\definecolor{layer2}{rgb}{1.0,0.0,0.0}
\definecolor{layer3}{rgb}{0.0,0.5,0.5}
\definecolor{layer4}{rgb}{1.0,0.78,0.0}
\definecolor{layer5}{rgb}{0.5,1.0,0.0}
\definecolor{layer6}{rgb}{0.0,1.0,1.0}
\definecolor{layer7}{rgb}{0.0,0.5,0.0}
\definecolor{layer8}{rgb}{0.6,0.8,0.2}
\definecolor{layer9}{rgb}{1.0,0.08,0.58}
\definecolor{layer10}{rgb}{0.71,0.61,0.05}
\definecolor{layer11}{rgb}{0.0,0.5,1.0}
\definecolor{layer12}{rgb}{0.88,0.88,0.88}
\definecolor{layer13}{rgb}{0.64,0.64,0.64}
\definecolor{layer14}{rgb}{0.37,0.37,0.37}
\definecolor{layer15}{rgb}{0.0,0.0,0.0}
\color{layer0}
\pgfsetlinewidth{1.0pt}
\pgfsetdash{}{0pt}
\pgfline{\pgfxy(96.0,36.0)}{\pgfxy(96.0,66.0)}
\pgfline{\pgfxy(116.0,26.0)}{\pgfxy(116.0,76.0)}
\pgfline{\pgfxy(156.0,26.0)}{\pgfxy(156.0,76.0)}
\pgfline{\pgfxy(176.0,36.0)}{\pgfxy(176.0,66.0)}
\pgfline{\pgfxy(156.0,46.0)}{\pgfxy(76.0,46.0)}
\pgfline{\pgfxy(116.0,56.0)}{\pgfxy(196.0,56.0)}
\pgfline{\pgfxy(96.0,86.0)}{\pgfxy(96.0,106.0)}
\pgfline{\pgfxy(176.0,86.0)}{\pgfxy(176.0,106.0)}
\pgfline{\pgfxy(176.0,126.0)}{\pgfxy(176.0,136.0)}
\pgfline{\pgfxy(176.0,136.0)}{\pgfxy(96.0,136.0)}
\pgfline{\pgfxy(96.0,136.0)}{\pgfxy(96.0,126.0)}
\pgfline{\pgfxy(136.0,156.0)}{\pgfxy(136.0,136.0)}
\pgfmoveto{\pgfxy(130.0,180.0)}
\pgflineto{\pgfxy(142.0,180.0)}
\pgflineto{\pgfxy(136.0,186.0)}
\pgflineto{\pgfxy(136.0,186.0)}
\pgfclosepath 
\pgfqstroke 
\pgfline{\pgfxy(136.0,176.0)}{\pgfxy(136.0,180.0)}
\pgfline{\pgfxy(96.0,16.0)}{\pgfxy(96.0,6.0)}
\pgfline{\pgfxy(116.0,6.0)}{\pgfxy(66.0,6.0)}
\pgfline{\pgfxy(76.0,6.0)}{\pgfxy(216.0,6.0)}
\pgfline{\pgfxy(176.0,16.0)}{\pgfxy(176.0,6.0)}
\pgfsetlinewidth{0.33pt}
\pgfcircle[fill]{\pgfxy(176,56)}{2.0pt}\pgfcircle[fill]{\pgfxy(116,56)}{2.0pt}\pgfcircle[fill]{\pgfxy(156,46)}{2.0pt}\pgfcircle[fill]{\pgfxy(96,46)}{2.0pt}\pgfsetlinewidth{1.0pt}
\pgfline{\pgfxy(76.0,116.0)}{\pgfxy(66.0,116.0)}
\pgfline{\pgfxy(196.0,116.0)}{\pgfxy(206.0,116.0)}
\pgfline{\pgfxy(76.0,16.0)}{\pgfxy(76.0,6.0)}
\pgfline{\pgfxy(196.0,16.0)}{\pgfxy(196.0,6.0)}
\pgfline{\pgfxy(46.0,46.0)}{\pgfxy(116.0,46.0)}
\pgfline{\pgfxy(176.0,56.0)}{\pgfxy(226.0,56.0)}
\pgfline{\pgfxy(196.0,36.0)}{\pgfxy(196.0,56.0)}
\pgfline{\pgfxy(76.0,36.0)}{\pgfxy(76.0,46.0)}
\pgfline{\pgfxy(226.0,56.0)}{\pgfxy(228.0,56.0)}
\pgfmoveto{\pgfxy(236.0,56.0)}
\pgflineto{\pgfxy(236.0,56.0)}
\pgflineto{\pgfxy(234.0,54.0)}
\pgflineto{\pgfxy(228.0,54.0)}
\pgflineto{\pgfxy(228.0,58.0)}
\pgflineto{\pgfxy(234.0,58.0)}
\pgfclosepath 
\pgfqstroke 
\pgfline{\pgfxy(46.0,46.0)}{\pgfxy(44.0,46.0)}
\pgfmoveto{\pgfxy(36.0,46.0)}
\pgflineto{\pgfxy(36.0,46.0)}
\pgflineto{\pgfxy(38.0,48.0)}
\pgflineto{\pgfxy(44.0,48.0)}
\pgflineto{\pgfxy(44.0,44.0)}
\pgflineto{\pgfxy(38.0,44.0)}
\pgfclosepath 
\pgfqstroke 
\pgfline{\pgfxy(208.0,116.0)}{\pgfxy(206.0,116.0)}
\pgfmoveto{\pgfxy(208.0,116.0)}
\pgflineto{\pgfxy(208.0,116.0)}
\pgflineto{\pgfxy(210.0,118.0)}
\pgflineto{\pgfxy(216.0,118.0)}
\pgflineto{\pgfxy(216.0,114.0)}
\pgflineto{\pgfxy(210.0,114.0)}
\pgfclosepath 
\pgfqstroke 
\pgfline{\pgfxy(64.0,116.0)}{\pgfxy(66.0,116.0)}
\pgfmoveto{\pgfxy(64.0,116.0)}
\pgflineto{\pgfxy(64.0,116.0)}
\pgflineto{\pgfxy(62.0,114.0)}
\pgflineto{\pgfxy(56.0,114.0)}
\pgflineto{\pgfxy(56.0,118.0)}
\pgflineto{\pgfxy(62.0,118.0)}
\pgfclosepath 
\pgfqstroke 
\begin{pgfmagnify}{1}{-1}
\pgfputat{\pgfxy(26,-16)}{\pgfbox[left,top]{EN}}
\end{pgfmagnify}
\begin{pgfmagnify}{1}{-1}
\pgfputat{\pgfxy(236,-16)}{\pgfbox[left,top]{EN}}
\end{pgfmagnify}
\begin{pgfmagnify}{1}{-1}
\pgfputat{\pgfxy(236,-46)}{\pgfbox[left,top]{SAO}}
\end{pgfmagnify}
\begin{pgfmagnify}{1}{-1}
\pgfputat{\pgfxy(6,-36)}{\pgfbox[left,top]{$\overline{\textrm{SAO}}$}}
\end{pgfmagnify}
\begin{pgfmagnify}{1}{-1}
\pgfputat{\pgfxy(226,-106)}{\pgfbox[left,top]{$\overline{\textrm{BL}}$}}
\end{pgfmagnify}
\begin{pgfmagnify}{1}{-1}
\pgfputat{\pgfxy(36,-106)}{\pgfbox[left,top]{BL}}
\end{pgfmagnify}
\begin{pgfmagnify}{1}{-1}
\pgfputat{\pgfxy(76,-156)}{\pgfbox[left,top]{EN}}
\end{pgfmagnify}
\pgfline{\pgfxy(86.0,122.0)}{\pgfxy(96.0,122.0)}
\pgfline{\pgfxy(86.0,110.0)}{\pgfxy(96.0,110.0)}
\pgfline{\pgfxy(96.0,110.0)}{\pgfxy(96.0,106.0)}
\pgfline{\pgfxy(96.0,122.0)}{\pgfxy(96.0,126.0)}
\pgfline{\pgfxy(76.0,116.0)}{\pgfxy(84.0,116.0)}
\pgfline{\pgfxy(84.0,110.0)}{\pgfxy(84.0,122.0)}
\pgfline{\pgfxy(86.0,108.0)}{\pgfxy(86.0,124.0)}
\pgfline{\pgfxy(186.0,110.0)}{\pgfxy(176.0,110.0)}
\pgfline{\pgfxy(186.0,122.0)}{\pgfxy(176.0,122.0)}
\pgfline{\pgfxy(176.0,122.0)}{\pgfxy(176.0,126.0)}
\pgfline{\pgfxy(176.0,110.0)}{\pgfxy(176.0,106.0)}
\pgfline{\pgfxy(196.0,116.0)}{\pgfxy(188.0,116.0)}
\pgfline{\pgfxy(188.0,122.0)}{\pgfxy(188.0,110.0)}
\pgfline{\pgfxy(186.0,124.0)}{\pgfxy(186.0,108.0)}
\pgfline{\pgfxy(126.0,172.0)}{\pgfxy(136.0,172.0)}
\pgfline{\pgfxy(126.0,160.0)}{\pgfxy(136.0,160.0)}
\pgfline{\pgfxy(136.0,160.0)}{\pgfxy(136.0,156.0)}
\pgfline{\pgfxy(136.0,172.0)}{\pgfxy(136.0,176.0)}
\pgfline{\pgfxy(116.0,166.0)}{\pgfxy(124.0,166.0)}
\pgfline{\pgfxy(124.0,160.0)}{\pgfxy(124.0,172.0)}
\pgfline{\pgfxy(126.0,158.0)}{\pgfxy(126.0,174.0)}
\pgfline{\pgfxy(106.0,70.0)}{\pgfxy(96.0,70.0)}
\pgfline{\pgfxy(106.0,82.0)}{\pgfxy(96.0,82.0)}
\pgfline{\pgfxy(96.0,82.0)}{\pgfxy(96.0,86.0)}
\pgfline{\pgfxy(96.0,70.0)}{\pgfxy(96.0,66.0)}
\pgfline{\pgfxy(116.0,76.0)}{\pgfxy(108.0,76.0)}
\pgfline{\pgfxy(108.0,82.0)}{\pgfxy(108.0,70.0)}
\pgfline{\pgfxy(106.0,84.0)}{\pgfxy(106.0,68.0)}
\pgfline{\pgfxy(166.0,82.0)}{\pgfxy(176.0,82.0)}
\pgfline{\pgfxy(166.0,70.0)}{\pgfxy(176.0,70.0)}
\pgfline{\pgfxy(176.0,70.0)}{\pgfxy(176.0,66.0)}
\pgfline{\pgfxy(176.0,82.0)}{\pgfxy(176.0,86.0)}
\pgfline{\pgfxy(156.0,76.0)}{\pgfxy(164.0,76.0)}
\pgfline{\pgfxy(164.0,70.0)}{\pgfxy(164.0,82.0)}
\pgfline{\pgfxy(166.0,68.0)}{\pgfxy(166.0,84.0)}
\pgfline{\pgfxy(166.0,32.0)}{\pgfxy(176.0,32.0)}
\pgfline{\pgfxy(166.0,20.0)}{\pgfxy(176.0,20.0)}
\pgfline{\pgfxy(176.0,20.0)}{\pgfxy(176.0,16.0)}
\pgfline{\pgfxy(176.0,32.0)}{\pgfxy(176.0,36.0)}
\pgfline{\pgfxy(156.0,26.0)}{\pgfxy(160.0,26.0)}
\pgfline{\pgfxy(164.0,20.0)}{\pgfxy(164.0,32.0)}
\pgfline{\pgfxy(166.0,18.0)}{\pgfxy(166.0,34.0)}
\pgfellipse[stroke]{\pgfxy(162.0,26.0)}{\pgfxy(2.0,0)}{\pgfxy(0,2.0)}
\pgfline{\pgfxy(106.0,20.0)}{\pgfxy(96.0,20.0)}
\pgfline{\pgfxy(106.0,32.0)}{\pgfxy(96.0,32.0)}
\pgfline{\pgfxy(96.0,32.0)}{\pgfxy(96.0,36.0)}
\pgfline{\pgfxy(96.0,20.0)}{\pgfxy(96.0,16.0)}
\pgfline{\pgfxy(116.0,26.0)}{\pgfxy(112.0,26.0)}
\pgfline{\pgfxy(108.0,32.0)}{\pgfxy(108.0,20.0)}
\pgfline{\pgfxy(106.0,34.0)}{\pgfxy(106.0,18.0)}
\pgfellipse[stroke]{\pgfxy(110.0,26.0)}{\pgfxy(2.0,0)}{\pgfxy(0,2.0)}
\pgfline{\pgfxy(66.0,32.0)}{\pgfxy(76.0,32.0)}
\pgfline{\pgfxy(66.0,20.0)}{\pgfxy(76.0,20.0)}
\pgfline{\pgfxy(76.0,20.0)}{\pgfxy(76.0,16.0)}
\pgfline{\pgfxy(76.0,32.0)}{\pgfxy(76.0,36.0)}
\pgfline{\pgfxy(56.0,26.0)}{\pgfxy(60.0,26.0)}
\pgfline{\pgfxy(64.0,20.0)}{\pgfxy(64.0,32.0)}
\pgfline{\pgfxy(66.0,18.0)}{\pgfxy(66.0,34.0)}
\pgfellipse[stroke]{\pgfxy(62.0,26.0)}{\pgfxy(2.0,0)}{\pgfxy(0,2.0)}
\pgfline{\pgfxy(206.0,20.0)}{\pgfxy(196.0,20.0)}
\pgfline{\pgfxy(206.0,32.0)}{\pgfxy(196.0,32.0)}
\pgfline{\pgfxy(196.0,32.0)}{\pgfxy(196.0,36.0)}
\pgfline{\pgfxy(196.0,20.0)}{\pgfxy(196.0,16.0)}
\pgfline{\pgfxy(216.0,26.0)}{\pgfxy(212.0,26.0)}
\pgfline{\pgfxy(208.0,32.0)}{\pgfxy(208.0,20.0)}
\pgfline{\pgfxy(206.0,34.0)}{\pgfxy(206.0,18.0)}
\pgfellipse[stroke]{\pgfxy(210.0,26.0)}{\pgfxy(2.0,0)}{\pgfxy(0,2.0)}
\pgfsetlinewidth{0.33pt}
\pgfcircle[fill]{\pgfxy(76,46)}{2.0pt}\pgfcircle[fill]{\pgfxy(196,56)}{2.0pt}\pgfsetlinewidth{1.0pt}
\pgfline{\pgfxy(218.0,26.0)}{\pgfxy(216.0,26.0)}
\pgfmoveto{\pgfxy(218.0,26.0)}
\pgflineto{\pgfxy(218.0,26.0)}
\pgflineto{\pgfxy(220.0,28.0)}
\pgflineto{\pgfxy(226.0,28.0)}
\pgflineto{\pgfxy(226.0,24.0)}
\pgflineto{\pgfxy(220.0,24.0)}
\pgfclosepath 
\pgfqstroke 
\pgfline{\pgfxy(54.0,26.0)}{\pgfxy(56.0,26.0)}
\pgfmoveto{\pgfxy(54.0,26.0)}
\pgflineto{\pgfxy(54.0,26.0)}
\pgflineto{\pgfxy(52.0,24.0)}
\pgflineto{\pgfxy(46.0,24.0)}
\pgflineto{\pgfxy(46.0,28.0)}
\pgflineto{\pgfxy(52.0,28.0)}
\pgfclosepath 
\pgfqstroke 
\pgfline{\pgfxy(114.0,166.0)}{\pgfxy(116.0,166.0)}
\pgfmoveto{\pgfxy(114.0,166.0)}
\pgflineto{\pgfxy(114.0,166.0)}
\pgflineto{\pgfxy(112.0,164.0)}
\pgflineto{\pgfxy(106.0,164.0)}
\pgflineto{\pgfxy(106.0,168.0)}
\pgflineto{\pgfxy(112.0,168.0)}
\pgfclosepath 
\pgfqstroke 
\pgfsetlinewidth{0.33pt}
\pgfcircle[fill]{\pgfxy(136,136)}{2.0pt}\end{pgfmagnify}
\end{pgfpicture}}
    \vspace*{-10mm}
    \caption{}
    \label{fig:SRAM_SA}
  \end{subfigure}
  \begin{subfigure}[b]{\linewidth}
    \centering
    \vspace*{2mm}
    \includegraphics[width=0.5\linewidth]{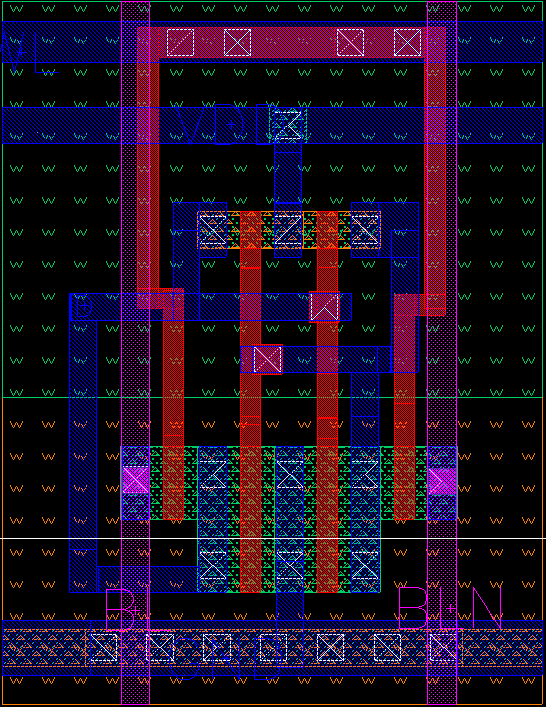}
    \caption{}
    \label{fig:SRAM_layout}
  \end{subfigure}
  \caption{The SRAM cell and SA. \textbf{(a)} The 6T cell. \textbf{(b)} The SRAM sense amplifer (SA) \cite{sram_sa}. \textbf{(c)} The SRAM cell layout.}
  \label{fig:SRAM_cell_and_SA}
\end{figure}

\medskip

For the SRAM array, the standard 6T cell memory cell is chosen (\autoref{fig:SRAM_cell}): since the aim of this work is to produce a memory architecture capable of performing logic operations, the cell area dedicated to the memory function is minimised by picking the design with the smallest cell footprint possible for the SRAM core. For what concerns the read sensing circuitry, a conventional voltage latch sense amplifier (SA) \cite{sram_sa} is chosen, which circuit is depicted in \autoref{fig:SRAM_SA}. A commonly adopted SA circuit topology is selected to compare the read operation performance among the memories, in order to understand how much the added complexity affects the standard memory operation of the array. This circuit provides a high sensing speed and low idle power consumption, which are due to the non linearity of the bi--stable ring used as latch.

\medskip

\begin{figure}[t!p]
  \begin{subfigure}[b]{\linewidth}
    \centering
    \scalebox{0.8}{\begin{pgfpicture}{0cm}{0cm}{182pt}{122pt}
\pgfsetxvec{\pgfpoint{1pt}{0pt}}
\pgfsetyvec{\pgfpoint{0pt}{1pt}}
\pgfsetroundjoin 
\pgfsetroundcap
\pgftranslateto{\pgfxy(0,122)}
\begin{pgfmagnify}{1}{-1}
\definecolor{layer0}{rgb}{0.0,0.0,0.0}
\definecolor{layer1}{rgb}{0.0,0.0,0.5}
\definecolor{layer2}{rgb}{1.0,0.0,0.0}
\definecolor{layer3}{rgb}{0.0,0.5,0.5}
\definecolor{layer4}{rgb}{1.0,0.78,0.0}
\definecolor{layer5}{rgb}{0.5,1.0,0.0}
\definecolor{layer6}{rgb}{0.0,1.0,1.0}
\definecolor{layer7}{rgb}{0.0,0.5,0.0}
\definecolor{layer8}{rgb}{0.6,0.8,0.2}
\definecolor{layer9}{rgb}{1.0,0.08,0.58}
\definecolor{layer10}{rgb}{0.71,0.61,0.05}
\definecolor{layer11}{rgb}{0.0,0.5,1.0}
\definecolor{layer12}{rgb}{0.88,0.88,0.88}
\definecolor{layer13}{rgb}{0.64,0.64,0.64}
\definecolor{layer14}{rgb}{0.37,0.37,0.37}
\definecolor{layer15}{rgb}{0.0,0.0,0.0}
\color{layer0}
\pgfsetlinewidth{1.0pt}
\pgfsetdash{}{0pt}
\pgfline{\pgfxy(108.0,46.0)}{\pgfxy(116.0,46.0)}
\pgfellipse[stroke]{\pgfxy(106.0,46.0)}{\pgfxy(2.0,0)}{\pgfxy(0,2.0)}
\pgfmoveto{\pgfxy(76.0,32.0)}
\pgflineto{\pgfxy(76.0,60.0)}
\pgflineto{\pgfxy(104.0,46.0)}
\pgfclosepath 
\pgfqstroke 
\pgfline{\pgfxy(66.0,46.0)}{\pgfxy(76.0,46.0)}
\pgfline{\pgfxy(74.0,76.0)}{\pgfxy(66.0,76.0)}
\pgfellipse[stroke]{\pgfxy(76.0,76.0)}{\pgfxy(2.0,0)}{\pgfxy(0,2.0)}
\pgfmoveto{\pgfxy(106.0,90.0)}
\pgflineto{\pgfxy(106.0,62.0)}
\pgflineto{\pgfxy(78.0,76.0)}
\pgfclosepath 
\pgfqstroke 
\pgfline{\pgfxy(116.0,76.0)}{\pgfxy(106.0,76.0)}
\pgfline{\pgfxy(66.0,46.0)}{\pgfxy(56.0,46.0)}
\pgfline{\pgfxy(56.0,46.0)}{\pgfxy(56.0,76.0)}
\pgfline{\pgfxy(56.0,76.0)}{\pgfxy(66.0,76.0)}
\pgfline{\pgfxy(116.0,46.0)}{\pgfxy(126.0,46.0)}
\pgfline{\pgfxy(126.0,46.0)}{\pgfxy(126.0,76.0)}
\pgfline{\pgfxy(126.0,76.0)}{\pgfxy(116.0,76.0)}
\pgfline{\pgfxy(136.0,52.0)}{\pgfxy(146.0,52.0)}
\pgfline{\pgfxy(136.0,40.0)}{\pgfxy(146.0,40.0)}
\pgfline{\pgfxy(146.0,40.0)}{\pgfxy(146.0,36.0)}
\pgfline{\pgfxy(146.0,52.0)}{\pgfxy(146.0,56.0)}
\pgfline{\pgfxy(126.0,46.0)}{\pgfxy(134.0,46.0)}
\pgfline{\pgfxy(134.0,40.0)}{\pgfxy(134.0,52.0)}
\pgfline{\pgfxy(136.0,38.0)}{\pgfxy(136.0,54.0)}
\pgfline{\pgfxy(46.0,40.0)}{\pgfxy(36.0,40.0)}
\pgfline{\pgfxy(46.0,52.0)}{\pgfxy(36.0,52.0)}
\pgfline{\pgfxy(36.0,52.0)}{\pgfxy(36.0,56.0)}
\pgfline{\pgfxy(36.0,40.0)}{\pgfxy(36.0,36.0)}
\pgfline{\pgfxy(56.0,46.0)}{\pgfxy(48.0,46.0)}
\pgfline{\pgfxy(48.0,52.0)}{\pgfxy(48.0,40.0)}
\pgfline{\pgfxy(46.0,54.0)}{\pgfxy(46.0,38.0)}
\pgfline{\pgfxy(26.0,92.0)}{\pgfxy(36.0,92.0)}
\pgfline{\pgfxy(26.0,80.0)}{\pgfxy(36.0,80.0)}
\pgfline{\pgfxy(36.0,80.0)}{\pgfxy(36.0,76.0)}
\pgfline{\pgfxy(36.0,92.0)}{\pgfxy(36.0,96.0)}
\pgfline{\pgfxy(16.0,86.0)}{\pgfxy(24.0,86.0)}
\pgfline{\pgfxy(24.0,80.0)}{\pgfxy(24.0,92.0)}
\pgfline{\pgfxy(26.0,78.0)}{\pgfxy(26.0,94.0)}
\pgfline{\pgfxy(156.0,80.0)}{\pgfxy(146.0,80.0)}
\pgfline{\pgfxy(156.0,92.0)}{\pgfxy(146.0,92.0)}
\pgfline{\pgfxy(146.0,92.0)}{\pgfxy(146.0,96.0)}
\pgfline{\pgfxy(146.0,80.0)}{\pgfxy(146.0,76.0)}
\pgfline{\pgfxy(166.0,86.0)}{\pgfxy(158.0,86.0)}
\pgfline{\pgfxy(158.0,92.0)}{\pgfxy(158.0,80.0)}
\pgfline{\pgfxy(156.0,94.0)}{\pgfxy(156.0,78.0)}
\pgfline{\pgfxy(36.0,56.0)}{\pgfxy(36.0,76.0)}
\pgfline{\pgfxy(36.0,36.0)}{\pgfxy(36.0,16.0)}
\pgfline{\pgfxy(146.0,16.0)}{\pgfxy(146.0,36.0)}
\pgfline{\pgfxy(166.0,16.0)}{\pgfxy(16.0,16.0)}
\pgfline{\pgfxy(146.0,76.0)}{\pgfxy(146.0,56.0)}
\pgfline{\pgfxy(166.0,66.0)}{\pgfxy(166.0,106.0)}
\pgfline{\pgfxy(16.0,106.0)}{\pgfxy(16.0,66.0)}
\pgfmoveto{\pgfxy(30.0,100.0)}
\pgflineto{\pgfxy(42.0,100.0)}
\pgflineto{\pgfxy(36.0,106.0)}
\pgflineto{\pgfxy(36.0,106.0)}
\pgfclosepath 
\pgfqstroke 
\pgfline{\pgfxy(36.0,96.0)}{\pgfxy(36.0,100.0)}
\pgfmoveto{\pgfxy(140.0,100.0)}
\pgflineto{\pgfxy(152.0,100.0)}
\pgflineto{\pgfxy(146.0,106.0)}
\pgflineto{\pgfxy(146.0,106.0)}
\pgfclosepath 
\pgfqstroke 
\pgfline{\pgfxy(146.0,96.0)}{\pgfxy(146.0,100.0)}
\begin{pgfmagnify}{1}{-1}
\pgfputat{\pgfxy(86,-6)}{\pgfbox[left,top]{ML}}
\end{pgfmagnify}
\begin{pgfmagnify}{1}{-1}
\pgfputat{\pgfxy(166,-46)}{\pgfbox[left,top]{$\overline{\textrm{BL}}$}}
\end{pgfmagnify}
\begin{pgfmagnify}{1}{-1}
\pgfputat{\pgfxy(126,-26)}{\pgfbox[left,top]{D}}
\end{pgfmagnify}
\begin{pgfmagnify}{1}{-1}
\pgfputat{\pgfxy(46,-26)}{\pgfbox[left,top]{$\overline{\textrm{D}}$}}
\end{pgfmagnify}
\begin{pgfmagnify}{1}{-1}
\pgfputat{\pgfxy(6,-46)}{\pgfbox[left,top]{BL}}
\end{pgfmagnify}
\pgfsetlinewidth{0.33pt}
\pgfcircle[fill]{\pgfxy(56,46)}{2.0pt}\pgfcircle[fill]{\pgfxy(126,46)}{2.0pt}\pgfcircle[fill]{\pgfxy(166,86)}{2.0pt}\pgfcircle[fill]{\pgfxy(16,86)}{2.0pt}\pgfcircle[fill]{\pgfxy(146,16)}{2.0pt}\pgfcircle[fill]{\pgfxy(36,16)}{2.0pt}\pgfsetlinewidth{1.0pt}
\pgfline{\pgfxy(166.0,60.0)}{\pgfxy(166.0,66.0)}
\pgfellipse[stroke]{\pgfxy(166.0,58.0)}{\pgfxy(2.0,0)}{\pgfxy(0,2.0)}
\pgfline{\pgfxy(166.0,112.0)}{\pgfxy(166.0,106.0)}
\pgfellipse[stroke]{\pgfxy(166.0,114.0)}{\pgfxy(2.0,0)}{\pgfxy(0,2.0)}
\pgfline{\pgfxy(16.0,112.0)}{\pgfxy(16.0,106.0)}
\pgfellipse[stroke]{\pgfxy(16.0,114.0)}{\pgfxy(2.0,0)}{\pgfxy(0,2.0)}
\pgfline{\pgfxy(16.0,60.0)}{\pgfxy(16.0,66.0)}
\pgfellipse[stroke]{\pgfxy(16.0,58.0)}{\pgfxy(2.0,0)}{\pgfxy(0,2.0)}
\pgfline{\pgfxy(10.0,16.0)}{\pgfxy(16.0,16.0)}
\pgfellipse[stroke]{\pgfxy(8.0,16.0)}{\pgfxy(2.0,0)}{\pgfxy(0,2.0)}
\pgfline{\pgfxy(172.0,16.0)}{\pgfxy(166.0,16.0)}
\pgfellipse[stroke]{\pgfxy(174.0,16.0)}{\pgfxy(2.0,0)}{\pgfxy(0,2.0)}
\end{pgfmagnify}
\end{pgfpicture}}
    \vspace*{-5mm}
    \caption{}
    \label{fig:CAM_cell}
  \end{subfigure}
  \begin{subfigure}[b]{\linewidth}
    \scalebox{0.8}{\input{Immagini/MLSA.pgf}}
    \vspace*{-5mm}
    \caption{}
    \label{fig:MLSA}
  \end{subfigure}
  \begin{subfigure}[b]{\linewidth}
    \centering
    \includegraphics[width=0.4\linewidth]{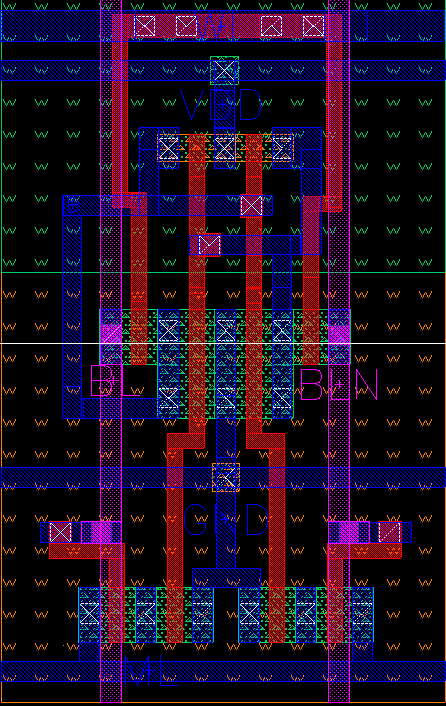}
    \caption{}
    \label{fig:CAM_layout}
  \end{subfigure}
  \caption{The CAM cell and MLSA. \textbf{(a)} Simplified schematic of the 10T NOR CAM cell \cite{cam_literature}. The access transistors of the SRAM core are omitted. \textbf{(b)} The matchline sense amplifier (MLSA) \cite{cam_sa}, that employs the current-save sensing scheme for the search operation of the CAM array. \textbf{(c)} The CAM cell layout.}
  \label{fig:CAM_cell_and_MLSA}
\end{figure}

For the CAM, a conventional NOR topology \cite{cam_literature} (\autoref{fig:CAM_cell}), is employed. For what concerns the CAM sensing circuitry, a current--saving scheme \cite{cam_sa} is selected among the possible ones \cite{cam_literature}. The correspondent matchline sense amplifier (MLSA) circuit is depicted in \autoref{fig:MLSA}. In CAM memories, this circuit is employed to reduce, with respect to the standard sensing scheme, the energy consumption associated to a search operation, thanks to the fact that the matchline (ML) is charged in case of match instead of being discharged when a mismatch occurs. In fact, it is well known that during a search operation in a NOR CAM array, the mismatch result is the most frequent one (only one or few words in the memory array match the searched one). By associating a matchline voltage commutation to the match result instead of the mismatch one, a large reduction in the energy consumption associated to the search operation is obtained, since only few lines experience a variation of their electric potential.

\smallskip

In \autoref{fig:MLSA}, an example of current--saving scheme \cite{cam_literature} is presented. This consists of a current source used to charge the matchline; when a match occurs, the matchline behaves as a capacitance; as a consequence, the capacitance gets charged resulting in a matchline voltage variation, and a match is registered in output. In case of a mismatch, instead, the ML connects the current source to ground and it does not get charged, preventing a variation in the matchline electric potential, which would lead to additional energy consumption. A feedback control circuit is employed to limit the current that is injected to ground in the mismatch case, in order to save power during the search operation; this circuit allows to deliver as few current as possible to the mismatch line, while providing the match ones with as much current as possible to speed up the match sensing. 

\medskip

\begin{figure}[t!p]
  \begin{subfigure}[b]{\linewidth}
    \centering
    \scalebox{0.7}{\begin{pgfpicture}{0cm}{0cm}{182pt}{202pt}
\pgfsetxvec{\pgfpoint{1pt}{0pt}}
\pgfsetyvec{\pgfpoint{0pt}{1pt}}
\pgfsetroundjoin 
\pgfsetroundcap
\pgftranslateto{\pgfxy(0,202)}
\begin{pgfmagnify}{1}{-1}
\definecolor{layer0}{rgb}{0.0,0.0,0.0}
\definecolor{layer1}{rgb}{0.0,0.0,0.5}
\definecolor{layer2}{rgb}{1.0,0.0,0.0}
\definecolor{layer3}{rgb}{0.0,0.5,0.5}
\definecolor{layer4}{rgb}{1.0,0.78,0.0}
\definecolor{layer5}{rgb}{0.5,1.0,0.0}
\definecolor{layer6}{rgb}{0.0,1.0,1.0}
\definecolor{layer7}{rgb}{0.0,0.5,0.0}
\definecolor{layer8}{rgb}{0.6,0.8,0.2}
\definecolor{layer9}{rgb}{1.0,0.08,0.58}
\definecolor{layer10}{rgb}{0.71,0.61,0.05}
\definecolor{layer11}{rgb}{0.0,0.5,1.0}
\definecolor{layer12}{rgb}{0.88,0.88,0.88}
\definecolor{layer13}{rgb}{0.64,0.64,0.64}
\definecolor{layer14}{rgb}{0.37,0.37,0.37}
\definecolor{layer15}{rgb}{0.0,0.0,0.0}
\color{layer0}
\pgfsetlinewidth{1.0pt}
\pgfsetdash{}{0pt}
\pgfline{\pgfxy(136.0,132.0)}{\pgfxy(146.0,132.0)}
\pgfline{\pgfxy(136.0,120.0)}{\pgfxy(146.0,120.0)}
\pgfline{\pgfxy(146.0,120.0)}{\pgfxy(146.0,116.0)}
\pgfline{\pgfxy(146.0,132.0)}{\pgfxy(146.0,136.0)}
\pgfline{\pgfxy(126.0,126.0)}{\pgfxy(134.0,126.0)}
\pgfline{\pgfxy(134.0,120.0)}{\pgfxy(134.0,132.0)}
\pgfline{\pgfxy(136.0,118.0)}{\pgfxy(136.0,134.0)}
\pgfline{\pgfxy(136.0,172.0)}{\pgfxy(146.0,172.0)}
\pgfline{\pgfxy(136.0,160.0)}{\pgfxy(146.0,160.0)}
\pgfline{\pgfxy(146.0,160.0)}{\pgfxy(146.0,156.0)}
\pgfline{\pgfxy(146.0,172.0)}{\pgfxy(146.0,176.0)}
\pgfline{\pgfxy(126.0,166.0)}{\pgfxy(134.0,166.0)}
\pgfline{\pgfxy(134.0,160.0)}{\pgfxy(134.0,172.0)}
\pgfline{\pgfxy(136.0,158.0)}{\pgfxy(136.0,174.0)}
\pgfline{\pgfxy(36.0,120.0)}{\pgfxy(26.0,120.0)}
\pgfline{\pgfxy(36.0,132.0)}{\pgfxy(26.0,132.0)}
\pgfline{\pgfxy(26.0,132.0)}{\pgfxy(26.0,136.0)}
\pgfline{\pgfxy(26.0,120.0)}{\pgfxy(26.0,116.0)}
\pgfline{\pgfxy(46.0,126.0)}{\pgfxy(38.0,126.0)}
\pgfline{\pgfxy(38.0,132.0)}{\pgfxy(38.0,120.0)}
\pgfline{\pgfxy(36.0,134.0)}{\pgfxy(36.0,118.0)}
\pgfline{\pgfxy(36.0,160.0)}{\pgfxy(26.0,160.0)}
\pgfline{\pgfxy(36.0,172.0)}{\pgfxy(26.0,172.0)}
\pgfline{\pgfxy(26.0,172.0)}{\pgfxy(26.0,176.0)}
\pgfline{\pgfxy(26.0,160.0)}{\pgfxy(26.0,156.0)}
\pgfline{\pgfxy(46.0,166.0)}{\pgfxy(38.0,166.0)}
\pgfline{\pgfxy(38.0,172.0)}{\pgfxy(38.0,160.0)}
\pgfline{\pgfxy(36.0,174.0)}{\pgfxy(36.0,158.0)}
\pgfline{\pgfxy(26.0,176.0)}{\pgfxy(26.0,186.0)}
\pgfline{\pgfxy(26.0,186.0)}{\pgfxy(146.0,186.0)}
\pgfline{\pgfxy(146.0,186.0)}{\pgfxy(146.0,176.0)}
\pgfline{\pgfxy(26.0,156.0)}{\pgfxy(26.0,136.0)}
\pgfline{\pgfxy(26.0,116.0)}{\pgfxy(26.0,96.0)}
\pgfline{\pgfxy(16.0,96.0)}{\pgfxy(156.0,96.0)}
\pgfline{\pgfxy(146.0,116.0)}{\pgfxy(146.0,96.0)}
\pgfline{\pgfxy(146.0,156.0)}{\pgfxy(146.0,136.0)}
\pgfsetlinewidth{0.33pt}
\pgfcircle[fill]{\pgfxy(146,96)}{2.0pt}\pgfcircle[fill]{\pgfxy(26,96)}{2.0pt}\pgfsetlinewidth{1.0pt}
\pgfmoveto{\pgfxy(80.0,190.0)}
\pgflineto{\pgfxy(92.0,190.0)}
\pgflineto{\pgfxy(86.0,196.0)}
\pgflineto{\pgfxy(86.0,196.0)}
\pgfclosepath 
\pgfqstroke 
\pgfline{\pgfxy(86.0,186.0)}{\pgfxy(86.0,190.0)}
\pgfline{\pgfxy(46.0,166.0)}{\pgfxy(56.0,166.0)}
\pgfline{\pgfxy(46.0,156.0)}{\pgfxy(66.0,156.0)}
\pgfline{\pgfxy(126.0,166.0)}{\pgfxy(116.0,166.0)}
\pgfline{\pgfxy(126.0,156.0)}{\pgfxy(106.0,156.0)}
\begin{pgfmagnify}{1}{-1}
\pgfputat{\pgfxy(86,-86)}{\pgfbox[left,top]{ML}}
\end{pgfmagnify}
\pgfline{\pgfxy(162.0,96.0)}{\pgfxy(156.0,96.0)}
\pgfellipse[stroke]{\pgfxy(164.0,96.0)}{\pgfxy(2.0,0)}{\pgfxy(0,2.0)}
\pgfline{\pgfxy(10.0,96.0)}{\pgfxy(16.0,96.0)}
\pgfellipse[stroke]{\pgfxy(8.0,96.0)}{\pgfxy(2.0,0)}{\pgfxy(0,2.0)}
\pgfline{\pgfxy(116.0,72.0)}{\pgfxy(126.0,72.0)}
\pgfline{\pgfxy(116.0,60.0)}{\pgfxy(126.0,60.0)}
\pgfline{\pgfxy(126.0,60.0)}{\pgfxy(126.0,56.0)}
\pgfline{\pgfxy(126.0,72.0)}{\pgfxy(126.0,76.0)}
\pgfline{\pgfxy(106.0,66.0)}{\pgfxy(114.0,66.0)}
\pgfline{\pgfxy(114.0,60.0)}{\pgfxy(114.0,72.0)}
\pgfline{\pgfxy(116.0,58.0)}{\pgfxy(116.0,74.0)}
\pgfline{\pgfxy(150.0,46.0)}{\pgfxy(150.0,56.0)}
\pgfline{\pgfxy(162.0,46.0)}{\pgfxy(162.0,56.0)}
\pgfline{\pgfxy(162.0,56.0)}{\pgfxy(166.0,56.0)}
\pgfline{\pgfxy(150.0,56.0)}{\pgfxy(146.0,56.0)}
\pgfline{\pgfxy(156.0,36.0)}{\pgfxy(156.0,44.0)}
\pgfline{\pgfxy(162.0,44.0)}{\pgfxy(150.0,44.0)}
\pgfline{\pgfxy(164.0,46.0)}{\pgfxy(148.0,46.0)}
\pgfline{\pgfxy(20.0,46.0)}{\pgfxy(20.0,56.0)}
\pgfline{\pgfxy(32.0,46.0)}{\pgfxy(32.0,56.0)}
\pgfline{\pgfxy(32.0,56.0)}{\pgfxy(36.0,56.0)}
\pgfline{\pgfxy(20.0,56.0)}{\pgfxy(16.0,56.0)}
\pgfline{\pgfxy(26.0,36.0)}{\pgfxy(26.0,44.0)}
\pgfline{\pgfxy(32.0,44.0)}{\pgfxy(20.0,44.0)}
\pgfline{\pgfxy(34.0,46.0)}{\pgfxy(18.0,46.0)}
\pgfline{\pgfxy(66.0,40.0)}{\pgfxy(56.0,40.0)}
\pgfline{\pgfxy(66.0,52.0)}{\pgfxy(56.0,52.0)}
\pgfline{\pgfxy(56.0,52.0)}{\pgfxy(56.0,56.0)}
\pgfline{\pgfxy(56.0,40.0)}{\pgfxy(56.0,36.0)}
\pgfline{\pgfxy(76.0,46.0)}{\pgfxy(72.0,46.0)}
\pgfline{\pgfxy(68.0,52.0)}{\pgfxy(68.0,40.0)}
\pgfline{\pgfxy(66.0,54.0)}{\pgfxy(66.0,38.0)}
\pgfellipse[stroke]{\pgfxy(70.0,46.0)}{\pgfxy(2.0,0)}{\pgfxy(0,2.0)}
\pgfline{\pgfxy(46.0,26.0)}{\pgfxy(66.0,26.0)}
\pgfline{\pgfxy(56.0,36.0)}{\pgfxy(56.0,26.0)}
\pgfline{\pgfxy(56.0,56.0)}{\pgfxy(36.0,56.0)}
\pgfline{\pgfxy(126.0,56.0)}{\pgfxy(146.0,56.0)}
\pgfline{\pgfxy(26.0,36.0)}{\pgfxy(26.0,16.0)}
\pgfline{\pgfxy(26.0,16.0)}{\pgfxy(156.0,16.0)}
\pgfline{\pgfxy(156.0,16.0)}{\pgfxy(156.0,36.0)}
\pgfline{\pgfxy(156.0,16.0)}{\pgfxy(166.0,16.0)}
\pgfline{\pgfxy(26.0,16.0)}{\pgfxy(16.0,16.0)}
\pgfsetlinewidth{0.33pt}
\pgfcircle[fill]{\pgfxy(26,16)}{2.0pt}\pgfcircle[fill]{\pgfxy(156,16)}{2.0pt}\pgfsetlinewidth{1.0pt}
\pgfline{\pgfxy(172.0,56.0)}{\pgfxy(166.0,56.0)}
\pgfellipse[stroke]{\pgfxy(174.0,56.0)}{\pgfxy(2.0,0)}{\pgfxy(0,2.0)}
\pgfline{\pgfxy(10.0,56.0)}{\pgfxy(16.0,56.0)}
\pgfellipse[stroke]{\pgfxy(8.0,56.0)}{\pgfxy(2.0,0)}{\pgfxy(0,2.0)}
\pgfmoveto{\pgfxy(120.0,80.0)}
\pgflineto{\pgfxy(132.0,80.0)}
\pgflineto{\pgfxy(126.0,86.0)}
\pgflineto{\pgfxy(126.0,86.0)}
\pgfclosepath 
\pgfqstroke 
\pgfline{\pgfxy(126.0,76.0)}{\pgfxy(126.0,80.0)}
\pgfmoveto{\pgfxy(80.0,52.0)}
\pgflineto{\pgfxy(80.0,40.0)}
\pgflineto{\pgfxy(86.0,46.0)}
\pgflineto{\pgfxy(86.0,46.0)}
\pgfclosepath 
\pgfqstroke 
\pgfline{\pgfxy(76.0,46.0)}{\pgfxy(80.0,46.0)}
\pgfline{\pgfxy(106.0,56.0)}{\pgfxy(86.0,56.0)}
\pgfline{\pgfxy(106.0,66.0)}{\pgfxy(96.0,66.0)}
\pgfmoveto{\pgfxy(122.0,120.0)}
\pgflineto{\pgfxy(122.0,132.0)}
\pgflineto{\pgfxy(116.0,126.0)}
\pgflineto{\pgfxy(116.0,126.0)}
\pgfclosepath 
\pgfqstroke 
\pgfline{\pgfxy(126.0,126.0)}{\pgfxy(122.0,126.0)}
\pgfmoveto{\pgfxy(50.0,132.0)}
\pgflineto{\pgfxy(50.0,120.0)}
\pgflineto{\pgfxy(56.0,126.0)}
\pgflineto{\pgfxy(56.0,126.0)}
\pgfclosepath 
\pgfqstroke 
\pgfline{\pgfxy(46.0,126.0)}{\pgfxy(50.0,126.0)}
\pgfline{\pgfxy(96.0,66.0)}{\pgfxy(96.0,56.0)}
\pgfline{\pgfxy(116.0,166.0)}{\pgfxy(116.0,156.0)}
\pgfline{\pgfxy(56.0,166.0)}{\pgfxy(56.0,156.0)}
\pgfline{\pgfxy(172.0,16.0)}{\pgfxy(166.0,16.0)}
\pgfellipse[stroke]{\pgfxy(174.0,16.0)}{\pgfxy(2.0,0)}{\pgfxy(0,2.0)}
\pgfline{\pgfxy(10.0,16.0)}{\pgfxy(16.0,16.0)}
\pgfellipse[stroke]{\pgfxy(8.0,16.0)}{\pgfxy(2.0,0)}{\pgfxy(0,2.0)}
\begin{pgfmagnify}{1}{-1}
\pgfputat{\pgfxy(86,-6)}{\pgfbox[left,top]{WL}}
\end{pgfmagnify}
\end{pgfmagnify}
\end{pgfpicture}}
    \caption{}
    \label{fig:CAM_dummy_cell}
  \end{subfigure}
  \begin{subfigure}[b]{\linewidth}
    \centering
    \hspace*{3mm}
    \scalebox{0.7}{\input{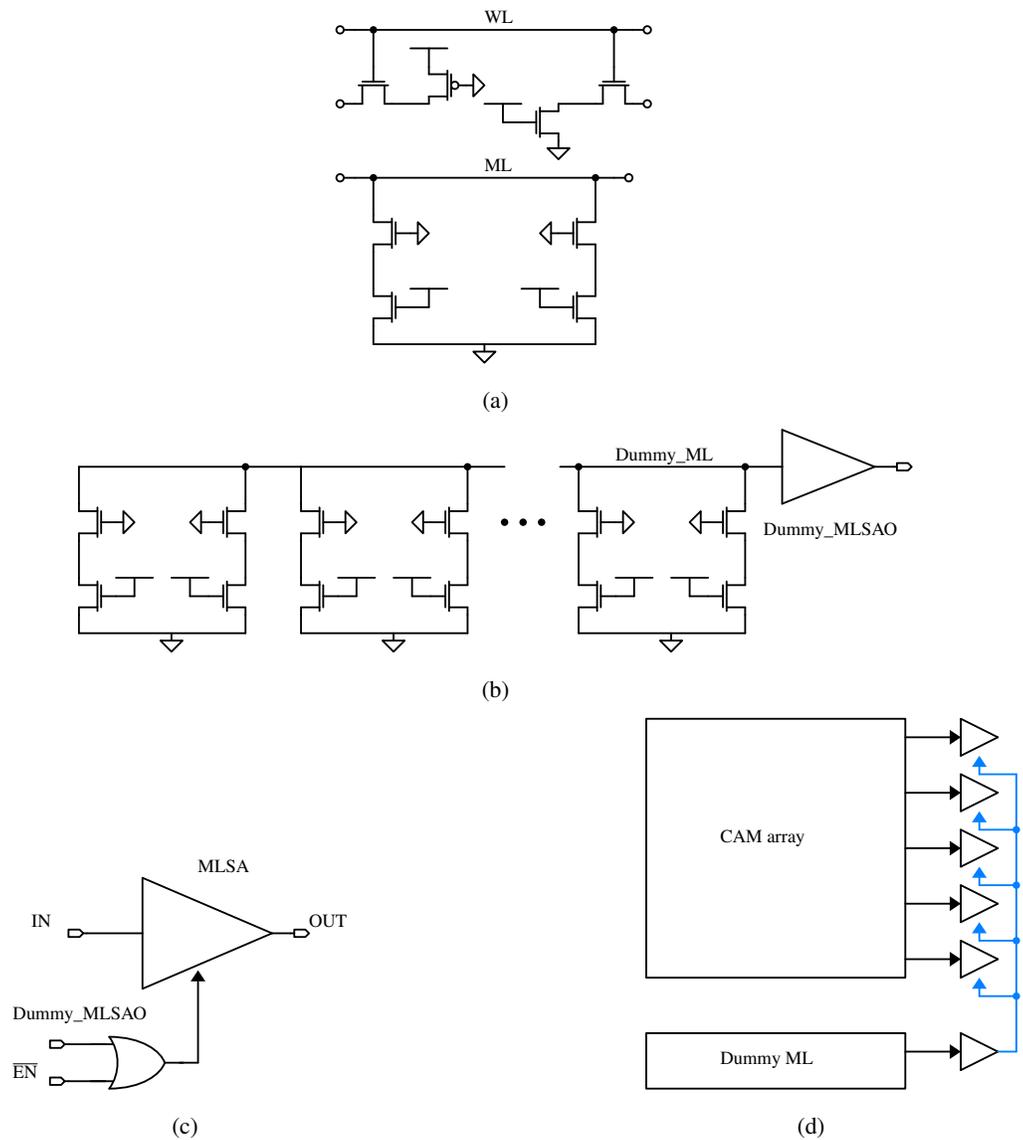}}
    \caption{}
    \label{fig:CAM_dummy_ML}
  \end{subfigure}
  \begin{subfigure}[b]{0.5\linewidth}
    \centering
    \scalebox{0.7}{\begin{pgfpicture}{0cm}{0cm}{196pt}{136pt}
\pgfsetxvec{\pgfpoint{1pt}{0pt}}
\pgfsetyvec{\pgfpoint{0pt}{1pt}}
\pgfsetroundjoin 
\pgfsetroundcap
\pgftranslateto{\pgfxy(0,136)}
\begin{pgfmagnify}{1}{-1}
\definecolor{layer0}{rgb}{0.0,0.0,0.0}
\definecolor{layer1}{rgb}{0.0,0.0,0.5}
\definecolor{layer2}{rgb}{1.0,0.0,0.0}
\definecolor{layer3}{rgb}{0.0,0.5,0.5}
\definecolor{layer4}{rgb}{1.0,0.78,0.0}
\definecolor{layer5}{rgb}{0.5,1.0,0.0}
\definecolor{layer6}{rgb}{0.0,1.0,1.0}
\definecolor{layer7}{rgb}{0.0,0.5,0.0}
\definecolor{layer8}{rgb}{0.6,0.8,0.2}
\definecolor{layer9}{rgb}{1.0,0.08,0.58}
\definecolor{layer10}{rgb}{0.71,0.61,0.05}
\definecolor{layer11}{rgb}{0.0,0.5,1.0}
\definecolor{layer12}{rgb}{0.88,0.88,0.88}
\definecolor{layer13}{rgb}{0.64,0.64,0.64}
\definecolor{layer14}{rgb}{0.37,0.37,0.37}
\definecolor{layer15}{rgb}{0.0,0.0,0.0}
\color{layer0}
\pgfsetlinewidth{1.0pt}
\pgfsetdash{}{0pt}
\pgfline{\pgfxy(76.0,16.0)}{\pgfxy(76.0,76.0)}
\pgfline{\pgfxy(76.0,76.0)}{\pgfxy(146.0,46.0)}
\pgfline{\pgfxy(146.0,46.0)}{\pgfxy(76.0,16.0)}
\pgfline{\pgfxy(96.0,116.0)}{\pgfxy(106.0,116.0)}
\pgfmoveto{\pgfxy(106.0,66.0)}
\pgflineto{\pgfxy(110.0,72.0)}
\pgflineto{\pgfxy(102.0,72.0)}
\pgfclosepath 
\pgffill 
\pgfline{\pgfxy(106.0,116.0)}{\pgfxy(106.0,72.0)}
\pgfline{\pgfxy(56.0,126.0)}{\pgfxy(36.0,126.0)}
\pgfline{\pgfxy(56.0,106.0)}{\pgfxy(36.0,106.0)}
\begin{pgfmagnify}{1}{-1}
\pgfputat{\pgfxy(6,-86)}{\pgfbox[left,top]{Dummy\_MLSAO}}
\end{pgfmagnify}
\begin{pgfmagnify}{1}{-1}
\pgfputat{\pgfxy(6,-116)}{\pgfbox[left,top]{$\overline{\textrm{EN}}$}}
\end{pgfmagnify}
\pgfline{\pgfxy(46.0,46.0)}{\pgfxy(76.0,46.0)}
\begin{pgfmagnify}{1}{-1}
\pgfputat{\pgfxy(16,-36)}{\pgfbox[left,top]{IN}}
\end{pgfmagnify}
\begin{pgfmagnify}{1}{-1}
\pgfputat{\pgfxy(166,-36)}{\pgfbox[left,top]{OUT}}
\end{pgfmagnify}
\pgfmoveto{\pgfxy(58,102)} 
\pgfcurveto{\pgfxy(76,102)}{\pgfxy(86,110)}{\pgfxy(88,116)}
\pgfstroke
\pgfmoveto{\pgfxy(58,130)} 
\pgfcurveto{\pgfxy(76,130)}{\pgfxy(86,122)}{\pgfxy(88,116)}
\pgfstroke
\pgfmoveto{\pgfxy(58,102)} 
\pgfcurveto{\pgfxy(64,110)}{\pgfxy(64,122)}{\pgfxy(58,130)}
\pgfstroke
\pgfline{\pgfxy(48.0,106.0)}{\pgfxy(60.0,106.0)}
\pgfline{\pgfxy(48.0,126.0)}{\pgfxy(60.0,126.0)}
\pgfline{\pgfxy(88.0,116.0)}{\pgfxy(98.0,116.0)}
\begin{pgfmagnify}{1}{-1}
\pgfputat{\pgfxy(106,-6)}{\pgfbox[left,top]{MLSA}}
\end{pgfmagnify}
\pgfline{\pgfxy(44.0,46.0)}{\pgfxy(46.0,46.0)}
\pgfmoveto{\pgfxy(44.0,46.0)}
\pgflineto{\pgfxy(44.0,46.0)}
\pgflineto{\pgfxy(42.0,44.0)}
\pgflineto{\pgfxy(36.0,44.0)}
\pgflineto{\pgfxy(36.0,48.0)}
\pgflineto{\pgfxy(42.0,48.0)}
\pgfclosepath 
\pgfqstroke 
\pgfline{\pgfxy(34.0,106.0)}{\pgfxy(36.0,106.0)}
\pgfmoveto{\pgfxy(34.0,106.0)}
\pgflineto{\pgfxy(34.0,106.0)}
\pgflineto{\pgfxy(32.0,104.0)}
\pgflineto{\pgfxy(26.0,104.0)}
\pgflineto{\pgfxy(26.0,108.0)}
\pgflineto{\pgfxy(32.0,108.0)}
\pgfclosepath 
\pgfqstroke 
\pgfline{\pgfxy(34.0,126.0)}{\pgfxy(36.0,126.0)}
\pgfmoveto{\pgfxy(34.0,126.0)}
\pgflineto{\pgfxy(34.0,126.0)}
\pgflineto{\pgfxy(32.0,124.0)}
\pgflineto{\pgfxy(26.0,124.0)}
\pgflineto{\pgfxy(26.0,128.0)}
\pgflineto{\pgfxy(32.0,128.0)}
\pgfclosepath 
\pgfqstroke 
\pgfline{\pgfxy(156.0,46.0)}{\pgfxy(158.0,46.0)}
\pgfmoveto{\pgfxy(166.0,46.0)}
\pgflineto{\pgfxy(166.0,46.0)}
\pgflineto{\pgfxy(164.0,44.0)}
\pgflineto{\pgfxy(158.0,44.0)}
\pgflineto{\pgfxy(158.0,48.0)}
\pgflineto{\pgfxy(164.0,48.0)}
\pgfclosepath 
\pgfqstroke 
\pgfline{\pgfxy(146.0,46.0)}{\pgfxy(156.0,46.0)}
\end{pgfmagnify}
\end{pgfpicture}}
    \caption{}
    \label{fig:CAM_dummy_enable}
  \end{subfigure}
  \begin{subfigure}[b]{0.5\linewidth}
    \centering
    \hspace*{5mm}
    \scalebox{0.7}{\begin{pgfpicture}{0cm}{0cm}{212pt}{212pt}
\pgfsetxvec{\pgfpoint{1pt}{0pt}}
\pgfsetyvec{\pgfpoint{0pt}{1pt}}
\pgfsetroundjoin 
\pgfsetroundcap
\pgftranslateto{\pgfxy(0,212)}
\begin{pgfmagnify}{1}{-1}
\definecolor{layer0}{rgb}{0.0,0.0,0.0}
\definecolor{layer1}{rgb}{0.0,0.0,0.5}
\definecolor{layer2}{rgb}{1.0,0.0,0.0}
\definecolor{layer3}{rgb}{0.0,0.5,0.5}
\definecolor{layer4}{rgb}{1.0,0.78,0.0}
\definecolor{layer5}{rgb}{0.5,1.0,0.0}
\definecolor{layer6}{rgb}{0.0,1.0,1.0}
\definecolor{layer7}{rgb}{0.0,0.5,0.0}
\definecolor{layer8}{rgb}{0.6,0.8,0.2}
\definecolor{layer9}{rgb}{1.0,0.08,0.58}
\definecolor{layer10}{rgb}{0.71,0.61,0.05}
\definecolor{layer11}{rgb}{0.0,0.5,1.0}
\definecolor{layer12}{rgb}{0.88,0.88,0.88}
\definecolor{layer13}{rgb}{0.64,0.64,0.64}
\definecolor{layer14}{rgb}{0.37,0.37,0.37}
\definecolor{layer15}{rgb}{0.0,0.0,0.0}
\color{layer0}
\pgfsetlinewidth{1.0pt}
\pgfsetdash{}{0pt}
\pgfmoveto{\pgfxy(6,6)}
\pgflineto{\pgfxy(146,6)}
\pgflineto{\pgfxy(146,146)}
\pgflineto{\pgfxy(6,146)}
\pgfclosepath 
\pgfqstroke 
\pgfmoveto{\pgfxy(6,176)}
\pgflineto{\pgfxy(146,176)}
\pgflineto{\pgfxy(146,206)}
\pgflineto{\pgfxy(6,206)}
\pgfclosepath 
\pgfqstroke 
\begin{pgfmagnify}{1}{-1}
\pgfputat{\pgfxy(46,-186)}{\pgfbox[left,top]{Dummy ML}}
\end{pgfmagnify}
\begin{pgfmagnify}{1}{-1}
\pgfputat{\pgfxy(46,-66)}{\pgfbox[left,top]{CAM array}}
\end{pgfmagnify}
\pgfline{\pgfxy(176.0,6.0)}{\pgfxy(176.0,26.0)}
\pgfline{\pgfxy(176.0,26.0)}{\pgfxy(196.0,16.0)}
\pgfline{\pgfxy(196.0,16.0)}{\pgfxy(176.0,6.0)}
\pgfline{\pgfxy(176.0,36.0)}{\pgfxy(176.0,56.0)}
\pgfline{\pgfxy(176.0,56.0)}{\pgfxy(196.0,46.0)}
\pgfline{\pgfxy(196.0,46.0)}{\pgfxy(176.0,36.0)}
\pgfline{\pgfxy(176.0,66.0)}{\pgfxy(196.0,76.0)}
\pgfline{\pgfxy(196.0,76.0)}{\pgfxy(176.0,86.0)}
\pgfline{\pgfxy(176.0,86.0)}{\pgfxy(176.0,66.0)}
\pgfline{\pgfxy(176.0,96.0)}{\pgfxy(196.0,106.0)}
\pgfline{\pgfxy(196.0,106.0)}{\pgfxy(176.0,116.0)}
\pgfline{\pgfxy(176.0,116.0)}{\pgfxy(176.0,96.0)}
\pgfline{\pgfxy(176.0,126.0)}{\pgfxy(196.0,136.0)}
\pgfline{\pgfxy(196.0,136.0)}{\pgfxy(176.0,146.0)}
\pgfline{\pgfxy(176.0,146.0)}{\pgfxy(176.0,126.0)}
\pgfmoveto{\pgfxy(176.0,16.0)}
\pgflineto{\pgfxy(170.0,20.0)}
\pgflineto{\pgfxy(170.0,12.0)}
\pgfclosepath 
\pgffill 
\pgfline{\pgfxy(146.0,16.0)}{\pgfxy(170.0,16.0)}
\pgfmoveto{\pgfxy(176.0,46.0)}
\pgflineto{\pgfxy(170.0,50.0)}
\pgflineto{\pgfxy(170.0,42.0)}
\pgfclosepath 
\pgffill 
\pgfline{\pgfxy(146.0,46.0)}{\pgfxy(170.0,46.0)}
\pgfmoveto{\pgfxy(176.0,76.0)}
\pgflineto{\pgfxy(170.0,80.0)}
\pgflineto{\pgfxy(170.0,72.0)}
\pgfclosepath 
\pgffill 
\pgfline{\pgfxy(146.0,76.0)}{\pgfxy(170.0,76.0)}
\pgfmoveto{\pgfxy(176.0,106.0)}
\pgflineto{\pgfxy(170.0,110.0)}
\pgflineto{\pgfxy(170.0,102.0)}
\pgfclosepath 
\pgffill 
\pgfline{\pgfxy(146.0,106.0)}{\pgfxy(170.0,106.0)}
\pgfmoveto{\pgfxy(176.0,136.0)}
\pgflineto{\pgfxy(170.0,140.0)}
\pgflineto{\pgfxy(170.0,132.0)}
\pgfclosepath 
\pgffill 
\pgfline{\pgfxy(146.0,136.0)}{\pgfxy(170.0,136.0)}
\pgfline{\pgfxy(176.0,176.0)}{\pgfxy(196.0,186.0)}
\pgfline{\pgfxy(196.0,186.0)}{\pgfxy(176.0,196.0)}
\pgfline{\pgfxy(176.0,196.0)}{\pgfxy(176.0,176.0)}
\pgfmoveto{\pgfxy(176.0,186.0)}
\pgflineto{\pgfxy(170.0,190.0)}
\pgflineto{\pgfxy(170.0,182.0)}
\pgfclosepath 
\pgffill 
\pgfline{\pgfxy(146.0,186.0)}{\pgfxy(170.0,186.0)}
\color{layer11}
\pgfmoveto{\pgfxy(186.0,146.0)}
\pgflineto{\pgfxy(190.0,152.0)}
\pgflineto{\pgfxy(182.0,152.0)}
\pgfclosepath 
\pgffill 
\pgfline{\pgfxy(186.0,156.0)}{\pgfxy(186.0,152.0)}
\pgfline{\pgfxy(206.0,156.0)}{\pgfxy(186.0,156.0)}
\pgfline{\pgfxy(206.0,186.0)}{\pgfxy(206.0,36.0)}
\pgfline{\pgfxy(196.0,186.0)}{\pgfxy(206.0,186.0)}
\pgfmoveto{\pgfxy(186.0,116.0)}
\pgflineto{\pgfxy(190.0,122.0)}
\pgflineto{\pgfxy(182.0,122.0)}
\pgfclosepath 
\pgffill 
\pgfline{\pgfxy(186.0,126.0)}{\pgfxy(186.0,122.0)}
\pgfline{\pgfxy(206.0,126.0)}{\pgfxy(186.0,126.0)}
\pgfmoveto{\pgfxy(186.0,86.0)}
\pgflineto{\pgfxy(190.0,92.0)}
\pgflineto{\pgfxy(182.0,92.0)}
\pgfclosepath 
\pgffill 
\pgfline{\pgfxy(186.0,96.0)}{\pgfxy(186.0,92.0)}
\pgfline{\pgfxy(206.0,96.0)}{\pgfxy(186.0,96.0)}
\pgfmoveto{\pgfxy(186.0,56.0)}
\pgflineto{\pgfxy(190.0,62.0)}
\pgflineto{\pgfxy(182.0,62.0)}
\pgfclosepath 
\pgffill 
\pgfline{\pgfxy(186.0,66.0)}{\pgfxy(186.0,62.0)}
\pgfline{\pgfxy(206.0,66.0)}{\pgfxy(186.0,66.0)}
\pgfmoveto{\pgfxy(186.0,26.0)}
\pgflineto{\pgfxy(190.0,32.0)}
\pgflineto{\pgfxy(182.0,32.0)}
\pgfclosepath 
\pgffill 
\pgfline{\pgfxy(186.0,36.0)}{\pgfxy(186.0,32.0)}
\pgfline{\pgfxy(206.0,36.0)}{\pgfxy(186.0,36.0)}
\pgfsetlinewidth{0.33pt}
\pgfcircle[fill]{\pgfxy(206,66)}{2.0pt}\pgfcircle[fill]{\pgfxy(206,96)}{2.0pt}\pgfcircle[fill]{\pgfxy(206,126)}{2.0pt}\pgfcircle[fill]{\pgfxy(206,156)}{2.0pt}\end{pgfmagnify}
\end{pgfpicture}}
    \caption{}
    \label{fig:Dummy_ML_scheme}
  \end{subfigure}
  \caption{The dummy matchline scheme. A dummy MLSA is used to disable the current sources of the real MLSAs to save power during a search operation. \textbf{(a)} The dummy cell of the CAM. Only the matchline and wordline transistors are kept in the circuit,  together with a dummy SRAM core that stores a logic `1'. \textbf{(b)} The dummy matchline. The dummy cells are arranged in a matchline of length equal to the memory width one, and that is connected to a dummy MLSA. Part of the dummy CAM cell is omitted for the sake of clarity. \textbf{(c)} The output of the dummy MLSA is used to disable the other MLSAs: as soon as the dummy MLSA output changes, it means that the time needed for the match sensing has passed, and the current sources of the real MLSAs can be disabled. In order to achieve this, an OR gate is added inside each MLSA, and its output is used as internal enable signal. \textbf{(d)} The output of the dummy MLSA is connected to all the other MLSAs. The position of the dummy matchline is critical: since the dummy MLSA determines the timing of the memory, the line position has to be associated to the worst case for the sensing delay.}
  \label{fig:Dummy_ML_architecture}
\end{figure}

In order to limit the conduction time of the MLSAs current sources, the circuit of the MLSA and the architecture are modified. To turn off all the current sources as soon as all the matchlines values are correctly sensed, i.e. all the matching lines are charged to the MLSAs input threshold, so that no current is wasted in the mismatch lines, the ``dummy matchline'' scheme, shown in \autoref{fig:Dummy_ML_architecture}, is employed.

In \autoref{fig:CAM_dummy_cell}, a dummy CAM cell is shown. This consists of a CAM cell from which all the transistors that are not connected to the matchline are removed. The gates electric potentials of the remaining MOSFETs are chosen so that the cell always provides a match, i.e. it behaves as a capacitance. In fact, since the result that involves a voltage variation on the line is the match one, the latter determines the search operation performance. 

In \autoref{fig:CAM_dummy_ML}, a dummy ML is shown. The dummy cells are arranged in a row, which is connected to an MLSA that provides in output a ``dummy match result'', denoted with $Dummy\_MLSAO$, at each search operation. This signal is used in the architecture to disable all the real MLSAs, as soon as a match is detected on the dummy ML.

In \autoref{fig:CAM_dummy_enable}, the circuit of the MLSA is depicted. An OR gate is added to each MLSA, and its output is used as an internal enable signal inside this. In particular, since the enable signal is low--active, the output of the OR gate should switch to `1' as soon as $Dummy\_MLSAO$ switches to `1', i.e. a match is detected on the matchline, in order to disable the MLSA current source. As a consequence, the global enable signal $\overline{EN}$ is connected using a logic OR with $Dummy\_MLSAO$. 

In \autoref{fig:Dummy_ML_scheme}, the whole CAM architecture is shown. As explained above, the output of the dummy MLSA is connected to all the MLSAs, together with the global enable signal. Since the dummy matchline sensing delay determines the time available to correctly sense the matchline potential for each MLSA, its position in the memory array is crucial for the circuit timing. This means that the worst-case delay has to be associated to the dummy matchline position, i.e. it has to be placed as far as possible from the enable signals drivers in the circuit. 

\begin{figure}
    \centering
    \includegraphics[width=\linewidth]{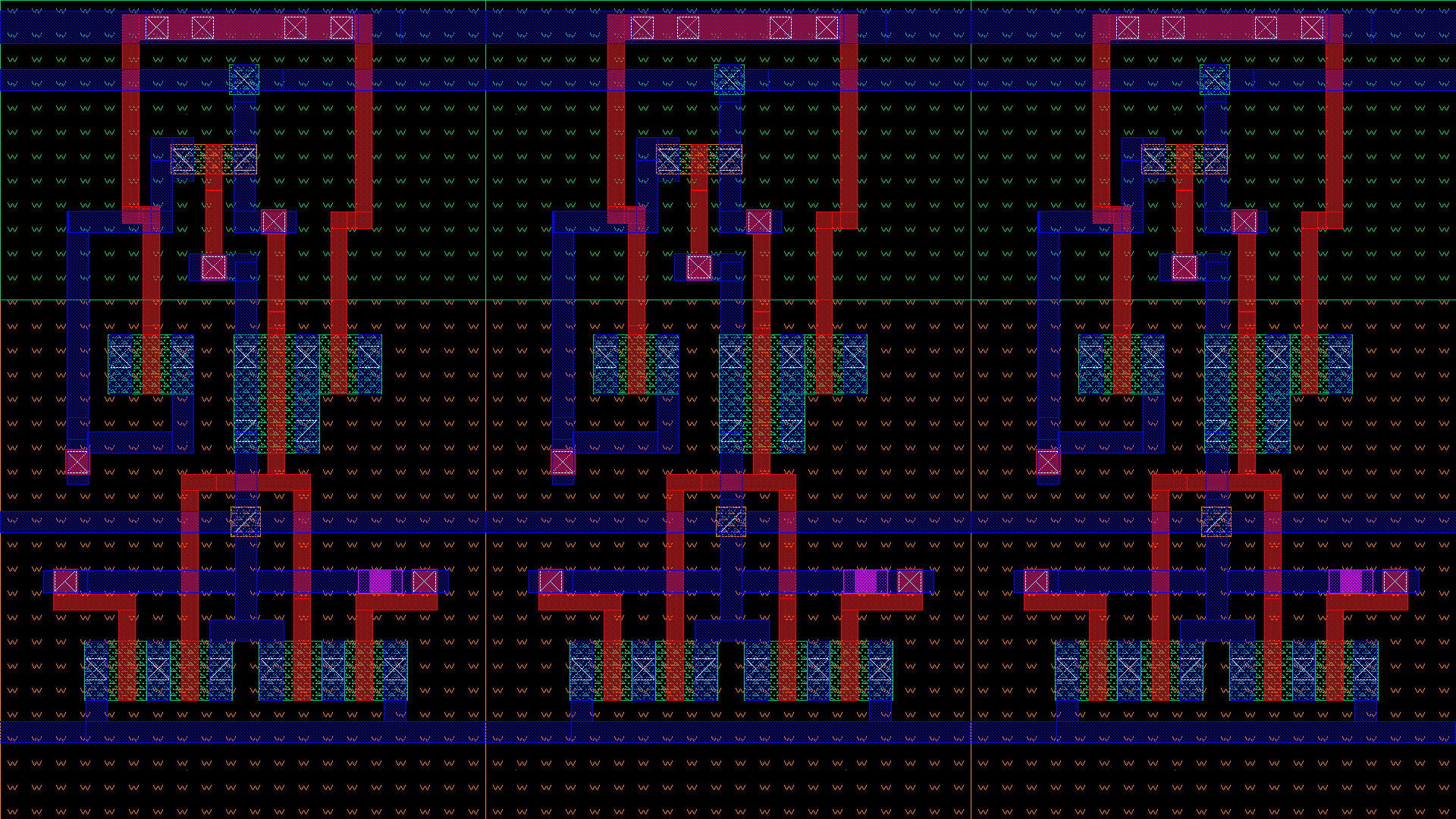}
    \caption{A layout section of the dummy line for the CAM architecture.}
    \label{fig:CAM_dummy_layout}
\end{figure}

In\autoref{fig:CAM_dummy_layout} it is shown a section of the layout for the dummy line. One can notice that some transistors are missing from the original layout depicted in \autoref{fig:CAM_layout}: in fact, the SRAM core is modified so that the cell stores a logic `1' without needing to explicitly write this value to each cell of the dummy line. 

\section{The LiM array}
\label{sec:LiM_array}

As a case of study, an architecture \cite{marco} for in--memory maximum/minimum computation designed by the authors is chosen, since it combines a general-purpose modification (bit wise in--memory AND logic operation) with a special--purpose near--memory logic circuitry for the maximum/minimum computation.

Therefore, it represents a good case of study to quantify the impact of this particular approach to in--memory computing, which is the goal of this work. The architecture is not intended as a CPU substitute, but as a hardware accelerator for particular tasks, such as the maximum/minimum computation or bit--wise memory operations.

\smallskip

The algorithm for in--memory maximum/minimum value search is based on the bitwise AND operation. All the words stored in memory are AND--ed with an external word called ``mask vector'', which is put on the memory bitlines one bit at a time until the whole word width is scanned; the results of these AND operations are then elaborated by the near--memory logic to choose the words to be discarded at each step, until only the maximum/minimum value remains. 

Consider the case in which unsigned words are stored in memory and the maximum value among these has to be found: in this case, at each iteration, only one bit of the mask is set to `1' starting from the MSB, and all the words for which the result of the AND is equal to `0' are discarded. In fact, if the bit of a word \textit{A} is equal to `0', while the same bit of a word \textit{B} is equal to `1', then \textit{B} is larger than \textit{A}; hence, \textit{A} is discarded from the search.

\begin{figure}[h]
  \centering
  \scalebox{0.95}{\input{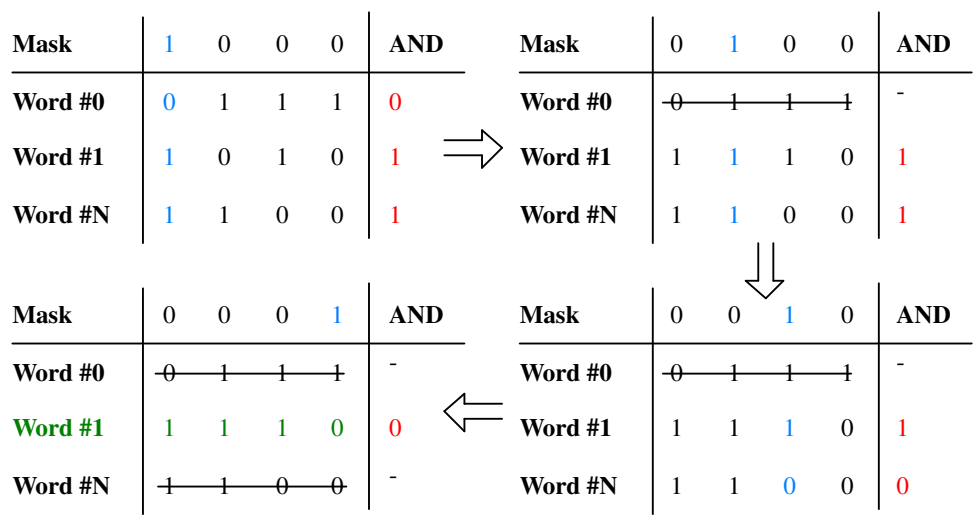}}
  \caption{All the words are scanned through a bitwise AND with an external word called ``mask vector''. The ones for which a logic `0' is obtained as result, are discarded; the remaining ones at the end are selected as maximum values, and a priority mechanism can be applied to choose among these. In the example, the selected word is highlighted in green. }
  \label{fig:Algorithm}
\end{figure}

An example of the maximum search for unsigned words is provided in \autoref{fig:Algorithm}. At each step, depending on the result of the AND operation, a word is discarded, until the whole memory width is processed or only one word remains. For minimum search and/or signed words, as well as other types of data encoding, it is enough to change the bits of the mask and to program the near--memory logic.

\medskip

\begin{figure}[h]
  \centering
  \includegraphics[scale=0.8]{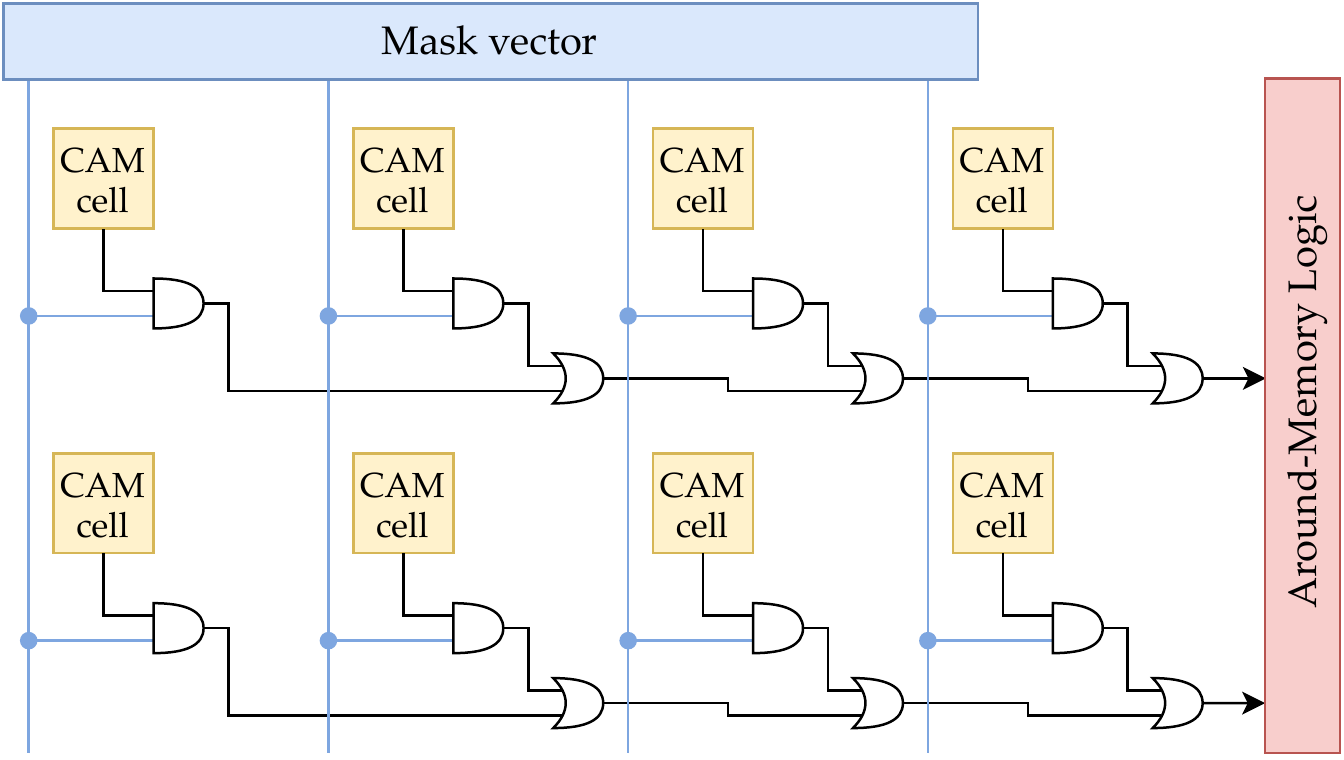}
  \caption{The LiM array. It consists of CAM cells to which the AND logic function capability is added; the results of the AND operations are OR--ed on the rows and provided to the near--memory logic.}
  \label{fig:LiM_array}
\end{figure}

The memory architecture consists of a standard NOR CAM, as the one presented in \autoref{sec:Reference_architectures}, to which the capability to perform the AND operation is added; the circuit is presented in \autoref{fig:LiM_array}. It has to be remarked that, in this work, only the LiM array schematic is presented, without including the near--memory logic circuitry that is described in \cite{marco}.

\begin{figure}[h]
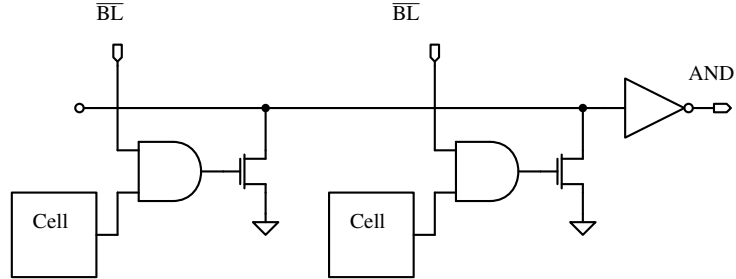

  \centering
  \scalebox{0.8}{\begin{pgfpicture}{0cm}{0cm}{352pt}{142pt}
\pgfsetxvec{\pgfpoint{1pt}{0pt}}
\pgfsetyvec{\pgfpoint{0pt}{1pt}}
\pgfsetroundjoin 
\pgfsetroundcap
\pgftranslateto{\pgfxy(0,142)}
\begin{pgfmagnify}{1}{-1}
\definecolor{layer0}{rgb}{0.0,0.0,0.0}
\definecolor{layer1}{rgb}{0.0,0.0,0.5}
\definecolor{layer2}{rgb}{1.0,0.0,0.0}
\definecolor{layer3}{rgb}{0.0,0.5,0.5}
\definecolor{layer4}{rgb}{1.0,0.78,0.0}
\definecolor{layer5}{rgb}{0.5,1.0,0.0}
\definecolor{layer6}{rgb}{0.0,1.0,1.0}
\definecolor{layer7}{rgb}{0.0,0.5,0.0}
\definecolor{layer8}{rgb}{0.6,0.8,0.2}
\definecolor{layer9}{rgb}{1.0,0.08,0.58}
\definecolor{layer10}{rgb}{0.71,0.61,0.05}
\definecolor{layer11}{rgb}{0.0,0.5,1.0}
\definecolor{layer12}{rgb}{0.88,0.88,0.88}
\definecolor{layer13}{rgb}{0.64,0.64,0.64}
\definecolor{layer14}{rgb}{0.37,0.37,0.37}
\definecolor{layer15}{rgb}{0.0,0.0,0.0}
\color{layer0}
\pgfsetlinewidth{1.0pt}
\pgfsetdash{}{0pt}
\pgfmoveto{\pgfxy(6,96)}
\pgflineto{\pgfxy(46,96)}
\pgflineto{\pgfxy(46,136)}
\pgflineto{\pgfxy(6,136)}
\pgfclosepath 
\pgfqstroke 
\pgfline{\pgfxy(56.0,76.0)}{\pgfxy(66.0,76.0)}
\pgfline{\pgfxy(66.0,72.0)}{\pgfxy(66.0,100.0)}
\pgfline{\pgfxy(56.0,96.0)}{\pgfxy(66.0,96.0)}
\pgfline{\pgfxy(66.0,100.0)}{\pgfxy(82.0,100.0)}
\pgfline{\pgfxy(66.0,72.0)}{\pgfxy(82.0,72.0)}
\pgfline{\pgfxy(96.0,86.0)}{\pgfxy(106.0,86.0)}
\pgfmoveto{\pgfxy(82,72)} 
\pgfcurveto{\pgfxy(100,72)}{\pgfxy(100,100)}{\pgfxy(82,100)}
\pgfstroke
\pgfline{\pgfxy(56.0,116.0)}{\pgfxy(46.0,116.0)}
\pgfline{\pgfxy(56.0,76.0)}{\pgfxy(56.0,36.0)}
\pgfline{\pgfxy(116.0,92.0)}{\pgfxy(126.0,92.0)}
\pgfline{\pgfxy(116.0,80.0)}{\pgfxy(126.0,80.0)}
\pgfline{\pgfxy(126.0,80.0)}{\pgfxy(126.0,76.0)}
\pgfline{\pgfxy(126.0,92.0)}{\pgfxy(126.0,96.0)}
\pgfline{\pgfxy(106.0,86.0)}{\pgfxy(114.0,86.0)}
\pgfline{\pgfxy(114.0,80.0)}{\pgfxy(114.0,92.0)}
\pgfline{\pgfxy(116.0,78.0)}{\pgfxy(116.0,94.0)}
\pgfline{\pgfxy(126.0,76.0)}{\pgfxy(126.0,56.0)}
\pgfline{\pgfxy(126.0,56.0)}{\pgfxy(286.0,56.0)}
\pgfmoveto{\pgfxy(156,96)}
\pgflineto{\pgfxy(196,96)}
\pgflineto{\pgfxy(196,136)}
\pgflineto{\pgfxy(156,136)}
\pgfclosepath 
\pgfqstroke 
\pgfline{\pgfxy(206.0,76.0)}{\pgfxy(216.0,76.0)}
\pgfline{\pgfxy(216.0,72.0)}{\pgfxy(216.0,100.0)}
\pgfline{\pgfxy(206.0,96.0)}{\pgfxy(216.0,96.0)}
\pgfline{\pgfxy(216.0,100.0)}{\pgfxy(232.0,100.0)}
\pgfline{\pgfxy(216.0,72.0)}{\pgfxy(232.0,72.0)}
\pgfline{\pgfxy(246.0,86.0)}{\pgfxy(256.0,86.0)}
\pgfmoveto{\pgfxy(232,72)} 
\pgfcurveto{\pgfxy(250,72)}{\pgfxy(250,100)}{\pgfxy(232,100)}
\pgfstroke
\pgfline{\pgfxy(206.0,96.0)}{\pgfxy(206.0,116.0)}
\pgfline{\pgfxy(206.0,76.0)}{\pgfxy(206.0,36.0)}
\pgfline{\pgfxy(266.0,92.0)}{\pgfxy(276.0,92.0)}
\pgfline{\pgfxy(266.0,80.0)}{\pgfxy(276.0,80.0)}
\pgfline{\pgfxy(276.0,80.0)}{\pgfxy(276.0,76.0)}
\pgfline{\pgfxy(276.0,92.0)}{\pgfxy(276.0,96.0)}
\pgfline{\pgfxy(256.0,86.0)}{\pgfxy(264.0,86.0)}
\pgfline{\pgfxy(264.0,80.0)}{\pgfxy(264.0,92.0)}
\pgfline{\pgfxy(266.0,78.0)}{\pgfxy(266.0,94.0)}
\pgfline{\pgfxy(276.0,76.0)}{\pgfxy(276.0,56.0)}
\pgfline{\pgfxy(276.0,96.0)}{\pgfxy(276.0,106.0)}
\pgfline{\pgfxy(126.0,96.0)}{\pgfxy(126.0,106.0)}
\pgfmoveto{\pgfxy(270.0,110.0)}
\pgflineto{\pgfxy(282.0,110.0)}
\pgflineto{\pgfxy(276.0,116.0)}
\pgflineto{\pgfxy(276.0,116.0)}
\pgfclosepath 
\pgfqstroke 
\pgfline{\pgfxy(276.0,106.0)}{\pgfxy(276.0,110.0)}
\pgfmoveto{\pgfxy(120.0,110.0)}
\pgflineto{\pgfxy(132.0,110.0)}
\pgflineto{\pgfxy(126.0,116.0)}
\pgflineto{\pgfxy(126.0,116.0)}
\pgfclosepath 
\pgfqstroke 
\pgfline{\pgfxy(126.0,106.0)}{\pgfxy(126.0,110.0)}
\pgfsetlinewidth{0.33pt}
\pgfcircle[fill]{\pgfxy(276,56)}{2.0pt}\pgfsetlinewidth{1.0pt}
\pgfline{\pgfxy(56.0,116.0)}{\pgfxy(56.0,96.0)}
\pgfline{\pgfxy(206.0,116.0)}{\pgfxy(196.0,116.0)}
\pgfline{\pgfxy(126.0,56.0)}{\pgfxy(46.0,56.0)}
\pgfsetlinewidth{0.33pt}
\pgfcircle[fill]{\pgfxy(126,56)}{2.0pt}\pgfsetlinewidth{1.0pt}
\pgfline{\pgfxy(328.0,56.0)}{\pgfxy(336.0,56.0)}
\pgfellipse[stroke]{\pgfxy(326.0,56.0)}{\pgfxy(2.0,0)}{\pgfxy(0,2.0)}
\pgfmoveto{\pgfxy(296.0,42.0)}
\pgflineto{\pgfxy(296.0,70.0)}
\pgflineto{\pgfxy(324.0,56.0)}
\pgfclosepath 
\pgfqstroke 
\pgfline{\pgfxy(286.0,56.0)}{\pgfxy(296.0,56.0)}
\pgfline{\pgfxy(336.0,56.0)}{\pgfxy(338.0,56.0)}
\pgfmoveto{\pgfxy(346.0,56.0)}
\pgflineto{\pgfxy(346.0,56.0)}
\pgflineto{\pgfxy(344.0,54.0)}
\pgflineto{\pgfxy(338.0,54.0)}
\pgflineto{\pgfxy(338.0,58.0)}
\pgflineto{\pgfxy(344.0,58.0)}
\pgfclosepath 
\pgfqstroke 
\begin{pgfmagnify}{1}{-1}
\pgfputat{\pgfxy(326,-36)}{\pgfbox[left,top]{AND}}
\end{pgfmagnify}
\pgfline{\pgfxy(206.0,34.0)}{\pgfxy(206.0,36.0)}
\pgfmoveto{\pgfxy(206.0,34.0)}
\pgflineto{\pgfxy(206.0,34.0)}
\pgflineto{\pgfxy(208.0,32.0)}
\pgflineto{\pgfxy(208.0,26.0)}
\pgflineto{\pgfxy(204.0,26.0)}
\pgflineto{\pgfxy(204.0,32.0)}
\pgfclosepath 
\pgfqstroke 
\pgfline{\pgfxy(56.0,34.0)}{\pgfxy(56.0,36.0)}
\pgfmoveto{\pgfxy(56.0,34.0)}
\pgflineto{\pgfxy(56.0,34.0)}
\pgflineto{\pgfxy(58.0,32.0)}
\pgflineto{\pgfxy(58.0,26.0)}
\pgflineto{\pgfxy(54.0,26.0)}
\pgflineto{\pgfxy(54.0,32.0)}
\pgfclosepath 
\pgfqstroke 
\begin{pgfmagnify}{1}{-1}
\pgfputat{\pgfxy(16,-106)}{\pgfbox[left,top]{Cell}}
\end{pgfmagnify}
\begin{pgfmagnify}{1}{-1}
\pgfputat{\pgfxy(166,-106)}{\pgfbox[left,top]{Cell}}
\end{pgfmagnify}
\begin{pgfmagnify}{1}{-1}
\pgfputat{\pgfxy(46,-6)}{\pgfbox[left,top]{$\overline{\textrm{BL}}$}}
\end{pgfmagnify}
\begin{pgfmagnify}{1}{-1}
\pgfputat{\pgfxy(186,-6)}{\pgfbox[left,top]{$\overline{\textrm{BL}}$}}
\end{pgfmagnify}
\pgfline{\pgfxy(40.0,56.0)}{\pgfxy(46.0,56.0)}
\pgfellipse[stroke]{\pgfxy(38.0,56.0)}{\pgfxy(2.0,0)}{\pgfxy(0,2.0)}
\end{pgfmagnify}
\end{pgfpicture}}
  \caption{The AND line. The AND gates outputs are connected to a wired--OR line through pull-down transistors. The signal on the line is then inverted to the AND result. The AND gates results are selected through $\overline{BL}$ by the mask.}
  \label{fig:AND_line}
\end{figure}

As previously explained, the AND operations between the memory content and the mask are performed in parallel on all the rows, one bit at time; then, the results of these are collected through OR operations on the rows by the sense amplifiers and provided to the peripheral logic circuits. Hence, the single cell includes two additional functionalities: AND and OR. The AND is a proper logic gate inserted into the cell, while the OR is implemented through a wired--OR line across the row, which result is handled on the periphery by a sense amplifier, denoted with ``ANDSA''. The AND line schematic is depicted in \autoref{fig:AND_line}.

\smallskip

To select the column on which the AND operation has to be performed, all the bits of the mask vector have to be set to `0' except the one corresponding to the selected column: in this way, all the AND operations on the other columns give `0' as result, disabling the corresponding pull--down transistors, and the logic value sensed on the line depends exclusively on the output of the AND on the selected cell. This can be clarified with an example. Denoting with $D_i$ the content of the cell on the $i$--th column, with $M_i$ the mask bit in the same position and with $O$ the result obtained in output, when considering the bitwise AND implemented on the row, \autoref{eq:Eq_selection} is obtained:

\begin{equation}
  O = \sum_{i=0}^{N-1}D_i \cdot M_i\\
  \label{eq:Eq_selection}
\end{equation}

A non--rigorous notation is used in the equation, associating to the sum sign `+' the OR operation and the product sign `$\cdot$' to the AND one. Indicating with the index $j$ the selected column, the formula can be rewritten in the following way:

\begin{equation*}
  M_i = 
  \begin{cases}
    1 & i = j\\
    0 & i\neq j
  \end{cases}
\end{equation*}

\begin{align*}
  O &= \sum_{i=0}^{N-1}D_i \cdot M_i\\
    &= \sum_{i=0, i \neq j}^{N-1}D_i \cdot M_i + D_j \cdot M_j\\
    &= \sum_{i=0, i \neq j}^{N-1}D_i \cdot 0 + D_j \cdot 1\\
    &= D_j
\end{align*}

Hence, the output of the OR operation is determined only by the selected cell content.

\medskip

The AND logic function is implemented by embedding a logic gate inside the memory cell. Three variants of this are presented:

\begin{itemize}
\item a dynamic CMOS logic AND gate, which is shown in \autoref{fig:Dynamic_AND}.
\item a static CMOS logic AND gate, which is depicted in \autoref{fig:Static_AND}.
\item a special purpose AND gate, designed appositely for the algorithm to be implemented in order to reduce as much as possible the cell area, which is presented in \autoref{fig:Special_AND}.
\end{itemize}

\subsection{Dynamic CMOS logic AND}
\label{sec:Dynamic_AND}

\begin{figure}[t!p]
  \begin{subfigure}[b]{0.5\linewidth}
    \centering
    \scalebox{0.8}{\begin{pgfpicture}{0cm}{0cm}{230pt}{132pt}
\pgfsetxvec{\pgfpoint{1pt}{0pt}}
\pgfsetyvec{\pgfpoint{0pt}{1pt}}
\pgfsetroundjoin 
\pgfsetroundcap
\pgftranslateto{\pgfxy(0,132)}
\begin{pgfmagnify}{1}{-1}
\definecolor{layer0}{rgb}{0.0,0.0,0.0}
\definecolor{layer1}{rgb}{0.0,0.0,0.5}
\definecolor{layer2}{rgb}{1.0,0.0,0.0}
\definecolor{layer3}{rgb}{0.0,0.5,0.5}
\definecolor{layer4}{rgb}{1.0,0.78,0.0}
\definecolor{layer5}{rgb}{0.5,1.0,0.0}
\definecolor{layer6}{rgb}{0.0,1.0,1.0}
\definecolor{layer7}{rgb}{0.0,0.5,0.0}
\definecolor{layer8}{rgb}{0.6,0.8,0.2}
\definecolor{layer9}{rgb}{1.0,0.08,0.58}
\definecolor{layer10}{rgb}{0.71,0.61,0.05}
\definecolor{layer11}{rgb}{0.0,0.5,1.0}
\definecolor{layer12}{rgb}{0.88,0.88,0.88}
\definecolor{layer13}{rgb}{0.64,0.64,0.64}
\definecolor{layer14}{rgb}{0.37,0.37,0.37}
\definecolor{layer15}{rgb}{0.0,0.0,0.0}
\color{layer0}
\pgfsetlinewidth{1.0pt}
\pgfsetdash{}{0pt}
\pgfline{\pgfxy(136.0,32.0)}{\pgfxy(146.0,32.0)}
\pgfline{\pgfxy(136.0,20.0)}{\pgfxy(146.0,20.0)}
\pgfline{\pgfxy(146.0,20.0)}{\pgfxy(146.0,16.0)}
\pgfline{\pgfxy(146.0,32.0)}{\pgfxy(146.0,36.0)}
\pgfline{\pgfxy(126.0,26.0)}{\pgfxy(130.0,26.0)}
\pgfline{\pgfxy(134.0,20.0)}{\pgfxy(134.0,32.0)}
\pgfline{\pgfxy(136.0,18.0)}{\pgfxy(136.0,34.0)}
\pgfellipse[stroke]{\pgfxy(132.0,26.0)}{\pgfxy(2.0,0)}{\pgfxy(0,2.0)}
\pgfline{\pgfxy(116.0,72.0)}{\pgfxy(126.0,72.0)}
\pgfline{\pgfxy(116.0,60.0)}{\pgfxy(126.0,60.0)}
\pgfline{\pgfxy(126.0,60.0)}{\pgfxy(126.0,56.0)}
\pgfline{\pgfxy(126.0,72.0)}{\pgfxy(126.0,76.0)}
\pgfline{\pgfxy(106.0,66.0)}{\pgfxy(114.0,66.0)}
\pgfline{\pgfxy(114.0,60.0)}{\pgfxy(114.0,72.0)}
\pgfline{\pgfxy(116.0,58.0)}{\pgfxy(116.0,74.0)}
\pgfline{\pgfxy(176.0,60.0)}{\pgfxy(166.0,60.0)}
\pgfline{\pgfxy(176.0,72.0)}{\pgfxy(166.0,72.0)}
\pgfline{\pgfxy(166.0,72.0)}{\pgfxy(166.0,76.0)}
\pgfline{\pgfxy(166.0,60.0)}{\pgfxy(166.0,56.0)}
\pgfline{\pgfxy(186.0,66.0)}{\pgfxy(178.0,66.0)}
\pgfline{\pgfxy(178.0,72.0)}{\pgfxy(178.0,60.0)}
\pgfline{\pgfxy(176.0,74.0)}{\pgfxy(176.0,58.0)}
\pgfline{\pgfxy(166.0,46.0)}{\pgfxy(126.0,46.0)}
\pgfline{\pgfxy(146.0,46.0)}{\pgfxy(146.0,36.0)}
\pgfline{\pgfxy(126.0,56.0)}{\pgfxy(126.0,46.0)}
\pgfline{\pgfxy(166.0,56.0)}{\pgfxy(166.0,46.0)}
\pgfline{\pgfxy(126.0,76.0)}{\pgfxy(126.0,86.0)}
\pgfline{\pgfxy(126.0,86.0)}{\pgfxy(166.0,86.0)}
\pgfline{\pgfxy(166.0,76.0)}{\pgfxy(166.0,86.0)}
\pgfsetlinewidth{0.33pt}
\pgfcircle[fill]{\pgfxy(146,46)}{2.0pt}\pgfcircle[fill]{\pgfxy(166,46)}{2.0pt}\pgfsetlinewidth{1.0pt}
\pgfline{\pgfxy(146.0,16.0)}{\pgfxy(146.0,6.0)}
\pgfline{\pgfxy(156.0,6.0)}{\pgfxy(136.0,6.0)}
\pgfmoveto{\pgfxy(140.0,120.0)}
\pgflineto{\pgfxy(152.0,120.0)}
\pgflineto{\pgfxy(146.0,126.0)}
\pgflineto{\pgfxy(146.0,126.0)}
\pgfclosepath 
\pgfqstroke 
\pgfline{\pgfxy(146.0,116.0)}{\pgfxy(146.0,120.0)}
\pgfsetlinewidth{0.33pt}
\pgfcircle[fill]{\pgfxy(146,86)}{2.0pt}\begin{pgfmagnify}{1}{-1}
\pgfputat{\pgfxy(36,-16)}{\pgfbox[left,top]{}}
\end{pgfmagnify}
\begin{pgfmagnify}{1}{-1}
\pgfputat{\pgfxy(6,-16)}{\pgfbox[left,top]{$\overline{\textrm{PRE}}$}}
\end{pgfmagnify}
\begin{pgfmagnify}{1}{-1}
\pgfputat{\pgfxy(76,-56)}{\pgfbox[left,top]{$\overline{\textrm{BL}}$}}
\end{pgfmagnify}
\begin{pgfmagnify}{1}{-1}
\pgfputat{\pgfxy(206,-56)}{\pgfbox[left,top]{$\overline{\textrm{D}}$}}
\end{pgfmagnify}
\pgfsetlinewidth{1.0pt}
\pgfline{\pgfxy(166.0,46.0)}{\pgfxy(196.0,46.0)}
\begin{pgfmagnify}{1}{-1}
\pgfputat{\pgfxy(216,-36)}{\pgfbox[left,top]{O}}
\end{pgfmagnify}
\pgfline{\pgfxy(54.0,26.0)}{\pgfxy(56.0,26.0)}
\pgfmoveto{\pgfxy(54.0,26.0)}
\pgflineto{\pgfxy(54.0,26.0)}
\pgflineto{\pgfxy(52.0,24.0)}
\pgflineto{\pgfxy(46.0,24.0)}
\pgflineto{\pgfxy(46.0,28.0)}
\pgflineto{\pgfxy(52.0,28.0)}
\pgfclosepath 
\pgfqstroke 
\pgfline{\pgfxy(104.0,66.0)}{\pgfxy(106.0,66.0)}
\pgfmoveto{\pgfxy(104.0,66.0)}
\pgflineto{\pgfxy(104.0,66.0)}
\pgflineto{\pgfxy(102.0,64.0)}
\pgflineto{\pgfxy(96.0,64.0)}
\pgflineto{\pgfxy(96.0,68.0)}
\pgflineto{\pgfxy(102.0,68.0)}
\pgfclosepath 
\pgfqstroke 
\pgfline{\pgfxy(188.0,66.0)}{\pgfxy(186.0,66.0)}
\pgfmoveto{\pgfxy(188.0,66.0)}
\pgflineto{\pgfxy(188.0,66.0)}
\pgflineto{\pgfxy(190.0,68.0)}
\pgflineto{\pgfxy(196.0,68.0)}
\pgflineto{\pgfxy(196.0,64.0)}
\pgflineto{\pgfxy(190.0,64.0)}
\pgfclosepath 
\pgfqstroke 
\pgfline{\pgfxy(196.0,46.0)}{\pgfxy(198.0,46.0)}
\pgfmoveto{\pgfxy(206.0,46.0)}
\pgflineto{\pgfxy(206.0,46.0)}
\pgflineto{\pgfxy(204.0,44.0)}
\pgflineto{\pgfxy(198.0,44.0)}
\pgflineto{\pgfxy(198.0,48.0)}
\pgflineto{\pgfxy(204.0,48.0)}
\pgfclosepath 
\pgfqstroke 
\pgfline{\pgfxy(136.0,112.0)}{\pgfxy(146.0,112.0)}
\pgfline{\pgfxy(136.0,100.0)}{\pgfxy(146.0,100.0)}
\pgfline{\pgfxy(146.0,100.0)}{\pgfxy(146.0,96.0)}
\pgfline{\pgfxy(146.0,112.0)}{\pgfxy(146.0,116.0)}
\pgfline{\pgfxy(126.0,106.0)}{\pgfxy(134.0,106.0)}
\pgfline{\pgfxy(134.0,100.0)}{\pgfxy(134.0,112.0)}
\pgfline{\pgfxy(136.0,98.0)}{\pgfxy(136.0,114.0)}
\pgfline{\pgfxy(146.0,96.0)}{\pgfxy(146.0,86.0)}
\pgfline{\pgfxy(56.0,26.0)}{\pgfxy(126.0,26.0)}
\pgfline{\pgfxy(126.0,106.0)}{\pgfxy(66.0,106.0)}
\pgfline{\pgfxy(66.0,106.0)}{\pgfxy(66.0,26.0)}
\pgfsetlinewidth{0.33pt}
\pgfcircle[fill]{\pgfxy(66,26)}{2.0pt}\end{pgfmagnify}
\end{pgfpicture}}
    \vspace*{-5mm}
    \caption{}
    \label{fig:Dynamic_AND}
  \end{subfigure}
  \hspace*{0.2mm}
  \begin{subfigure}[b]{0.5\linewidth}
    \scalebox{0.8}{\begin{pgfpicture}{0cm}{0cm}{182pt}{188pt}
\pgfsetxvec{\pgfpoint{1pt}{0pt}}
\pgfsetyvec{\pgfpoint{0pt}{1pt}}
\pgfsetroundjoin 
\pgfsetroundcap
\pgftranslateto{\pgfxy(0,188)}
\begin{pgfmagnify}{1}{-1}
\definecolor{layer0}{rgb}{0.0,0.0,0.0}
\definecolor{layer1}{rgb}{0.0,0.0,0.5}
\definecolor{layer2}{rgb}{1.0,0.0,0.0}
\definecolor{layer3}{rgb}{0.0,0.5,0.5}
\definecolor{layer4}{rgb}{1.0,0.78,0.0}
\definecolor{layer5}{rgb}{0.5,1.0,0.0}
\definecolor{layer6}{rgb}{0.0,1.0,1.0}
\definecolor{layer7}{rgb}{0.0,0.5,0.0}
\definecolor{layer8}{rgb}{0.6,0.8,0.2}
\definecolor{layer9}{rgb}{1.0,0.08,0.58}
\definecolor{layer10}{rgb}{0.71,0.61,0.05}
\definecolor{layer11}{rgb}{0.0,0.5,1.0}
\definecolor{layer12}{rgb}{0.88,0.88,0.88}
\definecolor{layer13}{rgb}{0.64,0.64,0.64}
\definecolor{layer14}{rgb}{0.37,0.37,0.37}
\definecolor{layer15}{rgb}{0.0,0.0,0.0}
\color{layer0}
\pgfsetlinewidth{1.0pt}
\pgfsetdash{}{0pt}
\pgfline{\pgfxy(106.0,72.0)}{\pgfxy(106.0,62.0)}
\pgfline{\pgfxy(106.0,62.0)}{\pgfxy(116.0,62.0)}
\pgfline{\pgfxy(106.0,62.0)}{\pgfxy(96.0,62.0)}
\pgfline{\pgfxy(106.0,92.0)}{\pgfxy(106.0,102.0)}
\pgfline{\pgfxy(96.0,102.0)}{\pgfxy(116.0,102.0)}
\pgfline{\pgfxy(116.0,102.0)}{\pgfxy(116.0,112.0)}
\pgfline{\pgfxy(96.0,102.0)}{\pgfxy(96.0,112.0)}
\pgfline{\pgfxy(86.0,128.0)}{\pgfxy(96.0,128.0)}
\pgfline{\pgfxy(86.0,116.0)}{\pgfxy(96.0,116.0)}
\pgfline{\pgfxy(96.0,116.0)}{\pgfxy(96.0,112.0)}
\pgfline{\pgfxy(96.0,128.0)}{\pgfxy(96.0,132.0)}
\pgfline{\pgfxy(76.0,122.0)}{\pgfxy(84.0,122.0)}
\pgfline{\pgfxy(84.0,116.0)}{\pgfxy(84.0,128.0)}
\pgfline{\pgfxy(86.0,114.0)}{\pgfxy(86.0,130.0)}
\pgfline{\pgfxy(126.0,116.0)}{\pgfxy(116.0,116.0)}
\pgfline{\pgfxy(126.0,128.0)}{\pgfxy(116.0,128.0)}
\pgfline{\pgfxy(116.0,128.0)}{\pgfxy(116.0,132.0)}
\pgfline{\pgfxy(116.0,116.0)}{\pgfxy(116.0,112.0)}
\pgfline{\pgfxy(136.0,122.0)}{\pgfxy(128.0,122.0)}
\pgfline{\pgfxy(128.0,128.0)}{\pgfxy(128.0,116.0)}
\pgfline{\pgfxy(126.0,130.0)}{\pgfxy(126.0,114.0)}
\pgfline{\pgfxy(96.0,168.0)}{\pgfxy(106.0,168.0)}
\pgfline{\pgfxy(96.0,156.0)}{\pgfxy(106.0,156.0)}
\pgfline{\pgfxy(106.0,156.0)}{\pgfxy(106.0,152.0)}
\pgfline{\pgfxy(106.0,168.0)}{\pgfxy(106.0,172.0)}
\pgfline{\pgfxy(86.0,162.0)}{\pgfxy(94.0,162.0)}
\pgfline{\pgfxy(94.0,156.0)}{\pgfxy(94.0,168.0)}
\pgfline{\pgfxy(96.0,154.0)}{\pgfxy(96.0,170.0)}
\pgfline{\pgfxy(146.0,108.0)}{\pgfxy(156.0,108.0)}
\pgfline{\pgfxy(146.0,96.0)}{\pgfxy(156.0,96.0)}
\pgfline{\pgfxy(156.0,96.0)}{\pgfxy(156.0,92.0)}
\pgfline{\pgfxy(156.0,108.0)}{\pgfxy(156.0,112.0)}
\pgfline{\pgfxy(136.0,102.0)}{\pgfxy(144.0,102.0)}
\pgfline{\pgfxy(144.0,96.0)}{\pgfxy(144.0,108.0)}
\pgfline{\pgfxy(146.0,94.0)}{\pgfxy(146.0,110.0)}
\pgfline{\pgfxy(116.0,142.0)}{\pgfxy(96.0,142.0)}
\pgfline{\pgfxy(96.0,88.0)}{\pgfxy(106.0,88.0)}
\pgfline{\pgfxy(96.0,76.0)}{\pgfxy(106.0,76.0)}
\pgfline{\pgfxy(106.0,76.0)}{\pgfxy(106.0,72.0)}
\pgfline{\pgfxy(106.0,88.0)}{\pgfxy(106.0,92.0)}
\pgfline{\pgfxy(86.0,82.0)}{\pgfxy(90.0,82.0)}
\pgfline{\pgfxy(94.0,76.0)}{\pgfxy(94.0,88.0)}
\pgfline{\pgfxy(96.0,74.0)}{\pgfxy(96.0,90.0)}
\pgfellipse[stroke]{\pgfxy(92.0,82.0)}{\pgfxy(2.0,0)}{\pgfxy(0,2.0)}
\pgfline{\pgfxy(76.0,122.0)}{\pgfxy(66.0,122.0)}
\pgfline{\pgfxy(116.0,102.0)}{\pgfxy(136.0,102.0)}
\pgfmoveto{\pgfxy(100.0,176.0)}
\pgflineto{\pgfxy(112.0,176.0)}
\pgflineto{\pgfxy(106.0,182.0)}
\pgflineto{\pgfxy(106.0,182.0)}
\pgfclosepath 
\pgfqstroke 
\pgfline{\pgfxy(106.0,172.0)}{\pgfxy(106.0,176.0)}
\pgfmoveto{\pgfxy(150.0,116.0)}
\pgflineto{\pgfxy(162.0,116.0)}
\pgflineto{\pgfxy(156.0,122.0)}
\pgflineto{\pgfxy(156.0,122.0)}
\pgfclosepath 
\pgfqstroke 
\pgfline{\pgfxy(156.0,112.0)}{\pgfxy(156.0,116.0)}
\pgfline{\pgfxy(116.0,142.0)}{\pgfxy(116.0,132.0)}
\pgfline{\pgfxy(96.0,142.0)}{\pgfxy(96.0,132.0)}
\pgfline{\pgfxy(106.0,152.0)}{\pgfxy(106.0,142.0)}
\pgfsetlinewidth{0.33pt}
\pgfcircle[fill]{\pgfxy(106,142)}{2.0pt}\pgfcircle[fill]{\pgfxy(106,102)}{2.0pt}\pgfcircle[fill]{\pgfxy(116,102)}{2.0pt}\pgfsetlinewidth{1.0pt}
\pgfline{\pgfxy(156.0,82.0)}{\pgfxy(166.0,82.0)}
\pgfline{\pgfxy(166.0,82.0)}{\pgfxy(168.0,82.0)}
\pgfmoveto{\pgfxy(176.0,82.0)}
\pgflineto{\pgfxy(176.0,82.0)}
\pgflineto{\pgfxy(174.0,80.0)}
\pgflineto{\pgfxy(168.0,80.0)}
\pgflineto{\pgfxy(168.0,84.0)}
\pgflineto{\pgfxy(174.0,84.0)}
\pgfclosepath 
\pgfqstroke 
\pgfline{\pgfxy(156.0,92.0)}{\pgfxy(156.0,82.0)}
\begin{pgfmagnify}{1}{-1}
\pgfputat{\pgfxy(86,-22)}{\pgfbox[left,top]{CAM}}
\end{pgfmagnify}
\pgfline{\pgfxy(66.0,32.0)}{\pgfxy(76.0,32.0)}
\pgfline{\pgfxy(136.0,32.0)}{\pgfxy(126.0,32.0)}
\pgfmoveto{\pgfxy(76,12)}
\pgflineto{\pgfxy(126,12)}
\pgflineto{\pgfxy(126,52)}
\pgflineto{\pgfxy(76,52)}
\pgfclosepath 
\pgfqstroke 
\pgfline{\pgfxy(136.0,122.0)}{\pgfxy(136.0,32.0)}
\pgfline{\pgfxy(66.0,122.0)}{\pgfxy(66.0,32.0)}
\pgfline{\pgfxy(44.0,82.0)}{\pgfxy(46.0,82.0)}
\pgfmoveto{\pgfxy(44.0,82.0)}
\pgflineto{\pgfxy(44.0,82.0)}
\pgflineto{\pgfxy(42.0,80.0)}
\pgflineto{\pgfxy(36.0,80.0)}
\pgflineto{\pgfxy(36.0,84.0)}
\pgflineto{\pgfxy(42.0,84.0)}
\pgfclosepath 
\pgfqstroke 
\pgfline{\pgfxy(86.0,162.0)}{\pgfxy(56.0,162.0)}
\pgfline{\pgfxy(56.0,162.0)}{\pgfxy(56.0,82.0)}
\pgfline{\pgfxy(46.0,82.0)}{\pgfxy(86.0,82.0)}
\begin{pgfmagnify}{1}{-1}
\pgfputat{\pgfxy(6,-72)}{\pgfbox[left,top]{$\overline{\textrm{PRE}}$}}
\end{pgfmagnify}
\begin{pgfmagnify}{1}{-1}
\pgfputat{\pgfxy(156,-62)}{\pgfbox[left,top]{\overlinerm{AND}}}
\end{pgfmagnify}
\begin{pgfmagnify}{1}{-1}
\pgfputat{\pgfxy(46,-22)}{\pgfbox[left,top]{$\overline{\textrm{D}}$}}
\end{pgfmagnify}
\begin{pgfmagnify}{1}{-1}
\pgfputat{\pgfxy(146,-22)}{\pgfbox[left,top]{$\overline{\textrm{BL}}$}}
\end{pgfmagnify}
\end{pgfmagnify}
\end{pgfpicture}}
    \vspace*{-5mm}
    \caption{}
    \label{fig:Dynamic_cell}
  \end{subfigure}
  \begin{subfigure}[b]{\linewidth}
    \vspace*{2mm}
    \centering
    \includegraphics[width=0.35\linewidth]{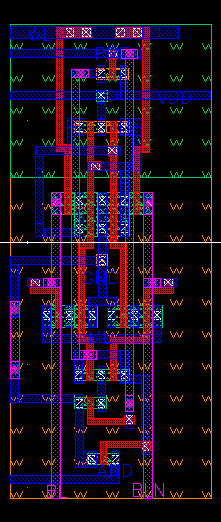}
    \vspace*{-5mm}
    \caption{}
    \label{fig:AND_DYN_layout}
  \end{subfigure}
  \caption{The dynamic AND gate and the cell in which it is integrated. \textbf{(a)} The dynamic AND gate. By using the negated values of the cell content, $\overline{D}$, and the mask bit, $\overline{BL}$, it is possible to take advantage of boolean logic laws to obtain an AND gate without adding an inverting stage on the output. \textbf{(b)} The memory cell that embeds the dynamic AND gate and the pull--down transistor of the AND line. It can be noticed that the output of the AND gate is negated using a single pull--down transistor, which output corresponds to the $\overline{AND}$ signal associated to the line. \textbf{(c)} The cell layout.}
  \label{fig:Dynamic_AND_gate_and_cell}
\end{figure}

In \autoref{fig:Dynamic_AND}, the circuit of the AND gate is shown. It takes in input the negated values of the cell content, $\overline{D}$, the mask bit on the bitlines, $\overline{BL}$, and an additional external signal, $\overline{PRE}$, used to precharge the output node of the gate, $O$. It can be noticed that an AND function is obtained without adding an inverting stage on the output of the inner gate: since the negated values of the cell content and mask bit are available, one can use De Morgan's laws to avoid the inverting stage. In fact, since the gate in \autoref{fig:Dynamic_AND} takes in input $\overline{D}$ and $\overline{BL}$, the logic NOR between these, implemented by the logic gate, can be rewritten in the following way:

\begin{equation*}
  \overline{\overline{D} + \overline{BL}} = D \cdot BL
\end{equation*}

Hence, the inverting stage is not needed to implement the AND function. This logic gate is embedded in the cell, obtaining the circuit show in \autoref{fig:Dynamic_cell}. One can notice that a pull--down transistor is added on the output of the AND gate and connected to the row line.

\smallskip

\begin{figure}[h]
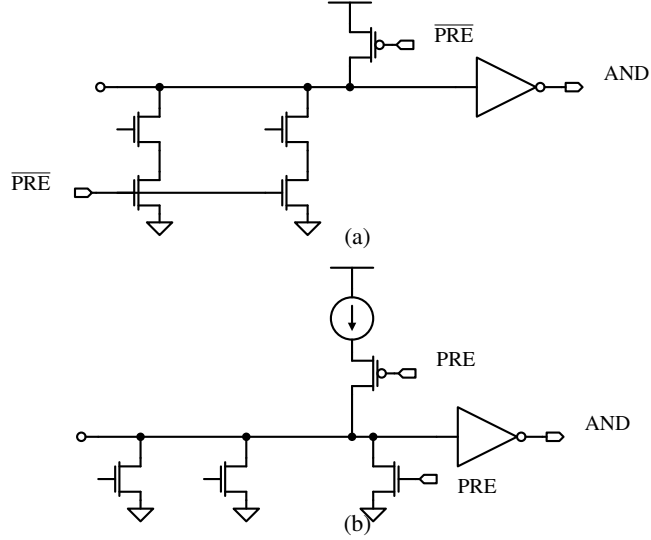

  \begin{subfigure}[b]{1.0\linewidth}
    \centering
    \scalebox{0.8}{\begin{pgfpicture}{0cm}{0cm}{334pt}{122pt}
\pgfsetxvec{\pgfpoint{1pt}{0pt}}
\pgfsetyvec{\pgfpoint{0pt}{1pt}}
\pgfsetroundjoin 
\pgfsetroundcap
\pgftranslateto{\pgfxy(0,122)}
\begin{pgfmagnify}{1}{-1}
\definecolor{layer0}{rgb}{0.0,0.0,0.0}
\definecolor{layer1}{rgb}{0.0,0.0,0.5}
\definecolor{layer2}{rgb}{1.0,0.0,0.0}
\definecolor{layer3}{rgb}{0.0,0.5,0.5}
\definecolor{layer4}{rgb}{1.0,0.78,0.0}
\definecolor{layer5}{rgb}{0.5,1.0,0.0}
\definecolor{layer6}{rgb}{0.0,1.0,1.0}
\definecolor{layer7}{rgb}{0.0,0.5,0.0}
\definecolor{layer8}{rgb}{0.6,0.8,0.2}
\definecolor{layer9}{rgb}{1.0,0.08,0.58}
\definecolor{layer10}{rgb}{0.71,0.61,0.05}
\definecolor{layer11}{rgb}{0.0,0.5,1.0}
\definecolor{layer12}{rgb}{0.88,0.88,0.88}
\definecolor{layer13}{rgb}{0.64,0.64,0.64}
\definecolor{layer14}{rgb}{0.37,0.37,0.37}
\definecolor{layer15}{rgb}{0.0,0.0,0.0}
\color{layer0}
\pgfsetlinewidth{1.0pt}
\pgfsetdash{}{0pt}
\pgfline{\pgfxy(66.0,72.0)}{\pgfxy(76.0,72.0)}
\pgfline{\pgfxy(66.0,60.0)}{\pgfxy(76.0,60.0)}
\pgfline{\pgfxy(76.0,60.0)}{\pgfxy(76.0,56.0)}
\pgfline{\pgfxy(76.0,72.0)}{\pgfxy(76.0,76.0)}
\pgfline{\pgfxy(56.0,66.0)}{\pgfxy(64.0,66.0)}
\pgfline{\pgfxy(64.0,60.0)}{\pgfxy(64.0,72.0)}
\pgfline{\pgfxy(66.0,58.0)}{\pgfxy(66.0,74.0)}
\pgfline{\pgfxy(66.0,102.0)}{\pgfxy(76.0,102.0)}
\pgfline{\pgfxy(66.0,90.0)}{\pgfxy(76.0,90.0)}
\pgfline{\pgfxy(76.0,90.0)}{\pgfxy(76.0,86.0)}
\pgfline{\pgfxy(76.0,102.0)}{\pgfxy(76.0,106.0)}
\pgfline{\pgfxy(56.0,96.0)}{\pgfxy(64.0,96.0)}
\pgfline{\pgfxy(64.0,90.0)}{\pgfxy(64.0,102.0)}
\pgfline{\pgfxy(66.0,88.0)}{\pgfxy(66.0,104.0)}
\pgfline{\pgfxy(136.0,72.0)}{\pgfxy(146.0,72.0)}
\pgfline{\pgfxy(136.0,60.0)}{\pgfxy(146.0,60.0)}
\pgfline{\pgfxy(146.0,60.0)}{\pgfxy(146.0,56.0)}
\pgfline{\pgfxy(146.0,72.0)}{\pgfxy(146.0,76.0)}
\pgfline{\pgfxy(126.0,66.0)}{\pgfxy(134.0,66.0)}
\pgfline{\pgfxy(134.0,60.0)}{\pgfxy(134.0,72.0)}
\pgfline{\pgfxy(136.0,58.0)}{\pgfxy(136.0,74.0)}
\pgfline{\pgfxy(136.0,102.0)}{\pgfxy(146.0,102.0)}
\pgfline{\pgfxy(136.0,90.0)}{\pgfxy(146.0,90.0)}
\pgfline{\pgfxy(146.0,90.0)}{\pgfxy(146.0,86.0)}
\pgfline{\pgfxy(146.0,102.0)}{\pgfxy(146.0,106.0)}
\pgfline{\pgfxy(126.0,96.0)}{\pgfxy(134.0,96.0)}
\pgfline{\pgfxy(134.0,90.0)}{\pgfxy(134.0,102.0)}
\pgfline{\pgfxy(136.0,88.0)}{\pgfxy(136.0,104.0)}
\pgfline{\pgfxy(46.0,96.0)}{\pgfxy(126.0,96.0)}
\pgfmoveto{\pgfxy(140.0,110.0)}
\pgflineto{\pgfxy(152.0,110.0)}
\pgflineto{\pgfxy(146.0,116.0)}
\pgflineto{\pgfxy(146.0,116.0)}
\pgfclosepath 
\pgfqstroke 
\pgfline{\pgfxy(146.0,106.0)}{\pgfxy(146.0,110.0)}
\pgfmoveto{\pgfxy(70.0,110.0)}
\pgflineto{\pgfxy(82.0,110.0)}
\pgflineto{\pgfxy(76.0,116.0)}
\pgflineto{\pgfxy(76.0,116.0)}
\pgfclosepath 
\pgfqstroke 
\pgfline{\pgfxy(76.0,106.0)}{\pgfxy(76.0,110.0)}
\pgfline{\pgfxy(146.0,86.0)}{\pgfxy(146.0,76.0)}
\pgfline{\pgfxy(76.0,86.0)}{\pgfxy(76.0,76.0)}
\pgfline{\pgfxy(56.0,46.0)}{\pgfxy(216.0,46.0)}
\pgfline{\pgfxy(146.0,56.0)}{\pgfxy(146.0,46.0)}
\pgfline{\pgfxy(76.0,56.0)}{\pgfxy(76.0,46.0)}
\pgfline{\pgfxy(176.0,20.0)}{\pgfxy(166.0,20.0)}
\pgfline{\pgfxy(176.0,32.0)}{\pgfxy(166.0,32.0)}
\pgfline{\pgfxy(166.0,32.0)}{\pgfxy(166.0,36.0)}
\pgfline{\pgfxy(166.0,20.0)}{\pgfxy(166.0,16.0)}
\pgfline{\pgfxy(186.0,26.0)}{\pgfxy(182.0,26.0)}
\pgfline{\pgfxy(178.0,32.0)}{\pgfxy(178.0,20.0)}
\pgfline{\pgfxy(176.0,34.0)}{\pgfxy(176.0,18.0)}
\pgfellipse[stroke]{\pgfxy(180.0,26.0)}{\pgfxy(2.0,0)}{\pgfxy(0,2.0)}
\pgfline{\pgfxy(156.0,6.0)}{\pgfxy(176.0,6.0)}
\pgfline{\pgfxy(166.0,16.0)}{\pgfxy(166.0,6.0)}
\pgfline{\pgfxy(166.0,36.0)}{\pgfxy(166.0,46.0)}
\pgfline{\pgfxy(44.0,96.0)}{\pgfxy(46.0,96.0)}
\pgfmoveto{\pgfxy(44.0,96.0)}
\pgflineto{\pgfxy(44.0,96.0)}
\pgflineto{\pgfxy(42.0,94.0)}
\pgflineto{\pgfxy(36.0,94.0)}
\pgflineto{\pgfxy(36.0,98.0)}
\pgflineto{\pgfxy(42.0,98.0)}
\pgfclosepath 
\pgfqstroke 
\pgfline{\pgfxy(188.0,26.0)}{\pgfxy(186.0,26.0)}
\pgfmoveto{\pgfxy(188.0,26.0)}
\pgflineto{\pgfxy(188.0,26.0)}
\pgflineto{\pgfxy(190.0,28.0)}
\pgflineto{\pgfxy(196.0,28.0)}
\pgflineto{\pgfxy(196.0,24.0)}
\pgflineto{\pgfxy(190.0,24.0)}
\pgfclosepath 
\pgfqstroke 
\begin{pgfmagnify}{1}{-1}
\pgfputat{\pgfxy(206,-16)}{\pgfbox[left,top]{$\overline{\textrm{PRE}}$}}
\end{pgfmagnify}
\begin{pgfmagnify}{1}{-1}
\pgfputat{\pgfxy(6,-86)}{\pgfbox[left,top]{$\overline{\textrm{PRE}}$}}
\end{pgfmagnify}
\pgfline{\pgfxy(258.0,46.0)}{\pgfxy(266.0,46.0)}
\pgfellipse[stroke]{\pgfxy(256.0,46.0)}{\pgfxy(2.0,0)}{\pgfxy(0,2.0)}
\pgfmoveto{\pgfxy(226.0,32.0)}
\pgflineto{\pgfxy(226.0,60.0)}
\pgflineto{\pgfxy(254.0,46.0)}
\pgfclosepath 
\pgfqstroke 
\pgfline{\pgfxy(216.0,46.0)}{\pgfxy(226.0,46.0)}
\pgfline{\pgfxy(266.0,46.0)}{\pgfxy(268.0,46.0)}
\pgfmoveto{\pgfxy(276.0,46.0)}
\pgflineto{\pgfxy(276.0,46.0)}
\pgflineto{\pgfxy(274.0,44.0)}
\pgflineto{\pgfxy(268.0,44.0)}
\pgflineto{\pgfxy(268.0,48.0)}
\pgflineto{\pgfxy(274.0,48.0)}
\pgfclosepath 
\pgfqstroke 
\begin{pgfmagnify}{1}{-1}
\pgfputat{\pgfxy(286,-36)}{\pgfbox[left,top]{AND}}
\end{pgfmagnify}
\pgfline{\pgfxy(50.0,46.0)}{\pgfxy(56.0,46.0)}
\pgfellipse[stroke]{\pgfxy(48.0,46.0)}{\pgfxy(2.0,0)}{\pgfxy(0,2.0)}
\pgfsetlinewidth{0.33pt}
\pgfcircle[fill]{\pgfxy(166,46)}{2.0pt}\pgfcircle[fill]{\pgfxy(146,46)}{2.0pt}\pgfcircle[fill]{\pgfxy(76,46)}{2.0pt}\end{pgfmagnify}
\end{pgfpicture}}
    \vspace*{-5mm}
    \caption{}
    \label{fig:Classic_matchline}
  \end{subfigure}
  \begin{subfigure}[b]{1.0\linewidth}
    \centering
    \scalebox{0.8}{\begin{pgfpicture}{0cm}{0cm}{272pt}{132pt}
\pgfsetxvec{\pgfpoint{1pt}{0pt}}
\pgfsetyvec{\pgfpoint{0pt}{1pt}}
\pgfsetroundjoin 
\pgfsetroundcap
\pgftranslateto{\pgfxy(0,132)}
\begin{pgfmagnify}{1}{-1}
\definecolor{layer0}{rgb}{0.0,0.0,0.0}
\definecolor{layer1}{rgb}{0.0,0.0,0.5}
\definecolor{layer2}{rgb}{1.0,0.0,0.0}
\definecolor{layer3}{rgb}{0.0,0.5,0.5}
\definecolor{layer4}{rgb}{1.0,0.78,0.0}
\definecolor{layer5}{rgb}{0.5,1.0,0.0}
\definecolor{layer6}{rgb}{0.0,1.0,1.0}
\definecolor{layer7}{rgb}{0.0,0.5,0.0}
\definecolor{layer8}{rgb}{0.6,0.8,0.2}
\definecolor{layer9}{rgb}{1.0,0.08,0.58}
\definecolor{layer10}{rgb}{0.71,0.61,0.05}
\definecolor{layer11}{rgb}{0.0,0.5,1.0}
\definecolor{layer12}{rgb}{0.88,0.88,0.88}
\definecolor{layer13}{rgb}{0.64,0.64,0.64}
\definecolor{layer14}{rgb}{0.37,0.37,0.37}
\definecolor{layer15}{rgb}{0.0,0.0,0.0}
\color{layer0}
\pgfsetlinewidth{1.0pt}
\pgfsetdash{}{0pt}
\pgfline{\pgfxy(26.0,112.0)}{\pgfxy(36.0,112.0)}
\pgfline{\pgfxy(26.0,100.0)}{\pgfxy(36.0,100.0)}
\pgfline{\pgfxy(36.0,100.0)}{\pgfxy(36.0,96.0)}
\pgfline{\pgfxy(36.0,112.0)}{\pgfxy(36.0,116.0)}
\pgfline{\pgfxy(16.0,106.0)}{\pgfxy(24.0,106.0)}
\pgfline{\pgfxy(24.0,100.0)}{\pgfxy(24.0,112.0)}
\pgfline{\pgfxy(26.0,98.0)}{\pgfxy(26.0,114.0)}
\pgfline{\pgfxy(76.0,112.0)}{\pgfxy(86.0,112.0)}
\pgfline{\pgfxy(76.0,100.0)}{\pgfxy(86.0,100.0)}
\pgfline{\pgfxy(86.0,100.0)}{\pgfxy(86.0,96.0)}
\pgfline{\pgfxy(86.0,112.0)}{\pgfxy(86.0,116.0)}
\pgfline{\pgfxy(66.0,106.0)}{\pgfxy(74.0,106.0)}
\pgfline{\pgfxy(74.0,100.0)}{\pgfxy(74.0,112.0)}
\pgfline{\pgfxy(76.0,98.0)}{\pgfxy(76.0,114.0)}
\pgfline{\pgfxy(16.0,86.0)}{\pgfxy(176.0,86.0)}
\pgfline{\pgfxy(86.0,96.0)}{\pgfxy(86.0,86.0)}
\pgfline{\pgfxy(36.0,96.0)}{\pgfxy(36.0,86.0)}
\pgfline{\pgfxy(218.0,86.0)}{\pgfxy(226.0,86.0)}
\pgfellipse[stroke]{\pgfxy(216.0,86.0)}{\pgfxy(2.0,0)}{\pgfxy(0,2.0)}
\pgfmoveto{\pgfxy(186.0,72.0)}
\pgflineto{\pgfxy(186.0,100.0)}
\pgflineto{\pgfxy(214.0,86.0)}
\pgfclosepath 
\pgfqstroke 
\pgfline{\pgfxy(176.0,86.0)}{\pgfxy(186.0,86.0)}
\pgfline{\pgfxy(226.0,86.0)}{\pgfxy(228.0,86.0)}
\pgfmoveto{\pgfxy(236.0,86.0)}
\pgflineto{\pgfxy(236.0,86.0)}
\pgflineto{\pgfxy(234.0,84.0)}
\pgflineto{\pgfxy(228.0,84.0)}
\pgflineto{\pgfxy(228.0,88.0)}
\pgflineto{\pgfxy(234.0,88.0)}
\pgfclosepath 
\pgfqstroke 
\begin{pgfmagnify}{1}{-1}
\pgfputat{\pgfxy(246,-76)}{\pgfbox[left,top]{AND}}
\end{pgfmagnify}
\pgfline{\pgfxy(10.0,86.0)}{\pgfxy(16.0,86.0)}
\pgfellipse[stroke]{\pgfxy(8.0,86.0)}{\pgfxy(2.0,0)}{\pgfxy(0,2.0)}
\pgfsetlinewidth{0.33pt}
\pgfcircle[fill]{\pgfxy(86,86)}{2.0pt}\pgfcircle[fill]{\pgfxy(36,86)}{2.0pt}\pgfsetlinewidth{1.0pt}
\pgfmoveto{\pgfxy(30.0,120.0)}
\pgflineto{\pgfxy(42.0,120.0)}
\pgflineto{\pgfxy(36.0,126.0)}
\pgflineto{\pgfxy(36.0,126.0)}
\pgfclosepath 
\pgfqstroke 
\pgfline{\pgfxy(36.0,116.0)}{\pgfxy(36.0,120.0)}
\pgfmoveto{\pgfxy(80.0,120.0)}
\pgflineto{\pgfxy(92.0,120.0)}
\pgflineto{\pgfxy(86.0,126.0)}
\pgflineto{\pgfxy(86.0,126.0)}
\pgfclosepath 
\pgfqstroke 
\pgfline{\pgfxy(86.0,116.0)}{\pgfxy(86.0,120.0)}
\pgfline{\pgfxy(156.0,100.0)}{\pgfxy(146.0,100.0)}
\pgfline{\pgfxy(156.0,112.0)}{\pgfxy(146.0,112.0)}
\pgfline{\pgfxy(146.0,112.0)}{\pgfxy(146.0,116.0)}
\pgfline{\pgfxy(146.0,100.0)}{\pgfxy(146.0,96.0)}
\pgfline{\pgfxy(166.0,106.0)}{\pgfxy(158.0,106.0)}
\pgfline{\pgfxy(158.0,112.0)}{\pgfxy(158.0,100.0)}
\pgfline{\pgfxy(156.0,114.0)}{\pgfxy(156.0,98.0)}
\pgfmoveto{\pgfxy(140.0,120.0)}
\pgflineto{\pgfxy(152.0,120.0)}
\pgflineto{\pgfxy(146.0,126.0)}
\pgflineto{\pgfxy(146.0,126.0)}
\pgfclosepath 
\pgfqstroke 
\pgfline{\pgfxy(146.0,116.0)}{\pgfxy(146.0,120.0)}
\pgfline{\pgfxy(146.0,96.0)}{\pgfxy(146.0,86.0)}
\pgfsetlinewidth{0.33pt}
\pgfcircle[fill]{\pgfxy(146,86)}{2.0pt}\pgfsetlinewidth{1.0pt}
\pgfline{\pgfxy(168.0,106.0)}{\pgfxy(166.0,106.0)}
\pgfmoveto{\pgfxy(168.0,106.0)}
\pgflineto{\pgfxy(168.0,106.0)}
\pgflineto{\pgfxy(170.0,108.0)}
\pgflineto{\pgfxy(176.0,108.0)}
\pgflineto{\pgfxy(176.0,104.0)}
\pgflineto{\pgfxy(170.0,104.0)}
\pgfclosepath 
\pgfqstroke 
\begin{pgfmagnify}{1}{-1}
\pgfputat{\pgfxy(186,-106)}{\pgfbox[left,top]{PRE}}
\end{pgfmagnify}
\pgfellipse[stroke]{\pgfxy(136.0,30.0)}{\pgfxy(10.0,0)}{\pgfxy(0,10.0)}
\pgfline{\pgfxy(136.0,20.0)}{\pgfxy(136.0,16.0)}
\pgfmoveto{\pgfxy(136.0,36.0)}
\pgflineto{\pgfxy(134.0,32.0)}
\pgflineto{\pgfxy(138.0,32.0)}
\pgfclosepath 
\pgffill 
\pgfline{\pgfxy(136.0,34.0)}{\pgfxy(136.0,24.0)}
\pgfline{\pgfxy(136.0,46.0)}{\pgfxy(136.0,40.0)}
\pgfline{\pgfxy(126.0,6.0)}{\pgfxy(146.0,6.0)}
\pgfline{\pgfxy(136.0,16.0)}{\pgfxy(136.0,6.0)}
\pgfline{\pgfxy(136.0,66.0)}{\pgfxy(136.0,86.0)}
\pgfsetlinewidth{0.33pt}
\pgfcircle[fill]{\pgfxy(136,86)}{2.0pt}\pgfsetlinewidth{1.0pt}
\pgfline{\pgfxy(146.0,50.0)}{\pgfxy(136.0,50.0)}
\pgfline{\pgfxy(146.0,62.0)}{\pgfxy(136.0,62.0)}
\pgfline{\pgfxy(136.0,62.0)}{\pgfxy(136.0,66.0)}
\pgfline{\pgfxy(136.0,50.0)}{\pgfxy(136.0,46.0)}
\pgfline{\pgfxy(156.0,56.0)}{\pgfxy(152.0,56.0)}
\pgfline{\pgfxy(148.0,62.0)}{\pgfxy(148.0,50.0)}
\pgfline{\pgfxy(146.0,64.0)}{\pgfxy(146.0,48.0)}
\pgfellipse[stroke]{\pgfxy(150.0,56.0)}{\pgfxy(2.0,0)}{\pgfxy(0,2.0)}
\pgfline{\pgfxy(158.0,56.0)}{\pgfxy(156.0,56.0)}
\pgfmoveto{\pgfxy(158.0,56.0)}
\pgflineto{\pgfxy(158.0,56.0)}
\pgflineto{\pgfxy(160.0,58.0)}
\pgflineto{\pgfxy(166.0,58.0)}
\pgflineto{\pgfxy(166.0,54.0)}
\pgflineto{\pgfxy(160.0,54.0)}
\pgfclosepath 
\pgfqstroke 
\begin{pgfmagnify}{1}{-1}
\pgfputat{\pgfxy(176,-46)}{\pgfbox[left,top]{PRE}}
\end{pgfmagnify}
\end{pgfmagnify}
\end{pgfpicture}}
    \vspace*{-5mm}
    \caption{}
    \label{fig:Current_save_matchline}
  \end{subfigure}
  \caption{Standard and current--saving schemes. \textbf{(a)} A standard precharge line. The line is precharged to the logic `1' using a pull--up transistor, while all the pull--downs are disabled using footer transistors; then, these are enabled by deactivating $\overline{PRE}$ during the evaluation phase. \textbf{(b)} The current--saving line. In this scheme, footer transistors are not needed for disabling the pull--downs, since the line is pre--discharged instead of being pre--charged; then, during the evaluation phase, the line gets charged if there are not conducting pull--downs.}
  \label{fig:Classic_and_current_save_lines}
\end{figure}

The AND line is an implementation of dynamic CMOS logic: the line is precharged to the logic `1' and then, if one of the pull--down transistors connected to it is enabled, discharged during the evaluation phase. In order to properly carry out the precharge phase, all the pull--down transistors must be disabled. This is usually achieved by adding a footer transistor on the source of the pull--down of each cell, that is disabled during the precharge phase through a dedicated row signal, preventing the pull--downs from discharging the line independently of the output values of the AND gates. A possible circuit is highlighted in \autoref{fig:Classic_matchline}.

In this work, a different approach is used to disable the pull--down transistors during the precharge phase: the same current--saving sensing scheme of the CAM is adopted for the AND line. In this way, since the line is pre--\textit{discharged} and not pre--\textit{charged}, there is no need to disable the pull--downs and, hence, additional transistors and signals are not required, allowing for smaller cell and row footprints.

A circuit is proposed in \autoref{fig:Current_save_matchline}. A truth table for the logic gate, which takes into account the implementation of the current--saving scheme, is shown in \autoref{tab:AND_line}.

\begin{table}[h]
  \begin{tabular}{cccccc}
    \toprule
    \textit{\textbf{D}} & \textit{\textbf{BL}} & $\overline{D}$ & $\overline{BL}$ & \textit{\textbf{AND}} & $\overline{\textit{\textbf{AND}}}$\\
    \midrule
    0 & 0 & 1 & 1 & 1 $\rightarrow$ 0 & 0 $\rightarrow$ 1\\
    0 & 1 & 1 & 0 & 1 $\rightarrow$ 0 & 0 $\rightarrow$ 1\\
    1 & 0 & 0 & 1 & 1 $\rightarrow$ 0 & 0 $\rightarrow$ 1\\
    1 & 1 & 0 & 0 & 1 & 0\\
    \bottomrule
  \end{tabular}
   \smallskip
     \centering
  \caption{The truth table of the dynamic AND cell of \autoref{fig:Dynamic_cell}. One can notice that, using the current--saving scheme, the $\overline{AND}$ output is charged to `1' when \textit{AND} is discharged to `0', while it remains at the ground voltage when $AND$=`1'.}
  \label{tab:AND_line}

\end{table}

\medskip

\subsection{Static CMOS logic AND}
\label{sec:Static_AND}

\begin{figure}[t!p]
  \begin{subfigure}[b]{0.5\linewidth}
    \centering
    \scalebox{0.8}{\begin{pgfpicture}{0cm}{0cm}{142pt}{130pt}
\pgfsetxvec{\pgfpoint{1pt}{0pt}}
\pgfsetyvec{\pgfpoint{0pt}{1pt}}
\pgfsetroundjoin 
\pgfsetroundcap
\pgftranslateto{\pgfxy(0,130)}
\begin{pgfmagnify}{1}{-1}
\definecolor{layer0}{rgb}{0.0,0.0,0.0}
\definecolor{layer1}{rgb}{0.0,0.0,0.5}
\definecolor{layer2}{rgb}{1.0,0.0,0.0}
\definecolor{layer3}{rgb}{0.0,0.5,0.5}
\definecolor{layer4}{rgb}{1.0,0.78,0.0}
\definecolor{layer5}{rgb}{0.5,1.0,0.0}
\definecolor{layer6}{rgb}{0.0,1.0,1.0}
\definecolor{layer7}{rgb}{0.0,0.5,0.0}
\definecolor{layer8}{rgb}{0.6,0.8,0.2}
\definecolor{layer9}{rgb}{1.0,0.08,0.58}
\definecolor{layer10}{rgb}{0.71,0.61,0.05}
\definecolor{layer11}{rgb}{0.0,0.5,1.0}
\definecolor{layer12}{rgb}{0.88,0.88,0.88}
\definecolor{layer13}{rgb}{0.64,0.64,0.64}
\definecolor{layer14}{rgb}{0.37,0.37,0.37}
\definecolor{layer15}{rgb}{0.0,0.0,0.0}
\color{layer0}
\pgfsetlinewidth{1.0pt}
\pgfsetdash{}{0pt}
\pgfline{\pgfxy(56.0,110.0)}{\pgfxy(66.0,110.0)}
\pgfline{\pgfxy(56.0,98.0)}{\pgfxy(66.0,98.0)}
\pgfline{\pgfxy(66.0,98.0)}{\pgfxy(66.0,94.0)}
\pgfline{\pgfxy(66.0,110.0)}{\pgfxy(66.0,114.0)}
\pgfline{\pgfxy(46.0,104.0)}{\pgfxy(54.0,104.0)}
\pgfline{\pgfxy(54.0,98.0)}{\pgfxy(54.0,110.0)}
\pgfline{\pgfxy(56.0,96.0)}{\pgfxy(56.0,112.0)}
\pgfline{\pgfxy(96.0,98.0)}{\pgfxy(86.0,98.0)}
\pgfline{\pgfxy(96.0,110.0)}{\pgfxy(86.0,110.0)}
\pgfline{\pgfxy(86.0,110.0)}{\pgfxy(86.0,114.0)}
\pgfline{\pgfxy(86.0,98.0)}{\pgfxy(86.0,94.0)}
\pgfline{\pgfxy(106.0,104.0)}{\pgfxy(98.0,104.0)}
\pgfline{\pgfxy(98.0,110.0)}{\pgfxy(98.0,98.0)}
\pgfline{\pgfxy(96.0,112.0)}{\pgfxy(96.0,96.0)}
\pgfline{\pgfxy(66.0,70.0)}{\pgfxy(76.0,70.0)}
\pgfline{\pgfxy(66.0,58.0)}{\pgfxy(76.0,58.0)}
\pgfline{\pgfxy(76.0,58.0)}{\pgfxy(76.0,54.0)}
\pgfline{\pgfxy(76.0,70.0)}{\pgfxy(76.0,74.0)}
\pgfline{\pgfxy(56.0,64.0)}{\pgfxy(60.0,64.0)}
\pgfline{\pgfxy(64.0,58.0)}{\pgfxy(64.0,70.0)}
\pgfline{\pgfxy(66.0,56.0)}{\pgfxy(66.0,72.0)}
\pgfellipse[stroke]{\pgfxy(62.0,64.0)}{\pgfxy(2.0,0)}{\pgfxy(0,2.0)}
\pgfline{\pgfxy(86.0,28.0)}{\pgfxy(76.0,28.0)}
\pgfline{\pgfxy(86.0,40.0)}{\pgfxy(76.0,40.0)}
\pgfline{\pgfxy(76.0,40.0)}{\pgfxy(76.0,44.0)}
\pgfline{\pgfxy(76.0,28.0)}{\pgfxy(76.0,24.0)}
\pgfline{\pgfxy(96.0,34.0)}{\pgfxy(92.0,34.0)}
\pgfline{\pgfxy(88.0,40.0)}{\pgfxy(88.0,28.0)}
\pgfline{\pgfxy(86.0,42.0)}{\pgfxy(86.0,26.0)}
\pgfellipse[stroke]{\pgfxy(90.0,34.0)}{\pgfxy(2.0,0)}{\pgfxy(0,2.0)}
\pgfline{\pgfxy(66.0,14.0)}{\pgfxy(86.0,14.0)}
\pgfline{\pgfxy(76.0,24.0)}{\pgfxy(76.0,14.0)}
\pgfline{\pgfxy(76.0,44.0)}{\pgfxy(76.0,54.0)}
\pgfline{\pgfxy(76.0,74.0)}{\pgfxy(76.0,84.0)}
\pgfline{\pgfxy(66.0,84.0)}{\pgfxy(86.0,84.0)}
\pgfline{\pgfxy(86.0,84.0)}{\pgfxy(86.0,94.0)}
\pgfline{\pgfxy(66.0,84.0)}{\pgfxy(66.0,94.0)}
\pgfline{\pgfxy(46.0,104.0)}{\pgfxy(46.0,64.0)}
\pgfline{\pgfxy(56.0,64.0)}{\pgfxy(36.0,64.0)}
\pgfline{\pgfxy(96.0,34.0)}{\pgfxy(106.0,34.0)}
\pgfline{\pgfxy(106.0,34.0)}{\pgfxy(106.0,104.0)}
\pgfline{\pgfxy(106.0,34.0)}{\pgfxy(116.0,34.0)}
\pgfmoveto{\pgfxy(60.0,118.0)}
\pgflineto{\pgfxy(72.0,118.0)}
\pgflineto{\pgfxy(66.0,124.0)}
\pgflineto{\pgfxy(66.0,124.0)}
\pgfclosepath 
\pgfqstroke 
\pgfline{\pgfxy(66.0,114.0)}{\pgfxy(66.0,118.0)}
\pgfmoveto{\pgfxy(80.0,118.0)}
\pgflineto{\pgfxy(92.0,118.0)}
\pgflineto{\pgfxy(86.0,124.0)}
\pgflineto{\pgfxy(86.0,124.0)}
\pgfclosepath 
\pgfqstroke 
\pgfline{\pgfxy(86.0,114.0)}{\pgfxy(86.0,118.0)}
\pgfline{\pgfxy(34.0,64.0)}{\pgfxy(36.0,64.0)}
\pgfmoveto{\pgfxy(34.0,64.0)}
\pgflineto{\pgfxy(34.0,64.0)}
\pgflineto{\pgfxy(32.0,62.0)}
\pgflineto{\pgfxy(26.0,62.0)}
\pgflineto{\pgfxy(26.0,66.0)}
\pgflineto{\pgfxy(32.0,66.0)}
\pgfclosepath 
\pgfqstroke 
\pgfline{\pgfxy(118.0,34.0)}{\pgfxy(116.0,34.0)}
\pgfmoveto{\pgfxy(118.0,34.0)}
\pgflineto{\pgfxy(118.0,34.0)}
\pgflineto{\pgfxy(120.0,36.0)}
\pgflineto{\pgfxy(126.0,36.0)}
\pgflineto{\pgfxy(126.0,32.0)}
\pgflineto{\pgfxy(120.0,32.0)}
\pgfclosepath 
\pgfqstroke 
\begin{pgfmagnify}{1}{-1}
\pgfputat{\pgfxy(6,-54)}{\pgfbox[left,top]{$\overline{\textrm{D}}$}}
\end{pgfmagnify}
\begin{pgfmagnify}{1}{-1}
\pgfputat{\pgfxy(136,-24)}{\pgfbox[left,top]{$\overline{\textrm{BL}}$}}
\end{pgfmagnify}
\pgfsetlinewidth{0.33pt}
\pgfcircle[fill]{\pgfxy(46,64)}{2.0pt}\pgfcircle[fill]{\pgfxy(76,84)}{2.0pt}\pgfcircle[fill]{\pgfxy(106,34)}{2.0pt}\end{pgfmagnify}
\end{pgfpicture}}
    \vspace*{-5mm}
    \caption{}
    \label{fig:Static_AND}
  \end{subfigure}
  \hspace*{0.2mm}
  \begin{subfigure}[b]{0.5\linewidth}
    \centering
    \scalebox{0.8}{\begin{pgfpicture}{0cm}{0cm}{158pt}{178pt}
\pgfsetxvec{\pgfpoint{1pt}{0pt}}
\pgfsetyvec{\pgfpoint{0pt}{1pt}}
\pgfsetroundjoin 
\pgfsetroundcap
\pgftranslateto{\pgfxy(0,178)}
\begin{pgfmagnify}{1}{-1}
\definecolor{layer0}{rgb}{0.0,0.0,0.0}
\definecolor{layer1}{rgb}{0.0,0.0,0.5}
\definecolor{layer2}{rgb}{1.0,0.0,0.0}
\definecolor{layer3}{rgb}{0.0,0.5,0.5}
\definecolor{layer4}{rgb}{1.0,0.78,0.0}
\definecolor{layer5}{rgb}{0.5,1.0,0.0}
\definecolor{layer6}{rgb}{0.0,1.0,1.0}
\definecolor{layer7}{rgb}{0.0,0.5,0.0}
\definecolor{layer8}{rgb}{0.6,0.8,0.2}
\definecolor{layer9}{rgb}{1.0,0.08,0.58}
\definecolor{layer10}{rgb}{0.71,0.61,0.05}
\definecolor{layer11}{rgb}{0.0,0.5,1.0}
\definecolor{layer12}{rgb}{0.88,0.88,0.88}
\definecolor{layer13}{rgb}{0.64,0.64,0.64}
\definecolor{layer14}{rgb}{0.37,0.37,0.37}
\definecolor{layer15}{rgb}{0.0,0.0,0.0}
\color{layer0}
\pgfsetlinewidth{1.0pt}
\pgfsetdash{}{0pt}
\pgfline{\pgfxy(62.0,62.0)}{\pgfxy(82.0,62.0)}
\pgfline{\pgfxy(62.0,88.0)}{\pgfxy(72.0,88.0)}
\pgfline{\pgfxy(62.0,76.0)}{\pgfxy(72.0,76.0)}
\pgfline{\pgfxy(72.0,76.0)}{\pgfxy(72.0,72.0)}
\pgfline{\pgfxy(72.0,88.0)}{\pgfxy(72.0,92.0)}
\pgfline{\pgfxy(52.0,82.0)}{\pgfxy(56.0,82.0)}
\pgfline{\pgfxy(60.0,76.0)}{\pgfxy(60.0,88.0)}
\pgfline{\pgfxy(62.0,74.0)}{\pgfxy(62.0,90.0)}
\pgfellipse[stroke]{\pgfxy(58.0,82.0)}{\pgfxy(2.0,0)}{\pgfxy(0,2.0)}
\pgfline{\pgfxy(82.0,106.0)}{\pgfxy(72.0,106.0)}
\pgfline{\pgfxy(82.0,118.0)}{\pgfxy(72.0,118.0)}
\pgfline{\pgfxy(72.0,118.0)}{\pgfxy(72.0,122.0)}
\pgfline{\pgfxy(72.0,106.0)}{\pgfxy(72.0,102.0)}
\pgfline{\pgfxy(92.0,112.0)}{\pgfxy(88.0,112.0)}
\pgfline{\pgfxy(84.0,118.0)}{\pgfxy(84.0,106.0)}
\pgfline{\pgfxy(82.0,120.0)}{\pgfxy(82.0,104.0)}
\pgfellipse[stroke]{\pgfxy(86.0,112.0)}{\pgfxy(2.0,0)}{\pgfxy(0,2.0)}
\pgfline{\pgfxy(72.0,72.0)}{\pgfxy(72.0,62.0)}
\pgfline{\pgfxy(72.0,102.0)}{\pgfxy(72.0,92.0)}
\pgfline{\pgfxy(52.0,82.0)}{\pgfxy(42.0,82.0)}
\pgfline{\pgfxy(92.0,112.0)}{\pgfxy(112.0,112.0)}
\pgfline{\pgfxy(72.0,122.0)}{\pgfxy(72.0,132.0)}
\pgfline{\pgfxy(72.0,132.0)}{\pgfxy(62.0,132.0)}
\pgfline{\pgfxy(72.0,132.0)}{\pgfxy(92.0,132.0)}
\pgfline{\pgfxy(82.0,132.0)}{\pgfxy(82.0,142.0)}
\pgfline{\pgfxy(62.0,132.0)}{\pgfxy(62.0,142.0)}
\pgfline{\pgfxy(52.0,158.0)}{\pgfxy(62.0,158.0)}
\pgfline{\pgfxy(52.0,146.0)}{\pgfxy(62.0,146.0)}
\pgfline{\pgfxy(62.0,146.0)}{\pgfxy(62.0,142.0)}
\pgfline{\pgfxy(62.0,158.0)}{\pgfxy(62.0,162.0)}
\pgfline{\pgfxy(42.0,152.0)}{\pgfxy(50.0,152.0)}
\pgfline{\pgfxy(50.0,146.0)}{\pgfxy(50.0,158.0)}
\pgfline{\pgfxy(52.0,144.0)}{\pgfxy(52.0,160.0)}
\pgfline{\pgfxy(92.0,146.0)}{\pgfxy(82.0,146.0)}
\pgfline{\pgfxy(92.0,158.0)}{\pgfxy(82.0,158.0)}
\pgfline{\pgfxy(82.0,158.0)}{\pgfxy(82.0,162.0)}
\pgfline{\pgfxy(82.0,146.0)}{\pgfxy(82.0,142.0)}
\pgfline{\pgfxy(102.0,152.0)}{\pgfxy(94.0,152.0)}
\pgfline{\pgfxy(94.0,158.0)}{\pgfxy(94.0,146.0)}
\pgfline{\pgfxy(92.0,160.0)}{\pgfxy(92.0,144.0)}
\pgfmoveto{\pgfxy(76.0,166.0)}
\pgflineto{\pgfxy(88.0,166.0)}
\pgflineto{\pgfxy(82.0,172.0)}
\pgflineto{\pgfxy(82.0,172.0)}
\pgfclosepath 
\pgfqstroke 
\pgfline{\pgfxy(82.0,162.0)}{\pgfxy(82.0,166.0)}
\pgfmoveto{\pgfxy(56.0,166.0)}
\pgflineto{\pgfxy(68.0,166.0)}
\pgflineto{\pgfxy(62.0,172.0)}
\pgflineto{\pgfxy(62.0,172.0)}
\pgfclosepath 
\pgfqstroke 
\pgfline{\pgfxy(62.0,162.0)}{\pgfxy(62.0,166.0)}
\pgfline{\pgfxy(42.0,152.0)}{\pgfxy(42.0,82.0)}
\pgfsetlinewidth{0.33pt}
\pgfcircle[fill]{\pgfxy(42,82)}{2.0pt}\pgfsetlinewidth{1.0pt}
\pgfline{\pgfxy(112.0,152.0)}{\pgfxy(112.0,112.0)}
\pgfsetlinewidth{0.33pt}
\pgfcircle[fill]{\pgfxy(112,112)}{2.0pt}\pgfcircle[fill]{\pgfxy(72,132)}{2.0pt}\pgfcircle[fill]{\pgfxy(82,132)}{2.0pt}\pgfsetlinewidth{1.0pt}
\pgfline{\pgfxy(92.0,132.0)}{\pgfxy(112.0,132.0)}
\pgfline{\pgfxy(122.0,138.0)}{\pgfxy(132.0,138.0)}
\pgfline{\pgfxy(122.0,126.0)}{\pgfxy(132.0,126.0)}
\pgfline{\pgfxy(132.0,126.0)}{\pgfxy(132.0,122.0)}
\pgfline{\pgfxy(132.0,138.0)}{\pgfxy(132.0,142.0)}
\pgfline{\pgfxy(112.0,132.0)}{\pgfxy(120.0,132.0)}
\pgfline{\pgfxy(120.0,126.0)}{\pgfxy(120.0,138.0)}
\pgfline{\pgfxy(122.0,124.0)}{\pgfxy(122.0,140.0)}
\pgfmoveto{\pgfxy(126.0,146.0)}
\pgflineto{\pgfxy(138.0,146.0)}
\pgflineto{\pgfxy(132.0,152.0)}
\pgflineto{\pgfxy(132.0,152.0)}
\pgfclosepath 
\pgfqstroke 
\pgfline{\pgfxy(132.0,142.0)}{\pgfxy(132.0,146.0)}
\pgfline{\pgfxy(132.0,122.0)}{\pgfxy(132.0,112.0)}
\pgfline{\pgfxy(132.0,112.0)}{\pgfxy(142.0,112.0)}
\pgfline{\pgfxy(142.0,112.0)}{\pgfxy(144.0,112.0)}
\pgfmoveto{\pgfxy(152.0,112.0)}
\pgflineto{\pgfxy(152.0,112.0)}
\pgflineto{\pgfxy(150.0,110.0)}
\pgflineto{\pgfxy(144.0,110.0)}
\pgflineto{\pgfxy(144.0,114.0)}
\pgflineto{\pgfxy(150.0,114.0)}
\pgfclosepath 
\pgfqstroke 
\begin{pgfmagnify}{1}{-1}
\pgfputat{\pgfxy(62,-22)}{\pgfbox[left,top]{CAM}}
\end{pgfmagnify}
\pgfline{\pgfxy(42.0,32.0)}{\pgfxy(52.0,32.0)}
\pgfline{\pgfxy(112.0,32.0)}{\pgfxy(102.0,32.0)}
\pgfmoveto{\pgfxy(52,12)}
\pgflineto{\pgfxy(102,12)}
\pgflineto{\pgfxy(102,52)}
\pgflineto{\pgfxy(52,52)}
\pgfclosepath 
\pgfqstroke 
\pgfline{\pgfxy(112.0,112.0)}{\pgfxy(112.0,32.0)}
\pgfline{\pgfxy(42.0,82.0)}{\pgfxy(42.0,32.0)}
\pgfline{\pgfxy(112.0,152.0)}{\pgfxy(102.0,152.0)}
\begin{pgfmagnify}{1}{-1}
\pgfputat{\pgfxy(132,-92)}{\pgfbox[left,top]{\overlinerm{AND}}}
\end{pgfmagnify}
\begin{pgfmagnify}{1}{-1}
\pgfputat{\pgfxy(22,-22)}{\pgfbox[left,top]{$\overline{\textrm{D}}$}}
\end{pgfmagnify}
\begin{pgfmagnify}{1}{-1}
\pgfputat{\pgfxy(122,-22)}{\pgfbox[left,top]{$\overline{\textrm{BL}}$}}
\end{pgfmagnify}
\end{pgfmagnify}
\end{pgfpicture}}
    \vspace*{-5mm}
    \caption{} 
    \label{fig:Static_cell}
  \end{subfigure}
  \begin{subfigure}[b]{\linewidth}
    \vspace*{2mm}
    \centering
    \includegraphics[width=0.4\linewidth]{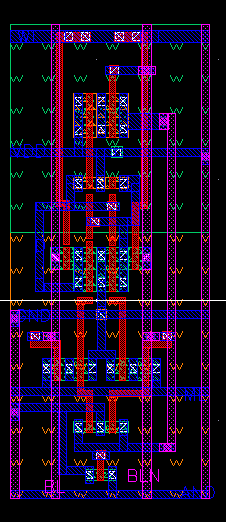}
    \vspace*{-5mm}
    \caption{}
    \label{fig:AND_ST_layout}
  \end{subfigure}
  \caption{The static AND gate and memory cell. \textbf{(a)} The static CMOS AND gate. \textbf{(b)} The memory cell that embeds the static AND gate. \textbf{(c)} The cell layout.}
  \label{fig:Static_AND_gate_and_cell}
\end{figure}

A second cell embedding a static CMOS logic AND gate is proposed. The circuits of the gate and the cell are depicted in \autoref{fig:Static_AND_gate_and_cell}.

The static AND gate is presented in \autoref{fig:Static_AND}. With respect to the dynamic AND (\autoref{sec:Dynamic_AND}), a larger cell footprint is required, since the additional pMOS transistors have to be sized with a width larger than the precharge transistor in \autoref{fig:Dynamic_AND}, following the rules of standard microeletronic design \cite{rabaey_book}. However, the addition of these allows to remove the precharge signal $\overline{PRE}$ of \autoref{fig:Dynamic_AND}, which is required for the dynamic logic functioning. The gate is embedded in the memory cell, as it is shown in \autoref{fig:Static_cell}, and its output is connected to the pull--down transistor of the AND line. The truth table for the gate is the same of the dynamic AND cell, which is reported in \autoref{tab:AND_line}, except for the fact that, for the static cell, the AND output signal is a static CMOS one.

\subsection{Special Purpose AND}
\label{sec:Special_AND}

\begin{figure}[t!p]
  \begin{subfigure}[b]{0.5\linewidth}
    \centering
    \scalebox{0.8}{\begin{pgfpicture}{0cm}{0cm}{150pt}{108pt}
\pgfsetxvec{\pgfpoint{1pt}{0pt}}
\pgfsetyvec{\pgfpoint{0pt}{1pt}}
\pgfsetroundjoin 
\pgfsetroundcap
\pgftranslateto{\pgfxy(0,108)}
\begin{pgfmagnify}{1}{-1}
\definecolor{layer0}{rgb}{0.0,0.0,0.0}
\definecolor{layer1}{rgb}{0.0,0.0,0.5}
\definecolor{layer2}{rgb}{1.0,0.0,0.0}
\definecolor{layer3}{rgb}{0.0,0.5,0.5}
\definecolor{layer4}{rgb}{1.0,0.78,0.0}
\definecolor{layer5}{rgb}{0.5,1.0,0.0}
\definecolor{layer6}{rgb}{0.0,1.0,1.0}
\definecolor{layer7}{rgb}{0.0,0.5,0.0}
\definecolor{layer8}{rgb}{0.6,0.8,0.2}
\definecolor{layer9}{rgb}{1.0,0.08,0.58}
\definecolor{layer10}{rgb}{0.71,0.61,0.05}
\definecolor{layer11}{rgb}{0.0,0.5,1.0}
\definecolor{layer12}{rgb}{0.88,0.88,0.88}
\definecolor{layer13}{rgb}{0.64,0.64,0.64}
\definecolor{layer14}{rgb}{0.37,0.37,0.37}
\definecolor{layer15}{rgb}{0.0,0.0,0.0}
\color{layer0}
\pgfsetlinewidth{1.0pt}
\pgfsetdash{}{0pt}
\pgfline{\pgfxy(72.0,58.0)}{\pgfxy(82.0,58.0)}
\pgfline{\pgfxy(72.0,46.0)}{\pgfxy(82.0,46.0)}
\pgfline{\pgfxy(82.0,46.0)}{\pgfxy(82.0,42.0)}
\pgfline{\pgfxy(82.0,58.0)}{\pgfxy(82.0,62.0)}
\pgfline{\pgfxy(62.0,52.0)}{\pgfxy(70.0,52.0)}
\pgfline{\pgfxy(70.0,46.0)}{\pgfxy(70.0,58.0)}
\pgfline{\pgfxy(72.0,44.0)}{\pgfxy(72.0,60.0)}
\pgfline{\pgfxy(92.0,76.0)}{\pgfxy(82.0,76.0)}
\pgfline{\pgfxy(92.0,88.0)}{\pgfxy(82.0,88.0)}
\pgfline{\pgfxy(82.0,88.0)}{\pgfxy(82.0,92.0)}
\pgfline{\pgfxy(82.0,76.0)}{\pgfxy(82.0,72.0)}
\pgfline{\pgfxy(102.0,82.0)}{\pgfxy(94.0,82.0)}
\pgfline{\pgfxy(94.0,88.0)}{\pgfxy(94.0,76.0)}
\pgfline{\pgfxy(92.0,90.0)}{\pgfxy(92.0,74.0)}
\pgfline{\pgfxy(82.0,62.0)}{\pgfxy(82.0,72.0)}
\pgfline{\pgfxy(82.0,42.0)}{\pgfxy(82.0,32.0)}
\pgfline{\pgfxy(82.0,32.0)}{\pgfxy(112.0,32.0)}
\pgfline{\pgfxy(60.0,52.0)}{\pgfxy(62.0,52.0)}
\pgfmoveto{\pgfxy(60.0,52.0)}
\pgflineto{\pgfxy(60.0,52.0)}
\pgflineto{\pgfxy(58.0,50.0)}
\pgflineto{\pgfxy(52.0,50.0)}
\pgflineto{\pgfxy(52.0,54.0)}
\pgflineto{\pgfxy(58.0,54.0)}
\pgfclosepath 
\pgfqstroke 
\pgfline{\pgfxy(104.0,82.0)}{\pgfxy(102.0,82.0)}
\pgfmoveto{\pgfxy(104.0,82.0)}
\pgflineto{\pgfxy(104.0,82.0)}
\pgflineto{\pgfxy(106.0,84.0)}
\pgflineto{\pgfxy(112.0,84.0)}
\pgflineto{\pgfxy(112.0,80.0)}
\pgflineto{\pgfxy(106.0,80.0)}
\pgfclosepath 
\pgfqstroke 
\pgfmoveto{\pgfxy(76.0,96.0)}
\pgflineto{\pgfxy(88.0,96.0)}
\pgflineto{\pgfxy(82.0,102.0)}
\pgflineto{\pgfxy(82.0,102.0)}
\pgfclosepath 
\pgfqstroke 
\pgfline{\pgfxy(82.0,92.0)}{\pgfxy(82.0,96.0)}
\begin{pgfmagnify}{1}{-1}
\pgfputat{\pgfxy(122,-72)}{\pgfbox[left,top]{$\overline{\textrm{BL}}$}}
\end{pgfmagnify}
\begin{pgfmagnify}{1}{-1}
\pgfputat{\pgfxy(32,-42)}{\pgfbox[left,top]{$\overline{\textrm{D}}$}}
\end{pgfmagnify}
\pgfline{\pgfxy(112.0,32.0)}{\pgfxy(114.0,32.0)}
\pgfmoveto{\pgfxy(122.0,32.0)}
\pgflineto{\pgfxy(122.0,32.0)}
\pgflineto{\pgfxy(120.0,30.0)}
\pgflineto{\pgfxy(114.0,30.0)}
\pgflineto{\pgfxy(114.0,34.0)}
\pgflineto{\pgfxy(120.0,34.0)}
\pgfclosepath 
\pgfqstroke 
\begin{pgfmagnify}{1}{-1}
\pgfputat{\pgfxy(132,-22)}{\pgfbox[left,top]{AND}}
\end{pgfmagnify}
\end{pgfmagnify}
\end{pgfpicture}}
    \vspace*{-5mm}
    \caption{}
    \label{fig:Special_AND}
  \end{subfigure}
  \hspace*{0.2mm}
  \begin{subfigure}[b]{0.5\linewidth}
    \centering
    \scalebox{0.8}{\begin{pgfpicture}{0cm}{0cm}{148pt}{148pt}
\pgfsetxvec{\pgfpoint{1pt}{0pt}}
\pgfsetyvec{\pgfpoint{0pt}{1pt}}
\pgfsetroundjoin 
\pgfsetroundcap
\pgftranslateto{\pgfxy(0,148)}
\begin{pgfmagnify}{1}{-1}
\definecolor{layer0}{rgb}{0.0,0.0,0.0}
\definecolor{layer1}{rgb}{0.0,0.0,0.5}
\definecolor{layer2}{rgb}{1.0,0.0,0.0}
\definecolor{layer3}{rgb}{0.0,0.5,0.5}
\definecolor{layer4}{rgb}{1.0,0.78,0.0}
\definecolor{layer5}{rgb}{0.5,1.0,0.0}
\definecolor{layer6}{rgb}{0.0,1.0,1.0}
\definecolor{layer7}{rgb}{0.0,0.5,0.0}
\definecolor{layer8}{rgb}{0.6,0.8,0.2}
\definecolor{layer9}{rgb}{1.0,0.08,0.58}
\definecolor{layer10}{rgb}{0.71,0.61,0.05}
\definecolor{layer11}{rgb}{0.0,0.5,1.0}
\definecolor{layer12}{rgb}{0.88,0.88,0.88}
\definecolor{layer13}{rgb}{0.64,0.64,0.64}
\definecolor{layer14}{rgb}{0.37,0.37,0.37}
\definecolor{layer15}{rgb}{0.0,0.0,0.0}
\color{layer0}
\pgfsetlinewidth{1.0pt}
\pgfsetdash{}{0pt}
\pgfline{\pgfxy(82.0,102.0)}{\pgfxy(82.0,112.0)}
\pgfline{\pgfxy(92.0,116.0)}{\pgfxy(82.0,116.0)}
\pgfline{\pgfxy(92.0,128.0)}{\pgfxy(82.0,128.0)}
\pgfline{\pgfxy(82.0,128.0)}{\pgfxy(82.0,132.0)}
\pgfline{\pgfxy(82.0,116.0)}{\pgfxy(82.0,112.0)}
\pgfline{\pgfxy(102.0,122.0)}{\pgfxy(94.0,122.0)}
\pgfline{\pgfxy(94.0,128.0)}{\pgfxy(94.0,116.0)}
\pgfline{\pgfxy(92.0,130.0)}{\pgfxy(92.0,114.0)}
\pgfline{\pgfxy(72.0,98.0)}{\pgfxy(82.0,98.0)}
\pgfline{\pgfxy(72.0,86.0)}{\pgfxy(82.0,86.0)}
\pgfline{\pgfxy(82.0,86.0)}{\pgfxy(82.0,82.0)}
\pgfline{\pgfxy(82.0,98.0)}{\pgfxy(82.0,102.0)}
\pgfline{\pgfxy(62.0,92.0)}{\pgfxy(70.0,92.0)}
\pgfline{\pgfxy(70.0,86.0)}{\pgfxy(70.0,98.0)}
\pgfline{\pgfxy(72.0,84.0)}{\pgfxy(72.0,100.0)}
\pgfmoveto{\pgfxy(76.0,136.0)}
\pgflineto{\pgfxy(88.0,136.0)}
\pgflineto{\pgfxy(82.0,142.0)}
\pgflineto{\pgfxy(82.0,142.0)}
\pgfclosepath 
\pgfqstroke 
\pgfline{\pgfxy(82.0,132.0)}{\pgfxy(82.0,136.0)}
\pgfline{\pgfxy(122.0,72.0)}{\pgfxy(82.0,72.0)}
\pgfline{\pgfxy(122.0,72.0)}{\pgfxy(124.0,72.0)}
\pgfmoveto{\pgfxy(132.0,72.0)}
\pgflineto{\pgfxy(132.0,72.0)}
\pgflineto{\pgfxy(130.0,70.0)}
\pgflineto{\pgfxy(124.0,70.0)}
\pgflineto{\pgfxy(124.0,74.0)}
\pgflineto{\pgfxy(130.0,74.0)}
\pgfclosepath 
\pgfqstroke 
\pgfmoveto{\pgfxy(52,12)}
\pgflineto{\pgfxy(102,12)}
\pgflineto{\pgfxy(102,52)}
\pgflineto{\pgfxy(52,52)}
\pgfclosepath 
\pgfqstroke 
\begin{pgfmagnify}{1}{-1}
\pgfputat{\pgfxy(62,-22)}{\pgfbox[left,top]{CAM}}
\end{pgfmagnify}
\pgfline{\pgfxy(42.0,32.0)}{\pgfxy(52.0,32.0)}
\pgfline{\pgfxy(112.0,32.0)}{\pgfxy(102.0,32.0)}
\pgfline{\pgfxy(62.0,92.0)}{\pgfxy(42.0,92.0)}
\pgfline{\pgfxy(42.0,32.0)}{\pgfxy(42.0,92.0)}
\pgfline{\pgfxy(102.0,122.0)}{\pgfxy(112.0,122.0)}
\pgfline{\pgfxy(112.0,122.0)}{\pgfxy(112.0,32.0)}
\pgfline{\pgfxy(82.0,82.0)}{\pgfxy(82.0,72.0)}
\begin{pgfmagnify}{1}{-1}
\pgfputat{\pgfxy(142,-62)}{\pgfbox[left,top]{AND}}
\end{pgfmagnify}
\begin{pgfmagnify}{1}{-1}
\pgfputat{\pgfxy(122,-22)}{\pgfbox[left,top]{$\overline{\textrm{BL}}$}}
\end{pgfmagnify}
\begin{pgfmagnify}{1}{-1}
\pgfputat{\pgfxy(22,-22)}{\pgfbox[left,top]{$\overline{\textrm{D}}$}}
\end{pgfmagnify}
\end{pgfmagnify}
\end{pgfpicture}}
    \vspace*{-5mm}
    \caption{} 
    \label{fig:Special_cell}
  \end{subfigure}
  \begin{subfigure}[b]{\linewidth}
    \vspace*{2mm}
    \centering
    \includegraphics[width=0.5\linewidth]{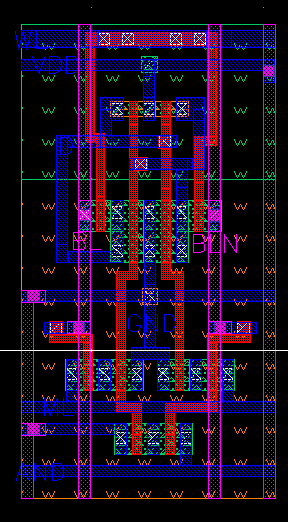}
    \vspace*{-5mm}
    \caption{}
    \label{fig:AND_SP_layout}
  \end{subfigure}
  \caption{The special--purpose AND gate and memory cell. Taking into account the algorithm characteristics, it is possible to implement the AND operation by using only two additional transistors. \textbf{(a)} The special--purpose AND gate. \textbf{(b)} The memory cell that embeds the logic gate. \textbf{(c)} The cell layout.}
  \label{fig:Special_AND_gate_and_cell}
\end{figure}

A third variant of the cell is proposed. The objective of this cell design is to reduce as much as possible the cell area overhead resulting from the addition of the AND gate, by making design choices tuned on the characteristics of the algorithm. The schematics of the gate and the cell are depicted in \autoref{fig:Special_AND_gate_and_cell}.

\smallskip

As it is highlighted in \autoref{fig:Algorithm}, the mask vector is used to select a memory column at each iteration by setting the corresponding bit in the mask to `1', while all the other cells are disabled. Since the AND operation is computed between a bit equal to `1' and the cell content, the result of this is determined by the cell, as it is shown in \autoref{eq:Eq_selection}; hence, it is more a selection operation than an AND one. For this reason, the cell circuit can be simplified to only implement the cell selection operation using the bitlines on which the mask vector is put, instead of a proper AND function, and to allow the selected cell content to be reflected on the AND line. This result can be achieved by connecting a single pull--down transistor with the input on the cell content and the output on the AND line, as it is depicted in \autoref{fig:Special_AND}.

Since the cell has to be selected only when the mask bit $M$ is equal to the logic `1' (i.e. $BL$=`1', $\overline{BL}$=`0'), it should be disconnected from the AND line when $M$=`0' (i.e. $BL$=`0', $\overline{BL}$=`1'); hence, it would be enough to add a footer transistor, which gate is connected to $BL$, on the source of the pull-down one in order to disable this. However, since the static (\autoref{fig:Static_AND}) and dynamic (\autoref{fig:Dynamic_AND}) gates have one of their inputs connected to $\overline{BL}$ instead of $BL$, a different encoding of the mask vector is used in this case, using the logic `0' as active value for the mask bit instead of the logic `1'; in this way, the footer transistor in \autoref{fig:Special_AND} can be connected to $\overline{BL}$; therefore, the three variants are equivalent in terms of connections to the memory signal lines and, hence, can be properly compared.

\smallskip

For what concerns the pull--down transistor, its gate is connected to the output of an AND logic gate in the static (\autoref{fig:Static_AND}) and dynamic (\autoref{fig:Dynamic_AND}) gates; in \autoref{fig:Special_AND}, instead, it is connected to the negated value of the cell content $\overline{D}$; in fact, once the cell is selected, the algorithm needs only to know if the cell content is equal to `0' or `1', and the latter can be connected directly to the pull--down transistor gate. In this way, when $D$=`1' ($\overline{D}$=`0'), the AND logic gate is disabled, the line is charged to the logic `1'; when $D$=`0' ($\overline{D}$=`1'), the pull--down transistor is enabled, the line is not charged and a logic `0' is sensed.

\smallskip

One can notice that the output pin of the cell is denoted with $AND$ instead of $\overline{AND}$, in \autoref{fig:Special_cell}: this is due to the fact that the AND result is not inverted by the pull--down transistor. In fact, the pull--down transistors of the unselected columns are disabled using the mechanism presented in \autoref{fig:Current_save_matchline} and, hence, the AND result on the selected column can be directly reported on the line. If the selected cell content $D_i$ is equal to `1', the line is charged and $D_i \cdot M_i=$`1' ($M_i$ is the active mask bit) is registered in output; otherwise, the line does not get charged and $D_i \cdot M_i=$`0'. Hence, there is no need for an additional separation stage between cell core and AND line, while there is for the static and dynamic implementations of \autoref{fig:Dynamic_cell} and \autoref{fig:Static_cell}, respectively, which logic gates outputs have to be disconnected from the line when the corresponding cells are not selected. The truth table for the special-purpose AND cell of \autoref{fig:Special_cell} is shown in \autoref{tab:AND_sp_line}.

\begin{table}[h]
  \centering
  \begin{tabular}{ccccc}
    \toprule
    \textit{\textbf{D}} & $BL$ & $\overline{D}$ & $\overline{\textit{\textbf{BL}}}$ & \textit{\textbf{AND}}\\
    \midrule
    0 & 0 & 1 & 1 & 0\\
    - & 1 & - & 0 & 0 $\rightarrow$ 1\\
    1 & 0 & 0 & 1 & 0 $\rightarrow$ 1\\
    \bottomrule
  \end{tabular}
   \smallskip
    \caption{The truth table of the special--purpose AND cell of \autoref{fig:Special_cell}. When evaluating this function, one needs to remember that $\overline{BL}$ is not a proper data signal but a selection one that allows to report the cell content $D$ on the line. Every time $\overline{BL}$=`0', the AND logic gate is disabled and the line is charged to `1' (in particular, the pull--down is prevented from discharging the line in the case in which $D$=`0').}
  \label{tab:AND_sp_line}
\end{table}

\medskip

The special--purpose cell in \autoref{fig:Special_cell} is characterised by the lowest area overhead (lowest number of additional transistors) among the cells. However, these are able to perform a proper AND logic operation, which can be useful for implementing other algorithms; nevertheless, in the special--purpose cell circuit it is demonstrated that, with proper optimisations, it is possible to greatly reduce the area overhead introduced by the logic circuits. 

The dynamic and static cells, in \autoref{fig:Dynamic_cell} and \autoref{fig:Static_cell} respectively, are characterised by the same number of transistors, but the static one occupies a larger area due to the pull--up pMOS transistors in the logic gate, that are much larger than the precharge pMOS of the dynamic cell; however, the static cell does not require the ($\overline{PRE}$) signal for its functioning, which leads to smaller cell and row areas.

\medskip

\subsection{Dummy line sensing scheme}
\label{sec:Dummy_line_scheme}

For the LiM array, the same dummy line sensing scheme of the CAM is adopted: dummy cells are used to create a dummy memory line that acts as reference for all the AND sense amplifiers (ANDSAs). 

\begin{figure}[h]
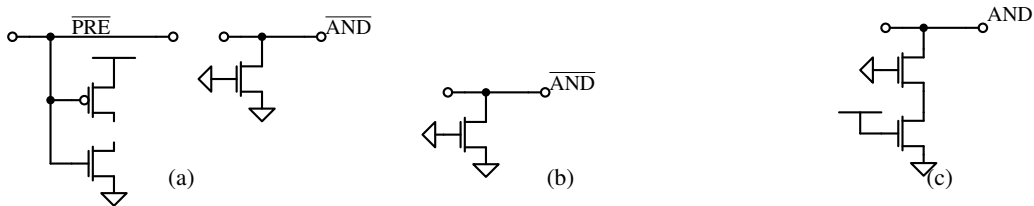

  \centering
  \begin{subfigure}[b]{0.3\linewidth}
    \scalebox{0.8}{\begin{pgfpicture}{0cm}{0cm}{162pt}{120pt}
\pgfsetxvec{\pgfpoint{1pt}{0pt}}
\pgfsetyvec{\pgfpoint{0pt}{1pt}}
\pgfsetroundjoin 
\pgfsetroundcap
\pgftranslateto{\pgfxy(0,120)}
\begin{pgfmagnify}{1}{-1}
\definecolor{layer0}{rgb}{0.0,0.0,0.0}
\definecolor{layer1}{rgb}{0.0,0.0,0.5}
\definecolor{layer2}{rgb}{1.0,0.0,0.0}
\definecolor{layer3}{rgb}{0.0,0.5,0.5}
\definecolor{layer4}{rgb}{1.0,0.78,0.0}
\definecolor{layer5}{rgb}{0.5,1.0,0.0}
\definecolor{layer6}{rgb}{0.0,1.0,1.0}
\definecolor{layer7}{rgb}{0.0,0.5,0.0}
\definecolor{layer8}{rgb}{0.6,0.8,0.2}
\definecolor{layer9}{rgb}{1.0,0.08,0.58}
\definecolor{layer10}{rgb}{0.71,0.61,0.05}
\definecolor{layer11}{rgb}{0.0,0.5,1.0}
\definecolor{layer12}{rgb}{0.88,0.88,0.88}
\definecolor{layer13}{rgb}{0.64,0.64,0.64}
\definecolor{layer14}{rgb}{0.37,0.37,0.37}
\definecolor{layer15}{rgb}{0.0,0.0,0.0}
\color{layer0}
\pgfsetlinewidth{1.0pt}
\pgfsetdash{}{0pt}
\pgfline{\pgfxy(46.0,100.0)}{\pgfxy(56.0,100.0)}
\pgfline{\pgfxy(46.0,88.0)}{\pgfxy(56.0,88.0)}
\pgfline{\pgfxy(56.0,88.0)}{\pgfxy(56.0,84.0)}
\pgfline{\pgfxy(56.0,100.0)}{\pgfxy(56.0,104.0)}
\pgfline{\pgfxy(36.0,94.0)}{\pgfxy(44.0,94.0)}
\pgfline{\pgfxy(44.0,88.0)}{\pgfxy(44.0,100.0)}
\pgfline{\pgfxy(46.0,86.0)}{\pgfxy(46.0,102.0)}
\pgfmoveto{\pgfxy(50.0,108.0)}
\pgflineto{\pgfxy(62.0,108.0)}
\pgflineto{\pgfxy(56.0,114.0)}
\pgflineto{\pgfxy(56.0,114.0)}
\pgfclosepath 
\pgfqstroke 
\pgfline{\pgfxy(56.0,104.0)}{\pgfxy(56.0,108.0)}
\pgfline{\pgfxy(46.0,70.0)}{\pgfxy(56.0,70.0)}
\pgfline{\pgfxy(46.0,58.0)}{\pgfxy(56.0,58.0)}
\pgfline{\pgfxy(56.0,58.0)}{\pgfxy(56.0,54.0)}
\pgfline{\pgfxy(56.0,70.0)}{\pgfxy(56.0,74.0)}
\pgfline{\pgfxy(36.0,64.0)}{\pgfxy(40.0,64.0)}
\pgfline{\pgfxy(44.0,58.0)}{\pgfxy(44.0,70.0)}
\pgfline{\pgfxy(46.0,56.0)}{\pgfxy(46.0,72.0)}
\pgfellipse[stroke]{\pgfxy(42.0,64.0)}{\pgfxy(2.0,0)}{\pgfxy(0,2.0)}
\pgfline{\pgfxy(46.0,44.0)}{\pgfxy(66.0,44.0)}
\pgfline{\pgfxy(56.0,54.0)}{\pgfxy(56.0,44.0)}
\pgfline{\pgfxy(36.0,94.0)}{\pgfxy(26.0,94.0)}
\pgfline{\pgfxy(26.0,94.0)}{\pgfxy(26.0,34.0)}
\pgfline{\pgfxy(36.0,64.0)}{\pgfxy(26.0,64.0)}
\pgfsetlinewidth{0.33pt}
\pgfcircle[fill]{\pgfxy(26,64)}{2.0pt}\pgfsetlinewidth{1.0pt}
\pgfline{\pgfxy(16.0,34.0)}{\pgfxy(76.0,34.0)}
\pgfsetlinewidth{0.33pt}
\pgfcircle[fill]{\pgfxy(26,34)}{2.0pt}\pgfsetlinewidth{1.0pt}
\pgfline{\pgfxy(10.0,34.0)}{\pgfxy(16.0,34.0)}
\pgfellipse[stroke]{\pgfxy(8.0,34.0)}{\pgfxy(2.0,0)}{\pgfxy(0,2.0)}
\pgfline{\pgfxy(82.0,34.0)}{\pgfxy(76.0,34.0)}
\pgfellipse[stroke]{\pgfxy(84.0,34.0)}{\pgfxy(2.0,0)}{\pgfxy(0,2.0)}
\begin{pgfmagnify}{1}{-1}
\pgfputat{\pgfxy(36,-24)}{\pgfbox[left,top]{\overlinerm{PRE}}}
\end{pgfmagnify}
\pgfline{\pgfxy(116.0,60.0)}{\pgfxy(126.0,60.0)}
\pgfline{\pgfxy(116.0,48.0)}{\pgfxy(126.0,48.0)}
\pgfline{\pgfxy(126.0,48.0)}{\pgfxy(126.0,44.0)}
\pgfline{\pgfxy(126.0,60.0)}{\pgfxy(126.0,64.0)}
\pgfline{\pgfxy(106.0,54.0)}{\pgfxy(114.0,54.0)}
\pgfline{\pgfxy(114.0,48.0)}{\pgfxy(114.0,60.0)}
\pgfline{\pgfxy(116.0,46.0)}{\pgfxy(116.0,62.0)}
\pgfmoveto{\pgfxy(102.0,48.0)}
\pgflineto{\pgfxy(102.0,60.0)}
\pgflineto{\pgfxy(96.0,54.0)}
\pgflineto{\pgfxy(96.0,54.0)}
\pgfclosepath 
\pgfqstroke 
\pgfline{\pgfxy(106.0,54.0)}{\pgfxy(102.0,54.0)}
\pgfline{\pgfxy(116.0,34.0)}{\pgfxy(146.0,34.0)}
\pgfline{\pgfxy(110.0,34.0)}{\pgfxy(116.0,34.0)}
\pgfellipse[stroke]{\pgfxy(108.0,34.0)}{\pgfxy(2.0,0)}{\pgfxy(0,2.0)}
\pgfline{\pgfxy(152.0,34.0)}{\pgfxy(146.0,34.0)}
\pgfellipse[stroke]{\pgfxy(154.0,34.0)}{\pgfxy(2.0,0)}{\pgfxy(0,2.0)}
\pgfline{\pgfxy(126.0,44.0)}{\pgfxy(126.0,34.0)}
\pgfmoveto{\pgfxy(120.0,68.0)}
\pgflineto{\pgfxy(132.0,68.0)}
\pgflineto{\pgfxy(126.0,74.0)}
\pgflineto{\pgfxy(126.0,74.0)}
\pgfclosepath 
\pgfqstroke 
\pgfline{\pgfxy(126.0,64.0)}{\pgfxy(126.0,68.0)}
\pgfsetlinewidth{0.33pt}
\pgfcircle[fill]{\pgfxy(126,34)}{2.0pt}\begin{pgfmagnify}{1}{-1}
\pgfputat{\pgfxy(156,-24)}{\pgfbox[left,top]{\overlinerm{AND}}}
\end{pgfmagnify}
\end{pgfmagnify}
\end{pgfpicture}}
    \vspace*{-9mm}
    \caption{}
    \label{fig:Dummy_cell_dynamic_AND}
  \end{subfigure}
  \begin{subfigure}[b]{0.3\linewidth}
    \centering
    \hspace*{4mm}
    \scalebox{0.8}{\begin{pgfpicture}{0cm}{0cm}{166pt}{62pt}
\pgfsetxvec{\pgfpoint{1pt}{0pt}}
\pgfsetyvec{\pgfpoint{0pt}{1pt}}
\pgfsetroundjoin 
\pgfsetroundcap
\pgftranslateto{\pgfxy(0,62)}
\begin{pgfmagnify}{1}{-1}
\definecolor{layer0}{rgb}{0.0,0.0,0.0}
\definecolor{layer1}{rgb}{0.0,0.0,0.5}
\definecolor{layer2}{rgb}{1.0,0.0,0.0}
\definecolor{layer3}{rgb}{0.0,0.5,0.5}
\definecolor{layer4}{rgb}{1.0,0.78,0.0}
\definecolor{layer5}{rgb}{0.5,1.0,0.0}
\definecolor{layer6}{rgb}{0.0,1.0,1.0}
\definecolor{layer7}{rgb}{0.0,0.5,0.0}
\definecolor{layer8}{rgb}{0.6,0.8,0.2}
\definecolor{layer9}{rgb}{1.0,0.08,0.58}
\definecolor{layer10}{rgb}{0.71,0.61,0.05}
\definecolor{layer11}{rgb}{0.0,0.5,1.0}
\definecolor{layer12}{rgb}{0.88,0.88,0.88}
\definecolor{layer13}{rgb}{0.64,0.64,0.64}
\definecolor{layer14}{rgb}{0.37,0.37,0.37}
\definecolor{layer15}{rgb}{0.0,0.0,0.0}
\color{layer0}
\pgfsetlinewidth{1.0pt}
\pgfsetdash{}{0pt}
\pgfline{\pgfxy(26.0,42.0)}{\pgfxy(36.0,42.0)}
\pgfline{\pgfxy(26.0,30.0)}{\pgfxy(36.0,30.0)}
\pgfline{\pgfxy(36.0,30.0)}{\pgfxy(36.0,26.0)}
\pgfline{\pgfxy(36.0,42.0)}{\pgfxy(36.0,46.0)}
\pgfline{\pgfxy(16.0,36.0)}{\pgfxy(24.0,36.0)}
\pgfline{\pgfxy(24.0,30.0)}{\pgfxy(24.0,42.0)}
\pgfline{\pgfxy(26.0,28.0)}{\pgfxy(26.0,44.0)}
\pgfmoveto{\pgfxy(12.0,30.0)}
\pgflineto{\pgfxy(12.0,42.0)}
\pgflineto{\pgfxy(6.0,36.0)}
\pgflineto{\pgfxy(6.0,36.0)}
\pgfclosepath 
\pgfqstroke 
\pgfline{\pgfxy(16.0,36.0)}{\pgfxy(12.0,36.0)}
\pgfline{\pgfxy(26.0,16.0)}{\pgfxy(56.0,16.0)}
\pgfline{\pgfxy(20.0,16.0)}{\pgfxy(26.0,16.0)}
\pgfellipse[stroke]{\pgfxy(18.0,16.0)}{\pgfxy(2.0,0)}{\pgfxy(0,2.0)}
\pgfline{\pgfxy(62.0,16.0)}{\pgfxy(56.0,16.0)}
\pgfellipse[stroke]{\pgfxy(64.0,16.0)}{\pgfxy(2.0,0)}{\pgfxy(0,2.0)}
\pgfline{\pgfxy(36.0,26.0)}{\pgfxy(36.0,16.0)}
\pgfmoveto{\pgfxy(30.0,50.0)}
\pgflineto{\pgfxy(42.0,50.0)}
\pgflineto{\pgfxy(36.0,56.0)}
\pgflineto{\pgfxy(36.0,56.0)}
\pgfclosepath 
\pgfqstroke 
\pgfline{\pgfxy(36.0,46.0)}{\pgfxy(36.0,50.0)}
\pgfsetlinewidth{0.33pt}
\pgfcircle[fill]{\pgfxy(36,16)}{2.0pt}\begin{pgfmagnify}{1}{-1}
\pgfputat{\pgfxy(66,-6)}{\pgfbox[left,top]{\overlinerm{AND}}}
\end{pgfmagnify}
\end{pgfmagnify}
\end{pgfpicture}}
    \vspace*{-9mm}
    \caption{}
    \label{fig:Dummy_cell_static_AND}
  \end{subfigure}
  \begin{subfigure}[b]{0.3\linewidth}
    \centering
    \scalebox{0.8}{\begin{pgfpicture}{0cm}{0cm}{102pt}{92pt}
\pgfsetxvec{\pgfpoint{1pt}{0pt}}
\pgfsetyvec{\pgfpoint{0pt}{1pt}}
\pgfsetroundjoin 
\pgfsetroundcap
\pgftranslateto{\pgfxy(0,92)}
\begin{pgfmagnify}{1}{-1}
\definecolor{layer0}{rgb}{0.0,0.0,0.0}
\definecolor{layer1}{rgb}{0.0,0.0,0.5}
\definecolor{layer2}{rgb}{1.0,0.0,0.0}
\definecolor{layer3}{rgb}{0.0,0.5,0.5}
\definecolor{layer4}{rgb}{1.0,0.78,0.0}
\definecolor{layer5}{rgb}{0.5,1.0,0.0}
\definecolor{layer6}{rgb}{0.0,1.0,1.0}
\definecolor{layer7}{rgb}{0.0,0.5,0.0}
\definecolor{layer8}{rgb}{0.6,0.8,0.2}
\definecolor{layer9}{rgb}{1.0,0.08,0.58}
\definecolor{layer10}{rgb}{0.71,0.61,0.05}
\definecolor{layer11}{rgb}{0.0,0.5,1.0}
\definecolor{layer12}{rgb}{0.88,0.88,0.88}
\definecolor{layer13}{rgb}{0.64,0.64,0.64}
\definecolor{layer14}{rgb}{0.37,0.37,0.37}
\definecolor{layer15}{rgb}{0.0,0.0,0.0}
\color{layer0}
\pgfsetlinewidth{1.0pt}
\pgfsetdash{}{0pt}
\pgfline{\pgfxy(36.0,42.0)}{\pgfxy(46.0,42.0)}
\pgfline{\pgfxy(36.0,30.0)}{\pgfxy(46.0,30.0)}
\pgfline{\pgfxy(46.0,30.0)}{\pgfxy(46.0,26.0)}
\pgfline{\pgfxy(46.0,42.0)}{\pgfxy(46.0,46.0)}
\pgfline{\pgfxy(26.0,36.0)}{\pgfxy(34.0,36.0)}
\pgfline{\pgfxy(34.0,30.0)}{\pgfxy(34.0,42.0)}
\pgfline{\pgfxy(36.0,28.0)}{\pgfxy(36.0,44.0)}
\pgfmoveto{\pgfxy(22.0,30.0)}
\pgflineto{\pgfxy(22.0,42.0)}
\pgflineto{\pgfxy(16.0,36.0)}
\pgflineto{\pgfxy(16.0,36.0)}
\pgfclosepath 
\pgfqstroke 
\pgfline{\pgfxy(26.0,36.0)}{\pgfxy(22.0,36.0)}
\pgfline{\pgfxy(36.0,16.0)}{\pgfxy(66.0,16.0)}
\pgfline{\pgfxy(30.0,16.0)}{\pgfxy(36.0,16.0)}
\pgfellipse[stroke]{\pgfxy(28.0,16.0)}{\pgfxy(2.0,0)}{\pgfxy(0,2.0)}
\pgfline{\pgfxy(72.0,16.0)}{\pgfxy(66.0,16.0)}
\pgfellipse[stroke]{\pgfxy(74.0,16.0)}{\pgfxy(2.0,0)}{\pgfxy(0,2.0)}
\pgfline{\pgfxy(46.0,26.0)}{\pgfxy(46.0,16.0)}
\pgfmoveto{\pgfxy(40.0,80.0)}
\pgflineto{\pgfxy(52.0,80.0)}
\pgflineto{\pgfxy(46.0,86.0)}
\pgflineto{\pgfxy(46.0,86.0)}
\pgfclosepath 
\pgfqstroke 
\pgfline{\pgfxy(46.0,76.0)}{\pgfxy(46.0,80.0)}
\pgfsetlinewidth{0.33pt}
\pgfcircle[fill]{\pgfxy(46,16)}{2.0pt}\begin{pgfmagnify}{1}{-1}
\pgfputat{\pgfxy(76,-6)}{\pgfbox[left,top]{AND}}
\end{pgfmagnify}
\pgfsetlinewidth{1.0pt}
\pgfline{\pgfxy(36.0,72.0)}{\pgfxy(46.0,72.0)}
\pgfline{\pgfxy(36.0,60.0)}{\pgfxy(46.0,60.0)}
\pgfline{\pgfxy(46.0,60.0)}{\pgfxy(46.0,56.0)}
\pgfline{\pgfxy(46.0,72.0)}{\pgfxy(46.0,76.0)}
\pgfline{\pgfxy(26.0,66.0)}{\pgfxy(34.0,66.0)}
\pgfline{\pgfxy(34.0,60.0)}{\pgfxy(34.0,72.0)}
\pgfline{\pgfxy(36.0,58.0)}{\pgfxy(36.0,74.0)}
\pgfline{\pgfxy(46.0,56.0)}{\pgfxy(46.0,46.0)}
\pgfline{\pgfxy(6.0,56.0)}{\pgfxy(26.0,56.0)}
\pgfline{\pgfxy(26.0,66.0)}{\pgfxy(16.0,66.0)}
\pgfline{\pgfxy(16.0,66.0)}{\pgfxy(16.0,56.0)}
\end{pgfmagnify}
\end{pgfpicture}}
    \vspace*{-5mm}
    \caption{}
    \label{fig:Dummy_cell_special_AND}
  \end{subfigure}
  \caption{The dummy cells of the LiM array. These are used to mimic a dummy memory row, which is sensed by a special sense amplifier that drives the other ones in order to reduce the overall energy consumption involved in the AND operation, which is performed on the whole array. \textbf{(a)} The dynamic AND dummy cell. \textbf{(b)} The static AND dummy cell. \textbf{(c)} The special--purpose AND dummy cell.}
  \label{fig:Dummy_cells_AND}
\end{figure}

In \autoref{fig:Dummy_cells_AND}, the dummy cells for the LiM variants are presented:
\begin{itemize}
\item in \autoref{fig:Dummy_cell_dynamic_AND}, the dummy cell for the dynamic logic version is depicted. In this gate, two row signals are connected to each cell: the AND line $\overline{AND}$ signal and the precharge signal $\overline{PRE}$; for this reason, the transistors connected to these signals have to be included.  
\item in \autoref{fig:Dummy_cell_static_AND} and \autoref{fig:Dummy_cell_special_AND}, the static and special--purpose variants are presented. Since these do not require an additional row signal, only the AND line pin is present in the circuit. 
\end{itemize}

\section{Memory arrays characterisation}
\label{sec:Arrays_characterization}

The cells are organised in memory arrays in order to evaluate their performance. The memory circuits are simulated for different values of height and width of the array, in order to obtain measurements valid for a wide range of memory sizes. All the simulations are performed at schematic level in Cadence Virtuoso, using the SPECTRE simulation engine.

\medskip

In order to take into account the interconnections parasitics contributions, the layouts of the dummy rows and columns are produced and included in the simulated netlist.

  In particular, 32--bits wide rows and columns are used as basic blocks to create the array: their layouts are extracted and converted in netlists which are, then, included in the testbench.

\smallskip

\begin{figure}[h]
  \centering
  \scalebox{0.8}{\input{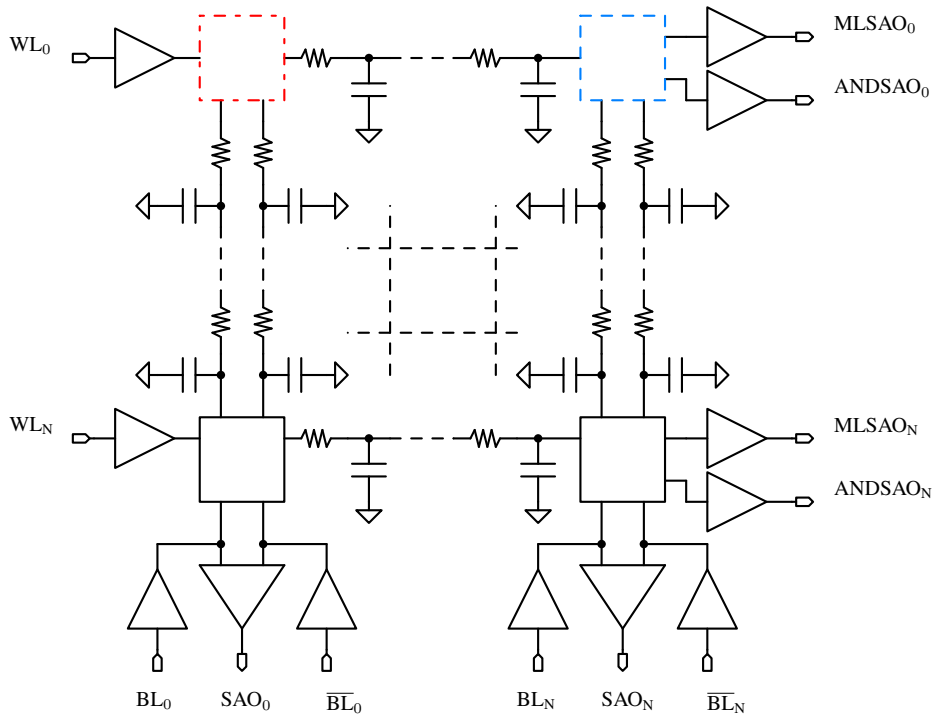}}
  \caption{Worst case delays for each memory operation. Most of the memory cells are omitted for the sake of clarity, and the interconnections are represented by the $RC$ circuits, that are substituted by the extracted rows/columns netlists in the testbench. The cell associated to the read and write operations, highlighted in blue and denoted with a dashed trait, is the farthest one from wordline and bitlines drivers, and sense amplifier; the cell associated to the worst case for the search and AND operation, highlighted in red and denoted with a dashed--and--dotted trait, is the farthest one from the MLSA and ANDSA.}
  \label{fig:Worst_case}
\end{figure}

When considering the read operation, the distances of the cell to be read from the wordline driver and the sense amplifier have to be taken into account to measure how much the cell position affects the performance. Consider the schematic shown in \autoref{fig:Worst_case}:

\begin{itemize}
\item when activating the wordline for selecting a cell, the farthest this is from the driver (i.e. on the last column in \autoref{fig:Worst_case}), the larger the selection delay results to be, due to the higher capacitive--resistive load that the driver has to supply; hence, the read delay associated this cell is the largest possible in the array.
\item when sensing the bitlines with the sense amplifier (SA), the farthest the cell is from the SA inputs (i.e. on the first row in \autoref{fig:Worst_case}), the longer the time needed by the cell to generate a voltage variation on the SA pins is. 
\end{itemize}

For these reasons, the cell to which the worst case read delay is associated is the one on the first row and last column in \autoref{fig:Worst_case} (highlighted in blue), and the read operation performance is evaluated on this cell. For what concerns the worst case for the write operation, a similar analysis can be conducted, referring to the schematic in \autoref{fig:Worst_case}:

\begin{itemize}
\item for the wordline activation and cell selection, the considerations made for the read operation hold true: the cell to which the largest selection delay is associated is the one on the last column.
\item when putting the datum to be written on the bitlines, to evaluate the worst case one needs to consider the farthest cells from the bitlines drivers outputs. In \autoref{fig:Worst_case}, these are the ones placed on the first row. 
\end{itemize}

For these reasons, the cell associated to the worst case sensing delay for the write operation is the one on the first row and last column (highlighted in blue) in \autoref{fig:Worst_case}.

\smallskip

For what concerns the AND and search operations, consider the schematic in \autoref{fig:Worst_case}: since both MLSA and ANDSA are placed at the end of the row, the farthest cell from these is the one on the first column, highlighted in red. Hence, to this cell it is associated the worst case for both AND and search operations. The row position does not affect the performance of the search and AND operations, even if these are associated to the bitline drivers: this is due to the particular sensing scheme employed for the architecture. In fact, since with the current--saving scheme the pull--down transistors of the cells do not require to be disabled during the pre--discharge phase, one can load the mask vector on the bitlines during this cycle, so that all the cells are already configured before the evaluation phase; in this way, the performance of the search and AND operations do not depend on the distance of the row from the bitline drivers outputs.

\medskip

\begin{figure}[h]
  \centering
  \scalebox{0.7}{\input{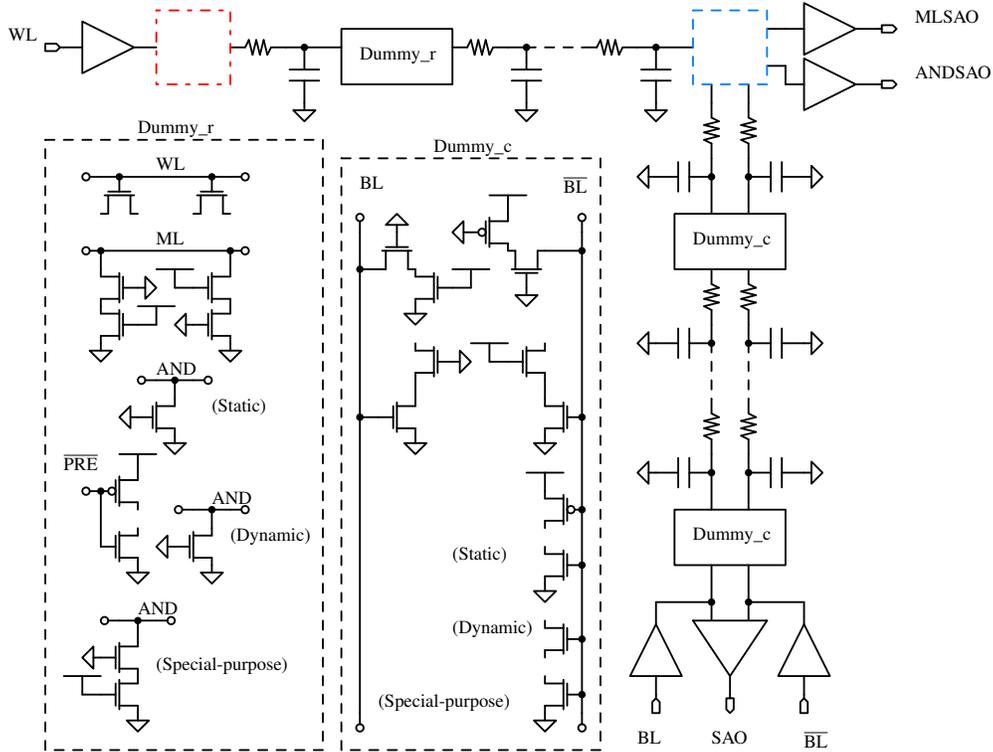}}
  \caption{The array model. The first row correspond to the top one, while the fist column with the leftmost one. Only the critical cells for the read, write, search and AND operations are actually included in the array, as layout--extracted circuits, while all the others are substituted by dummy rows and columns, extracted from the layout, that contain only the significant transistors (i.e. the ones connected to the row/column signals in the original array).}
  \label{fig:Array_model}
\end{figure}

Since the cells required to properly test the memory array are very few, it is not necessary to include all the memory cells in the simulation testbench: the array is reduced to a worst case model, based on the considerations made before, by removing all the cells that are not tested from the array, which leads to shorter simulation time and, hence, faster tuning of the design parameters; consecutively, the circuit model depicted in \autoref{fig:Array_model} is derived and used during the simulations.

Only two memory lines are considered in the model: the first row and the last column. This is due to the fact that the critical cells for all the memory operations are placed on these lines; moreover, since only two cells are tested, the remaining ones can be replaced with dummy versions, which circuits are depicted in \autoref{fig:Array_model}.

The dummy cells are distinguished in row and column ones: in the dummy row cells, only the transistors that are connected to the row signals (wordline; matchline; AND line; precharge line only for the dynamic AND cell) are included in the cell circuit; in the dummy column ones, instead, only the transistors that are connected to the bitlines are kept. In this way, the presence of a memory cell on the line signals is still taken into account while many transistors are removed from the circuit, which leads to a big reduction of the simulation time for large memory arrays.

\smallskip

\begin{figure}[h]
  \centering
  \scalebox{0.66}{\input{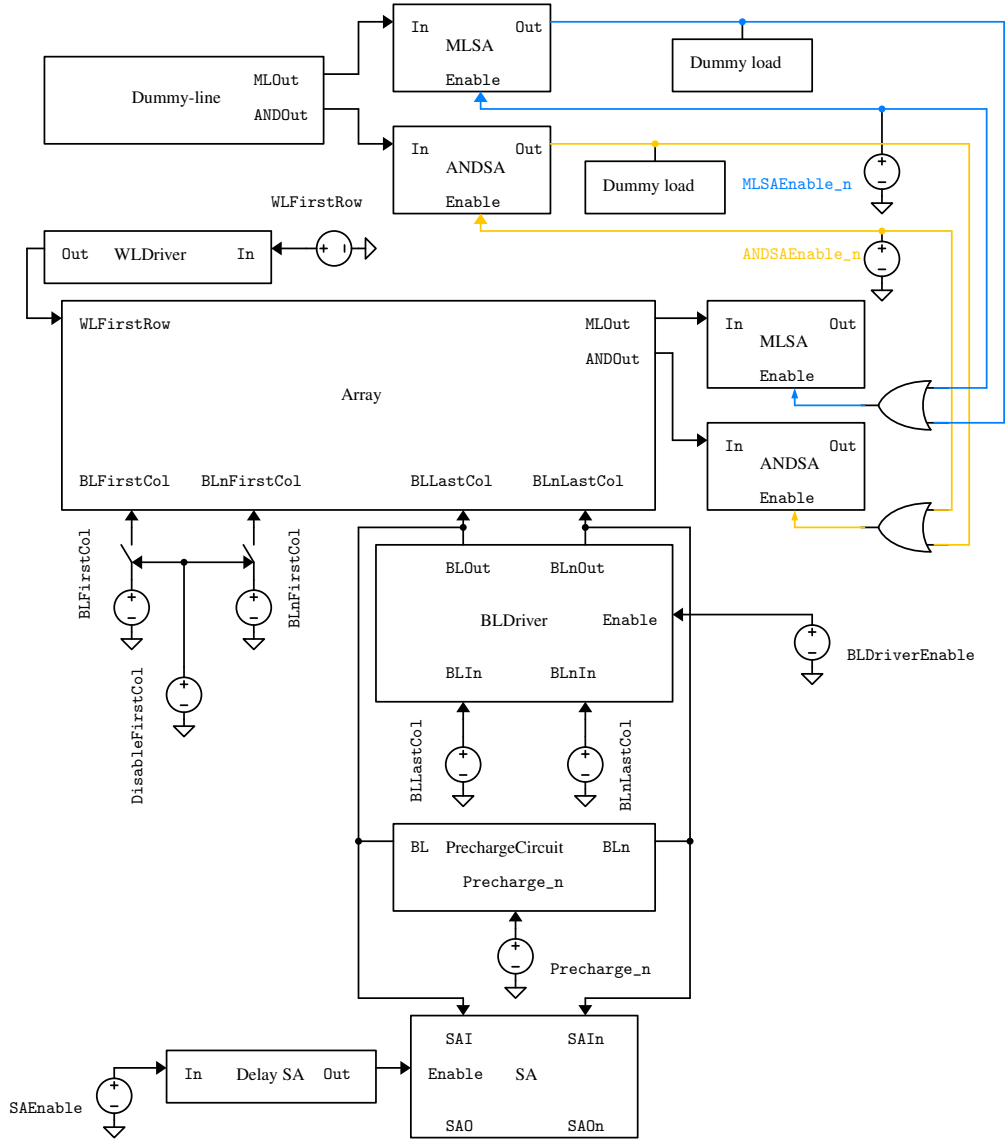}}
  \caption{The testbench. The wires related to the CAM and AND functionalities are highlighted in blue and orange, respectively, for the sake of clarity.}
  \label{fig:TB}
\end{figure}

In Cadence Virtuoso, the testbench shown in \autoref{fig:TB} is employed. This schematic is valid for the LiM array, but it can be simplified and adapted to the CAM and SRAM architectures, since the LiM memory embeds these functionalities, by removing some blocks and substituting the cells circuits.

In \autoref{fig:TB}, it can be noticed that the bitline drivers are included only for the last column, since only on this line the read and write operations and tested; for the first column, instead, ideal switches and voltage generators are employed to modify the cell content, since only row operations, such as the AND and search ones, are tested on it.

\smallskip

In the schematic shown in \autoref{fig:TB}, one can also notice that a block called ``dummy load'' is added on the output of each dummy sense amplifier: these blocks are needed to emulate the presence of all the sense amplifiers of the rows of an actual memory array. As it is discussed in \autoref{sec:LiM_array}, the dummy sense amplifier has to drive all the OR logic gates embedded in each real sense amplifier; since in the model presented in \autoref{fig:Array_model} only one row is equipped with MLSA and ANDSA, the other rows SAs have to be modeled to take into account their influence on performance in an actual memory array. For this reason, dummy loads made by OR gates input sections are connected to the output of the sense amplifiers. 

\begin{figure}[h]
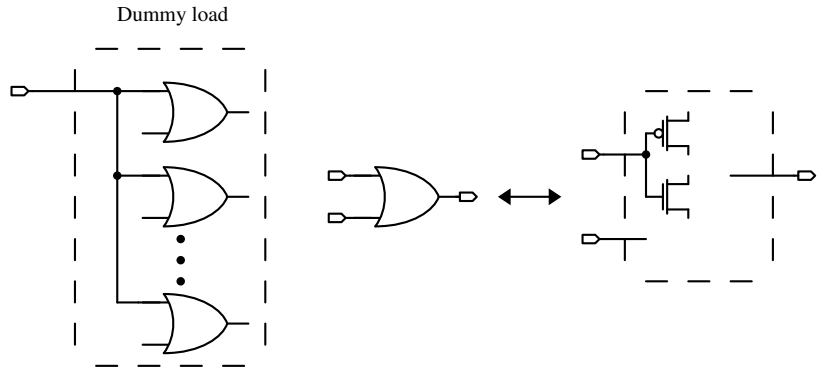

  \centering
  \scalebox{0.8}{\begin{pgfpicture}{0cm}{0cm}{392pt}{182pt}
\pgfsetxvec{\pgfpoint{1pt}{0pt}}
\pgfsetyvec{\pgfpoint{0pt}{1pt}}
\pgfsetroundjoin 
\pgfsetroundcap
\pgftranslateto{\pgfxy(0,182)}
\begin{pgfmagnify}{1}{-1}
\definecolor{layer0}{rgb}{0.0,0.0,0.0}
\definecolor{layer1}{rgb}{0.0,0.0,0.5}
\definecolor{layer2}{rgb}{1.0,0.0,0.0}
\definecolor{layer3}{rgb}{0.0,0.5,0.5}
\definecolor{layer4}{rgb}{1.0,0.78,0.0}
\definecolor{layer5}{rgb}{0.5,1.0,0.0}
\definecolor{layer6}{rgb}{0.0,1.0,1.0}
\definecolor{layer7}{rgb}{0.0,0.5,0.0}
\definecolor{layer8}{rgb}{0.6,0.8,0.2}
\definecolor{layer9}{rgb}{1.0,0.08,0.58}
\definecolor{layer10}{rgb}{0.71,0.61,0.05}
\definecolor{layer11}{rgb}{0.0,0.5,1.0}
\definecolor{layer12}{rgb}{0.88,0.88,0.88}
\definecolor{layer13}{rgb}{0.64,0.64,0.64}
\definecolor{layer14}{rgb}{0.37,0.37,0.37}
\definecolor{layer15}{rgb}{0.0,0.0,0.0}
\color{layer0}
\pgfsetlinewidth{1.0pt}
\pgfsetdash{}{0pt}
\pgfmoveto{\pgfxy(78,42)} 
\pgfcurveto{\pgfxy(96,42)}{\pgfxy(106,50)}{\pgfxy(108,56)}
\pgfstroke
\pgfmoveto{\pgfxy(78,70)} 
\pgfcurveto{\pgfxy(96,70)}{\pgfxy(106,62)}{\pgfxy(108,56)}
\pgfstroke
\pgfmoveto{\pgfxy(78,42)} 
\pgfcurveto{\pgfxy(84,50)}{\pgfxy(84,62)}{\pgfxy(78,70)}
\pgfstroke
\pgfline{\pgfxy(68.0,46.0)}{\pgfxy(80.0,46.0)}
\pgfline{\pgfxy(68.0,66.0)}{\pgfxy(80.0,66.0)}
\pgfline{\pgfxy(108.0,56.0)}{\pgfxy(118.0,56.0)}
\pgfmoveto{\pgfxy(78,82)} 
\pgfcurveto{\pgfxy(96,82)}{\pgfxy(106,90)}{\pgfxy(108,96)}
\pgfstroke
\pgfmoveto{\pgfxy(78,110)} 
\pgfcurveto{\pgfxy(96,110)}{\pgfxy(106,102)}{\pgfxy(108,96)}
\pgfstroke
\pgfmoveto{\pgfxy(78,82)} 
\pgfcurveto{\pgfxy(84,90)}{\pgfxy(84,102)}{\pgfxy(78,110)}
\pgfstroke
\pgfline{\pgfxy(68.0,86.0)}{\pgfxy(80.0,86.0)}
\pgfline{\pgfxy(68.0,106.0)}{\pgfxy(80.0,106.0)}
\pgfline{\pgfxy(108.0,96.0)}{\pgfxy(118.0,96.0)}
\pgfmoveto{\pgfxy(78,142)} 
\pgfcurveto{\pgfxy(96,142)}{\pgfxy(106,150)}{\pgfxy(108,156)}
\pgfstroke
\pgfmoveto{\pgfxy(78,170)} 
\pgfcurveto{\pgfxy(96,170)}{\pgfxy(106,162)}{\pgfxy(108,156)}
\pgfstroke
\pgfmoveto{\pgfxy(78,142)} 
\pgfcurveto{\pgfxy(84,150)}{\pgfxy(84,162)}{\pgfxy(78,170)}
\pgfstroke
\pgfline{\pgfxy(68.0,146.0)}{\pgfxy(80.0,146.0)}
\pgfline{\pgfxy(68.0,166.0)}{\pgfxy(80.0,166.0)}
\pgfline{\pgfxy(108.0,156.0)}{\pgfxy(118.0,156.0)}
\pgfsetlinewidth{0.33pt}
\pgfcircle[fill]{\pgfxy(86,116)}{2.0pt}\pgfcircle[fill]{\pgfxy(86,126)}{2.0pt}\pgfcircle[fill]{\pgfxy(86,136)}{2.0pt}\pgfsetlinewidth{1.0pt}
\pgfline{\pgfxy(76.0,46.0)}{\pgfxy(16.0,46.0)}
\pgfline{\pgfxy(56.0,146.0)}{\pgfxy(56.0,46.0)}
\pgfline{\pgfxy(76.0,86.0)}{\pgfxy(56.0,86.0)}
\pgfline{\pgfxy(66.0,146.0)}{\pgfxy(56.0,146.0)}
\pgfline{\pgfxy(76.0,146.0)}{\pgfxy(66.0,146.0)}
\pgfsetlinewidth{0.33pt}
\pgfcircle[fill]{\pgfxy(56,86)}{2.0pt}\pgfcircle[fill]{\pgfxy(56,46)}{2.0pt}\pgfsetlinewidth{1.0pt}
\pgfline{\pgfxy(36.0,36.0)}{\pgfxy(36.0,46.0)}
\pgfline{\pgfxy(36.0,56.0)}{\pgfxy(36.0,66.0)}
\pgfline{\pgfxy(36.0,76.0)}{\pgfxy(36.0,86.0)}
\pgfline{\pgfxy(36.0,96.0)}{\pgfxy(36.0,106.0)}
\pgfline{\pgfxy(36.0,116.0)}{\pgfxy(36.0,126.0)}
\pgfline{\pgfxy(36.0,136.0)}{\pgfxy(36.0,146.0)}
\pgfline{\pgfxy(36.0,156.0)}{\pgfxy(36.0,166.0)}
\pgfline{\pgfxy(46.0,176.0)}{\pgfxy(56.0,176.0)}
\pgfline{\pgfxy(66.0,176.0)}{\pgfxy(76.0,176.0)}
\pgfline{\pgfxy(86.0,176.0)}{\pgfxy(96.0,176.0)}
\pgfline{\pgfxy(106.0,176.0)}{\pgfxy(116.0,176.0)}
\pgfline{\pgfxy(126.0,166.0)}{\pgfxy(126.0,156.0)}
\pgfline{\pgfxy(126.0,136.0)}{\pgfxy(126.0,146.0)}
\pgfline{\pgfxy(126.0,126.0)}{\pgfxy(126.0,116.0)}
\pgfline{\pgfxy(126.0,106.0)}{\pgfxy(126.0,96.0)}
\pgfline{\pgfxy(126.0,86.0)}{\pgfxy(126.0,76.0)}
\pgfline{\pgfxy(126.0,66.0)}{\pgfxy(126.0,56.0)}
\pgfline{\pgfxy(126.0,46.0)}{\pgfxy(126.0,36.0)}
\pgfline{\pgfxy(116.0,26.0)}{\pgfxy(106.0,26.0)}
\pgfline{\pgfxy(96.0,26.0)}{\pgfxy(86.0,26.0)}
\pgfline{\pgfxy(76.0,26.0)}{\pgfxy(66.0,26.0)}
\pgfline{\pgfxy(56.0,26.0)}{\pgfxy(46.0,26.0)}
\begin{pgfmagnify}{1}{-1}
\pgfputat{\pgfxy(56,-6)}{\pgfbox[left,top]{Dummy load}}
\end{pgfmagnify}
\pgfline{\pgfxy(14.0,46.0)}{\pgfxy(16.0,46.0)}
\pgfmoveto{\pgfxy(14.0,46.0)}
\pgflineto{\pgfxy(14.0,46.0)}
\pgflineto{\pgfxy(12.0,44.0)}
\pgflineto{\pgfxy(6.0,44.0)}
\pgflineto{\pgfxy(6.0,48.0)}
\pgflineto{\pgfxy(12.0,48.0)}
\pgfclosepath 
\pgfqstroke 
\pgfmoveto{\pgfxy(178,82)} 
\pgfcurveto{\pgfxy(196,82)}{\pgfxy(206,90)}{\pgfxy(208,96)}
\pgfstroke
\pgfmoveto{\pgfxy(178,110)} 
\pgfcurveto{\pgfxy(196,110)}{\pgfxy(206,102)}{\pgfxy(208,96)}
\pgfstroke
\pgfmoveto{\pgfxy(178,82)} 
\pgfcurveto{\pgfxy(184,90)}{\pgfxy(184,102)}{\pgfxy(178,110)}
\pgfstroke
\pgfline{\pgfxy(168.0,86.0)}{\pgfxy(180.0,86.0)}
\pgfline{\pgfxy(168.0,106.0)}{\pgfxy(180.0,106.0)}
\pgfline{\pgfxy(208.0,96.0)}{\pgfxy(218.0,96.0)}
\pgfmoveto{\pgfxy(236.0,96.0)}
\pgflineto{\pgfxy(242.0,92.0)}
\pgflineto{\pgfxy(242.0,100.0)}
\pgfclosepath 
\pgffill 
\pgfmoveto{\pgfxy(266.0,96.0)}
\pgflineto{\pgfxy(260.0,100.0)}
\pgflineto{\pgfxy(260.0,92.0)}
\pgfclosepath 
\pgffill 
\pgfline{\pgfxy(242.0,96.0)}{\pgfxy(260.0,96.0)}
\pgfline{\pgfxy(316.0,102.0)}{\pgfxy(326.0,102.0)}
\pgfline{\pgfxy(316.0,90.0)}{\pgfxy(326.0,90.0)}
\pgfline{\pgfxy(326.0,90.0)}{\pgfxy(326.0,86.0)}
\pgfline{\pgfxy(326.0,102.0)}{\pgfxy(326.0,106.0)}
\pgfline{\pgfxy(306.0,96.0)}{\pgfxy(314.0,96.0)}
\pgfline{\pgfxy(314.0,90.0)}{\pgfxy(314.0,102.0)}
\pgfline{\pgfxy(316.0,88.0)}{\pgfxy(316.0,104.0)}
\pgfline{\pgfxy(316.0,72.0)}{\pgfxy(326.0,72.0)}
\pgfline{\pgfxy(316.0,60.0)}{\pgfxy(326.0,60.0)}
\pgfline{\pgfxy(326.0,60.0)}{\pgfxy(326.0,56.0)}
\pgfline{\pgfxy(326.0,72.0)}{\pgfxy(326.0,76.0)}
\pgfline{\pgfxy(306.0,66.0)}{\pgfxy(310.0,66.0)}
\pgfline{\pgfxy(314.0,60.0)}{\pgfxy(314.0,72.0)}
\pgfline{\pgfxy(316.0,58.0)}{\pgfxy(316.0,74.0)}
\pgfellipse[stroke]{\pgfxy(312.0,66.0)}{\pgfxy(2.0,0)}{\pgfxy(0,2.0)}
\pgfline{\pgfxy(306.0,96.0)}{\pgfxy(306.0,66.0)}
\pgfline{\pgfxy(306.0,76.0)}{\pgfxy(286.0,76.0)}
\pgfline{\pgfxy(306.0,116.0)}{\pgfxy(286.0,116.0)}
\pgfsetlinewidth{0.33pt}
\pgfcircle[fill]{\pgfxy(306,76)}{2.0pt}\pgfsetlinewidth{1.0pt}
\pgfline{\pgfxy(296.0,56.0)}{\pgfxy(296.0,66.0)}
\pgfline{\pgfxy(296.0,76.0)}{\pgfxy(296.0,86.0)}
\pgfline{\pgfxy(296.0,96.0)}{\pgfxy(296.0,106.0)}
\pgfline{\pgfxy(296.0,116.0)}{\pgfxy(296.0,126.0)}
\pgfline{\pgfxy(306.0,136.0)}{\pgfxy(316.0,136.0)}
\pgfline{\pgfxy(326.0,136.0)}{\pgfxy(336.0,136.0)}
\pgfline{\pgfxy(346.0,136.0)}{\pgfxy(356.0,136.0)}
\pgfline{\pgfxy(366.0,126.0)}{\pgfxy(366.0,116.0)}
\pgfline{\pgfxy(366.0,106.0)}{\pgfxy(366.0,96.0)}
\pgfline{\pgfxy(366.0,86.0)}{\pgfxy(366.0,76.0)}
\pgfline{\pgfxy(366.0,66.0)}{\pgfxy(366.0,56.0)}
\pgfline{\pgfxy(356.0,46.0)}{\pgfxy(346.0,46.0)}
\pgfline{\pgfxy(326.0,46.0)}{\pgfxy(336.0,46.0)}
\pgfline{\pgfxy(316.0,46.0)}{\pgfxy(306.0,46.0)}
\pgfline{\pgfxy(346.0,86.0)}{\pgfxy(376.0,86.0)}
\pgfline{\pgfxy(164.0,86.0)}{\pgfxy(166.0,86.0)}
\pgfmoveto{\pgfxy(164.0,86.0)}
\pgflineto{\pgfxy(164.0,86.0)}
\pgflineto{\pgfxy(162.0,84.0)}
\pgflineto{\pgfxy(156.0,84.0)}
\pgflineto{\pgfxy(156.0,88.0)}
\pgflineto{\pgfxy(162.0,88.0)}
\pgfclosepath 
\pgfqstroke 
\pgfline{\pgfxy(164.0,106.0)}{\pgfxy(166.0,106.0)}
\pgfmoveto{\pgfxy(164.0,106.0)}
\pgflineto{\pgfxy(164.0,106.0)}
\pgflineto{\pgfxy(162.0,104.0)}
\pgflineto{\pgfxy(156.0,104.0)}
\pgflineto{\pgfxy(156.0,108.0)}
\pgflineto{\pgfxy(162.0,108.0)}
\pgfclosepath 
\pgfqstroke 
\pgfline{\pgfxy(216.0,96.0)}{\pgfxy(218.0,96.0)}
\pgfmoveto{\pgfxy(226.0,96.0)}
\pgflineto{\pgfxy(226.0,96.0)}
\pgflineto{\pgfxy(224.0,94.0)}
\pgflineto{\pgfxy(218.0,94.0)}
\pgflineto{\pgfxy(218.0,98.0)}
\pgflineto{\pgfxy(224.0,98.0)}
\pgfclosepath 
\pgfqstroke 
\pgfline{\pgfxy(176.0,86.0)}{\pgfxy(166.0,86.0)}
\pgfline{\pgfxy(176.0,106.0)}{\pgfxy(166.0,106.0)}
\pgfline{\pgfxy(376.0,86.0)}{\pgfxy(378.0,86.0)}
\pgfmoveto{\pgfxy(386.0,86.0)}
\pgflineto{\pgfxy(386.0,86.0)}
\pgflineto{\pgfxy(384.0,84.0)}
\pgflineto{\pgfxy(378.0,84.0)}
\pgflineto{\pgfxy(378.0,88.0)}
\pgflineto{\pgfxy(384.0,88.0)}
\pgfclosepath 
\pgfqstroke 
\pgfline{\pgfxy(284.0,76.0)}{\pgfxy(286.0,76.0)}
\pgfmoveto{\pgfxy(284.0,76.0)}
\pgflineto{\pgfxy(284.0,76.0)}
\pgflineto{\pgfxy(282.0,74.0)}
\pgflineto{\pgfxy(276.0,74.0)}
\pgflineto{\pgfxy(276.0,78.0)}
\pgflineto{\pgfxy(282.0,78.0)}
\pgfclosepath 
\pgfqstroke 
\pgfline{\pgfxy(284.0,116.0)}{\pgfxy(286.0,116.0)}
\pgfmoveto{\pgfxy(284.0,116.0)}
\pgflineto{\pgfxy(284.0,116.0)}
\pgflineto{\pgfxy(282.0,114.0)}
\pgflineto{\pgfxy(276.0,114.0)}
\pgflineto{\pgfxy(276.0,118.0)}
\pgflineto{\pgfxy(282.0,118.0)}
\pgfclosepath 
\pgfqstroke 
\end{pgfmagnify}
\end{pgfpicture}}
  \caption{The dummy load for the dummy SA. This is used to emulate the input sections of multiple OR gates, which are embedded in each real MLSA/ANDSA, in order to take into account their influence on the sensing performance in the array. }
  \label{fig:Dummy_load}
\end{figure}

The circuit of the dummy load block is shown in \autoref{fig:Dummy_load}. It consists of multiple OR logic gates which share the same input, and the number of OR gates coincides with the number of rows in the array. These are not actual gates: only the transistors connected to the input are included, in order to reduce as much as possible the number of elements in the testbench netlist.

\smallskip

Some additional blocks are shown in \autoref{fig:TB}: the precharge circuit is used to precharge the bitlines before a read operation; the ``Delay SA'' circuit is used to delay the enable signal of the sense amplifier used to test the read operation, since a voltage latch SA \cite{sram_sa} is employed.
      
\section{The simulation framework}
\label{sec:Simulation_framework}

To characterise large memory arrays, a scripting approach is adopted, generating the circuit netlists automatically after an initial by--hand characterisation of the design.

\begin{figure}[h]
  \centering
  \includegraphics[scale=0.9]{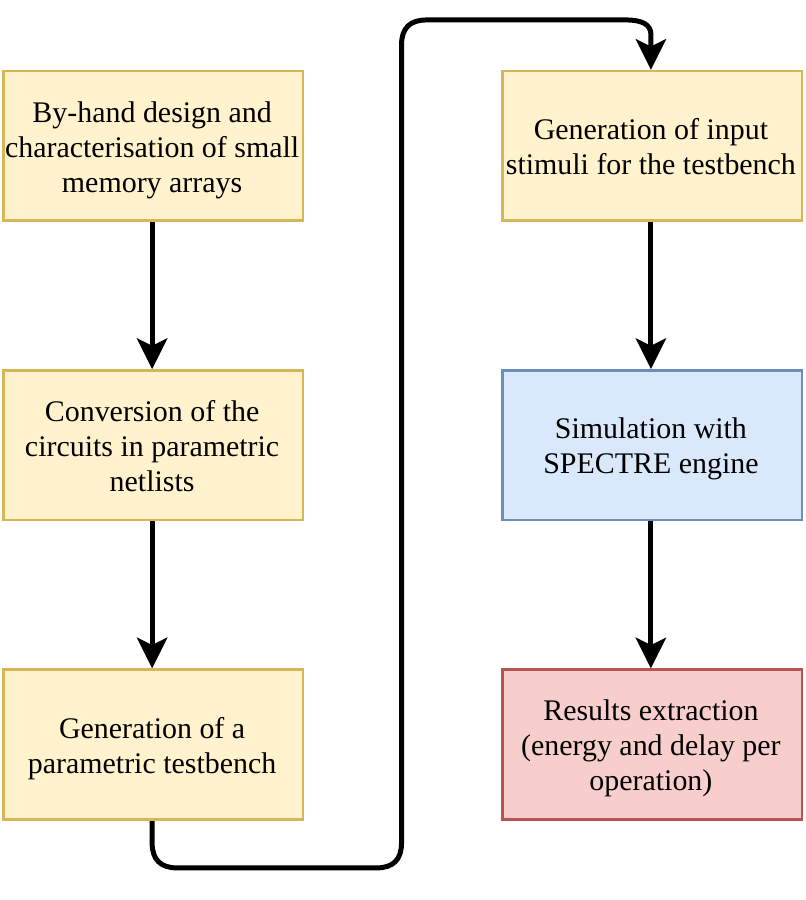}
  \caption{The simulation flow. Starting from the by--hand designed circuits of small arrays, the simulation of large ones is achieved through the algorithm shown in the figure.}
  \label{fig:Simulation_flow}  
\end{figure}
 
The approach adopted for the simulation of large arrays is presented in \autoref{fig:Simulation_flow}, and it consists of the following steps:
\begin{itemize}
\item the memory array and the sensing circuitry are designed by--hand and characterised by simulating small arrays (32x32 cells).
\item the cells and rows layouts are produced and extracted. 32--bits wide rows and columns are used as basic blocks to create the final array.
\item after the circuits netlists have been extracted, a script is written, following precise guidelines, to make the circuit parametric with respect to its size (array height and width).
\item a script is used to generate a parametric Cadence Virtuoso testbench that allows to characterise the circuit for arbitrary values of width and height, by using the SPECTRE simulation engine.
\item the input stimuli of the testbench are automatically generated, starting from the operations sequence to be simulated provided by the user.
\item the circuit is simulated using the SPECTRE engine of Cadence Virtuoso.
\item the array performance are extracted by measuring the energy consumption and the delay associated to each memory operation.
\end{itemize}

\begin{figure}[h]
  \centering
  \includegraphics[width=\linewidth]{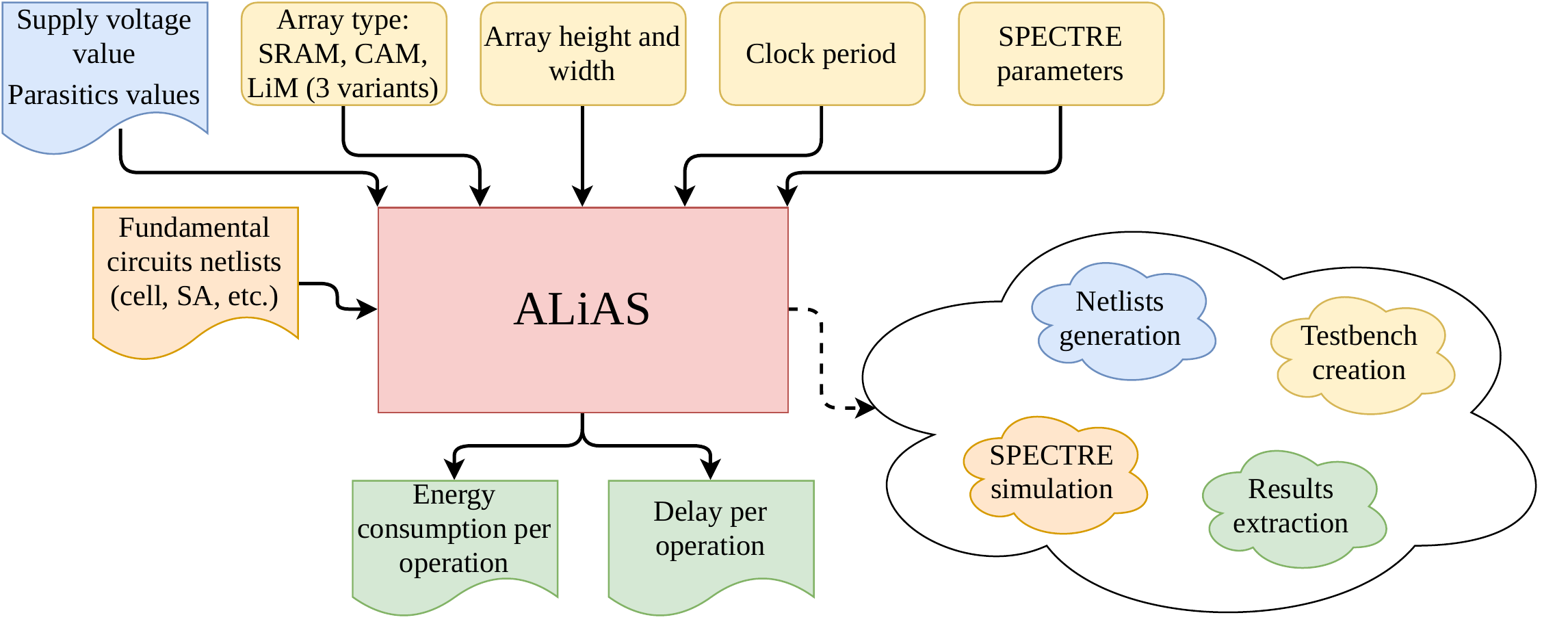}
  \caption{The scripting approach adopted, called ALiAS (Analog Logic--in--Memory Arrays Simulation). Starting from the array characteristics (type and dimensions), the simulation conditions (circuit parameters, clock period, SPECTRE configuration) and the layout extracted netlists of the basic circuits (memory cells, rows and columns), a simulation is performed in SPECTRE and the array performance is evaluated.}
  \label{fig:ALiAS}
\end{figure}

In \autoref{fig:ALiAS}, the scripting workflow, called ALiAS (Analog Logic--in--Memory Arrays Simulation), is presented. ALiAS takes in input:
\begin{itemize}
\item the netlists of the fundamental blocks, which are the memory cells and the sense amplifiers, that have to be designed by--hand.
\item the desired characteristics for the array to be simulated: type (SRAM, CAM, the three LiM variants) and size (width and height).
\item simulation parameters for SPECTRE (such as the maximum number of computational threads associated to the simulation, GUI mode etc.).
\item the clock period selected for the simulation, which is equal to $\SI{1}{\nano\second}$ by default. 
\end{itemize}

Given this information, the netlist of the array and the testbench are generated, the SPECTRE simulation is run, performance measurements are extracted (in particular, energy consumption and delay associated to each memory operation) and saved in different formats (bar diagrams and CSV files). With this approach, ALiAS allows to speed up the design and simulation of memory arrays with custom cell topologies at schematic level.

\section{Results and discussion}
\label{sec:Results}

To evaluate the memory arrays performance, energy consumption and latency of each memory operation are extracted from SPECTRE simulations. The energy consumption is measured by integrating the array instantaneous power consumption over each simulation cycle:

\begin{equation*}
  E_{operation} = \int_{cycle}{p(t)dt}
\end{equation*}

Each array is simulated with a supply voltage $V_{DD}=\SI{1}{\volt}$ and a clock period $t_{ck}=\SI{4}{\nano\second}$ using the SPECTRE simulator in Cadence Virtuoso.

\begin{figure}
    \hspace*{-10mm}
    \includegraphics[width=1.15\linewidth]{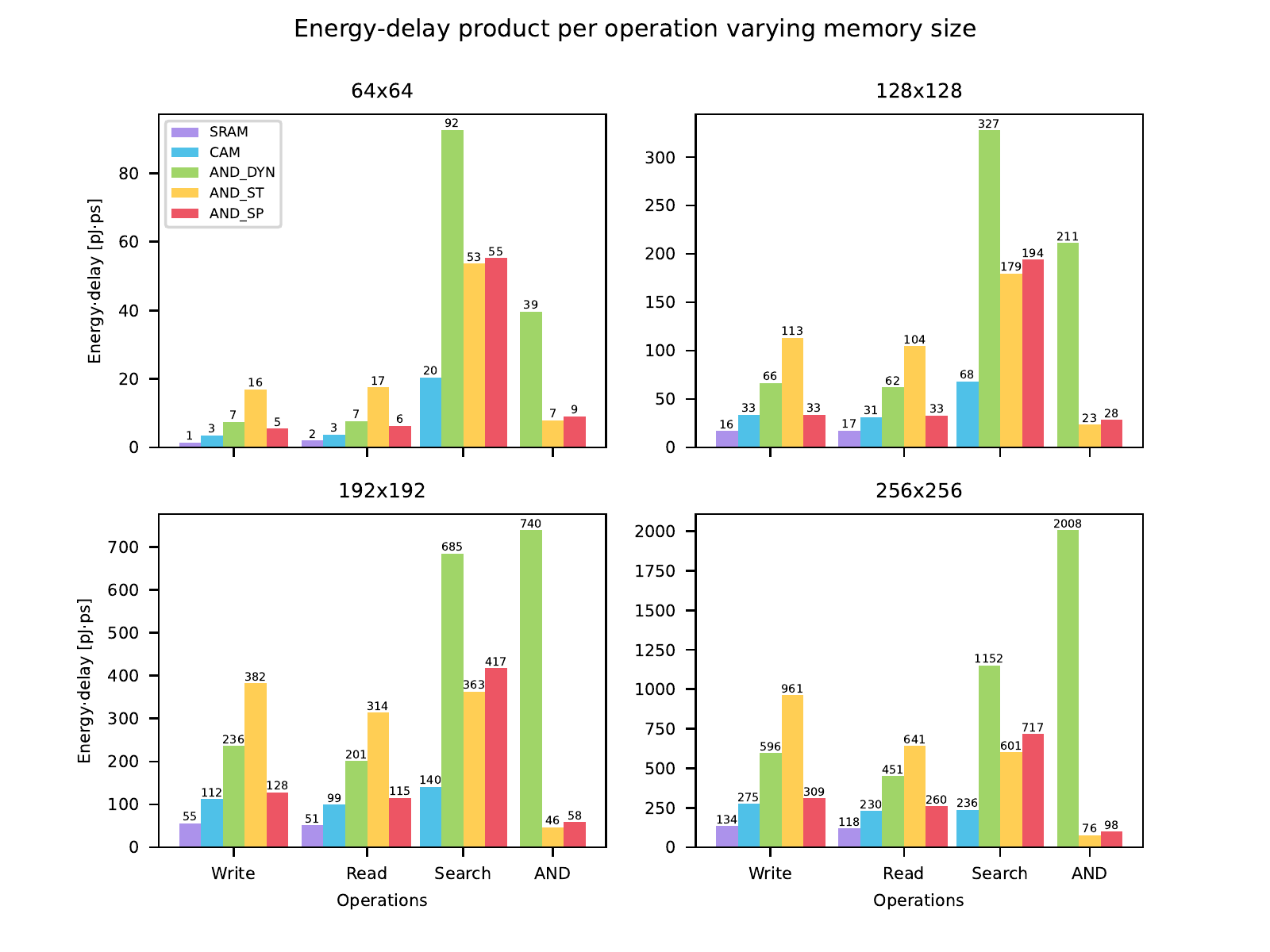}
    \vspace*{-10mm}
    \caption{The energy--delay product associated to each memory operation in each array, for different values of the memory size.}
    \label{fig:energy_delay}
\end{figure}

\begin{table}[h]
    \centering
    \begin{tabular}{ccccc}
        \toprule
        \multicolumn{5}{c}{\textbf{Energy--delay products $[\SI{}{\pico\joule\cdot\pico\second}]$}}\\
        \midrule
        \textbf{Memory} & \multicolumn{4}{c}{\textbf{Operations}}\\
        & \textbf{Write} & \textbf{Read} & \textbf{Search} & \textbf{AND}\\
        \textbf{SRAM} & 134 & 118 & --- & ---\\
        \textbf{CAM} & 275 & 230 & 236 & ---\\
        \textbf{AND SP} & 309 & 260 & 717 & 98 \\
        \textbf{AND DYN} & 596 & 451 & 1152 &  2008\\
        \textbf{AND ST} & 961 & 641 & 601 & 76 \\
        \bottomrule                     
    \end{tabular}
     \smallskip
    \captionsetup{justification=centering}
    \caption{The energy--products associated to each memory operation, for each memory array. Data are extracted from the 256x256 array of \autoref{fig:energy_delay}, which is used as case study.}
    \label{tab:energy_delay}
\end{table}

In \autoref{fig:energy_delay}, the energy--delays product per operation of each memory array, are presented. Four different memory sizes are considered: 64x64, 128x128, 192x192 and 256x256, intended as rows and columns. These values have been chosen to estimate how the array performance scales with its size, with size values usually adopted in literature for test chips \cite{cam_sizes_0, cam_sizes_1, cam_sizes_2, cam_literature}. In \autoref{tab:energy_delay}, the energy--delay products values are shown, using as reference case the 256x256 array of \autoref{fig:energy_delay}.

In the following, each operation is analysed and compared to the others.

\subsection{Read operation}
\label{sec:read_operation}

From \autoref{fig:energy_delay}, one can observe that the LiM (in the figure $AND\_SP$, $AND\_DYN$, $AND\_ST$ for special--purpose, dynamic and static AND cells, respectively) and CAM memories perform worse than the SRAM array for every value of the memory size. This is due to the fact that these architectures employ cell circuits that are much more complex (i.e. higher number of transistors, wider transistors and more interconnections) than the SRAM one. 

\begin{table}[h]
 \centering
 \begin{tabular}{cccccc}
    \toprule
    \multicolumn{6}{c}{\textbf{Energy--delay products relative variations for Read}}\\
    \midrule        
   & \textbf{SRAM} & \textbf{CAM} & \textbf{AND SP} & \textbf{AND DYN} & \textbf{AND ST}\\
    \textbf{SRAM} & --- & --- & --- & --- & ---\\
    \textbf{CAM} & +94.91\% & --- & --- & --- & ---\\
    \textbf{AND SP} & +120.34\% & +13.04\% & --- & --- & ---\\
    \textbf{AND DYN} & +282.2\% & +96.08\% & +73.46\% & --- & ---\\
    \textbf{AND ST} & +443.22\% & +178.69\% &  +146.54\% & +42.13\% & ---\\
    \bottomrule                     
 \end{tabular}
  \smallskip
  \captionsetup{justification=centering}
 \caption{Percentage differences in the read energy--delay product among the arrays. Each value corresponds to the increase, expressed in percentage, in the energy--delay product of the memory on the corresponding row with respect to the one of the memory on the corresponding column. Some values are omitted for the avoid ambiguities in the table interpretation (i.e. each percentage value is calculated using as reference the memory on the column and each comparison is made only one time per memory). The data are extracted from \autoref{tab:energy_delay}.}
 \label{tab:Read comparison}
\end{table}

In \autoref{tab:Read comparison}, the differences in the energy--delay products associated to the read operation, expressed in percentage, among the arrays, are shown. For instance, for the CAM memory an energy--delay product value 94.41\% higher than the SRAM one is measured; for the static AND memory, an energy--delay product value 40.57\% higher than the special--purpose AND one is obtained. The data are extracted from the 256x256 array of \autoref{fig:energy_delay}, which is used as case study in the following. 

\begin{figure}[h]
  \centering
  \scalebox{0.9}{\input{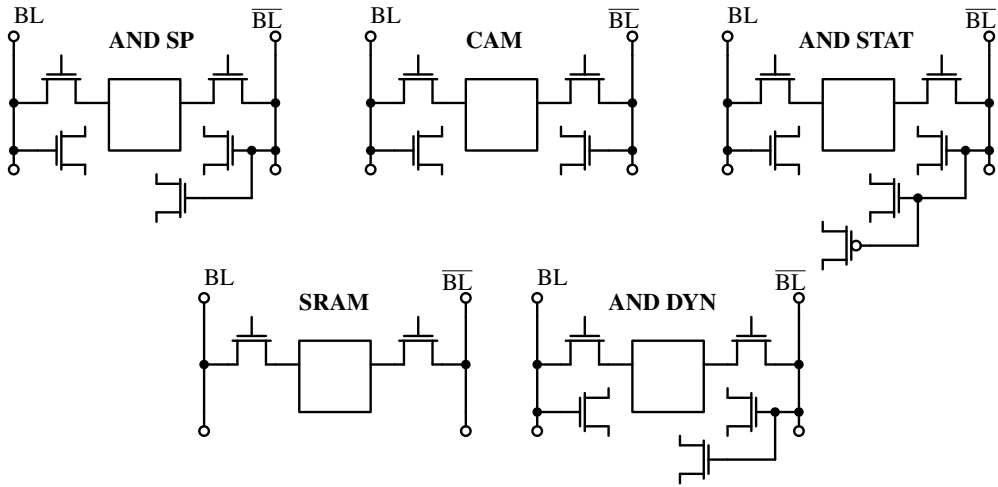}}
  \caption{The transistor load on bitlines for each cell type.}
  \label{fig:Capacitive_load}
\end{figure}

\smallskip

The differences among the memories performance can be explained by investigating their circuits. In \autoref{fig:Capacitive_load}, these are depicted showing only the cell transistors connected to the bitlines. In fact, it is well known that to read from a SRAM--like memory cell, one needs to access it through a wordline and to let the cell discharge one of the bitlines to determine its content; the higher the equivalent capacitive load corresponding to the bitlines, the longer the discharge time is, given the same discharge current. Since the bitlines capacitance is determined by the interconnections layout and the transistors connected to these, it follows that the higher the number of the cell transistors linked to the bitlines, the worse the read performance is.

Considering the data in \autoref{tab:Read comparison}, one can notice that the worst-performing memory is the static AND memory, which is also the one with the highest number of transistors connected to the bitlines (\autoref{fig:Capacitive_load}). This explains why the best performing memory is the SRAM: being the simplest from a circuital point of view, it has the lowest bitlines capacitance associated to it. Similar considerations can be made to explain the differences among the other cells.

\smallskip

One may notice from \autoref{fig:Capacitive_load} that even if the special--purpose and dynamic cell have the same number of transistors connected to the bitlines (in particular, to $\overline{BL}$), the second one performs worse than the first one; this is because one has to take into account also the layouts of these cells, depicted in \autoref{fig:AND_DYN_layout} and \autoref{fig:AND_SP_layout}, for the dynamic and special--purpose AND cell, respectively. It can be observed that the dynamic AND circuit is more complex, having a higher number of transistors and interconnection, which lead to more parasitics in the resultant circuit that slow down the cell read operation, increasing also the corresponding power consumption.

\subsection{Write operation}
\label{sec:write_operation}

\begin{table}[h]
 \centering
 \begin{tabular}{cccccc}
    \toprule
     \multicolumn{6}{c}{\textbf{Energy--delay products relative variations for Write}}\\   
    \midrule
   & \textbf{SRAM} & \textbf{CAM} & \textbf{AND SP} & \textbf{AND DYN} & \textbf{AND ST}\\
    \textbf{SRAM} & --- & --- & --- & --- & ---\\
    \textbf{CAM} & +105\% & --- & --- & --- & ---\\
    \textbf{AND SP} & +130.6\% & +12.36\% & --- & --- & ---\\
    \textbf{AND DYN} & +344.78\% & +116.72\% & +92.88\% & --- & ---\\
    \textbf{AND ST} & +617\% & 249.45\% & +211\% & +61.24\% & ---\\
    \bottomrule                     
 \end{tabular}
 \smallskip
  \captionsetup{justification=centering}
 \caption{Percentage differences in the write energy--delay products among the arrays. Each value corresponds to the increase, expressed in percentage, in the write energy--delay product of the memory on the corresponding row with respect to the one of the memory on the corresponding column. The data are extracted from \autoref{tab:energy_delay}.}
 \label{tab:Write_comparison}
\end{table}

In \autoref{tab:Write_comparison}, the differences in the write operation energy--delay products, expressed in percentage, among the arrays are shown. The same considerations made for the read operation apply, since write and read performance are both approximately determined by the memory circuit and layout.

\subsection{Search operation}
\label{sec:search_operation}

\begin{table}[h]
 \centering
 \begin{tabular}{ccccc}
    \toprule
     \multicolumn{5}{c}{\textbf{Energy--delay products relative variations for Search}}\\   
    \midrule
   & \textbf{CAM} & \textbf{AND SP} & \textbf{AND DYN} & \textbf{AND ST}\\
    \textbf{CAM} & --- & --- & --- & ---\\
    \textbf{AND SP} & +203.81\% & --- & --- & ---\\
    \textbf{AND DYN} & +388.13\% & +60.67\% & --- & ---\\
    \textbf{AND ST} & +154.66\% & -19.30\% & -91.68\% & ---\\    
    \bottomrule                     
 \end{tabular}
  \smallskip
  \captionsetup{justification=centering}
 \caption{Percentage differences in the search energy--delay product among the arrays. Each value corresponds to the increase, expressed in percentage, in the energy--delay product of the memory on the corresponding row with respect to the one of the memory on the corresponding column. The data are extracted from \autoref{tab:energy_delay}.}
 \label{tab:search_comparison}
\end{table}

In \autoref{tab:search_comparison}, the differences in the search operation energy--delay products, expressed in percentage, among the arrays, are shown.

One can notice that the LiM arrays perform worse than the CAM one in the search operation. This can be explained considering the layout of the cells: being the LiM cells more complex, their search functionalities are affected by more parasitics.

\smallskip

Consider the case of the dynamic AND cell, which layout lower section is shown in \autoref{fig:AND_DYN_layout}. One can notice that the CAM circuitry is placed very close to the AND one; as a consequence, the parasitics values associated with the matchline are increased with respect to the original CAM cell, which leads to higher latency and power consumption for the search operation. Similar considerations can be made for the special--purpose and static AND cells.

\smallskip

It can be observed that, among the LiM arrays, the best performing one for the search operation is the static AND array. This seems counter--intuitive, since the static AND gate is the most complex one among the AND cells; however, this can be explained by investigating the layout of the cells.

\begin{figure}
    \centering
    \includegraphics[width=\linewidth]{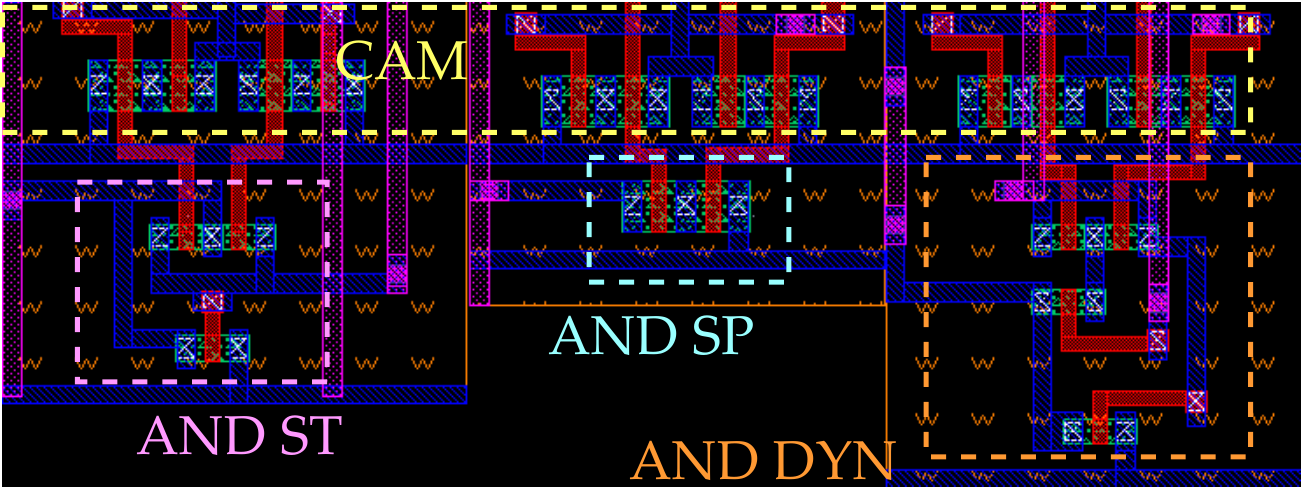}
    \caption{Layout bottom sections of static, special--purpose and dynamic AND cells.}
    \label{fig:comparison_search}
\end{figure}

In \autoref{fig:comparison_search}, the lower sections of the static, special--purpose and dynamic AND cells are shown side to side. By considering the AND gates regions in the layouts, which are highlighted in the figure, one can notice that the most complex layout (in terms of the number of transistors and local interconnections) is the dynamic AND one, highlighted in orange, followed by the special--purpose one (there are less transistors but these are wider), highlighted in cyan, and, then, the static one, highlighted in pink. For this reason, the worst performance is associated with this cell.

For what concerns the special--purpose cell, its circuit seems to be less complex than the static one, but it should be noted that the transistors of the special--purpose circuit are wider than the ones of the static cell; this leads to larger parasitic capacitances, that lead to a worsening in performance for the search operation, being these transistors connected through the gates to the CAM functionality ones.

\subsection{AND operation}
\label{sec:AND_operation}

\begin{table}[h]
 \centering
 \begin{tabular}{cccc}
    \toprule
    \multicolumn{4}{c}{\textbf{Energy--delay products relative variations for AND}}\\ 
    \midrule
   & \textbf{AND SP} & \textbf{AND DYN} & \textbf{AND ST}\\
    \textbf{AND SP} & --- & --- & --- \\
    \textbf{AND DYN} & +1948.98\% & --- & ---\\
    \textbf{AND ST} & -28.95\% & -2542.1\% & ---\\
    \bottomrule                     
 \end{tabular}
  \smallskip
  \captionsetup{justification=centering}
 \caption{Percentage differences in the AND energy--delay product among the arrays. Each value corresponds to the increase, expressed in percentage, in the energy--delay product of the memory on the corresponding row with respect to the one of the memory on the corresponding column. The data are extracted from \autoref{tab:energy_delay}.}
 \label{tab:AND_comparison}
\end{table}

In \autoref{tab:AND_comparison}, the differences in the AND operation energy--delay products, expressed in percentage, among the arrays are shown.

\smallskip

One can notice that the best performing array is the static AND one. This can be explained by referring to the cells circuits. 

The static cell performs better than the special--purpose one due to its simpler output circuit (\autoref{fig:Static_AND_gate_and_cell} for the static AND, \autoref{fig:Special_AND_gate_and_cell} for the special--purpose AND): while the static gate has only one transistor connected to the AND line, the special--purpose one has two NMOSFETs in series linked to it; this leads to higher latency and power consumption.

\smallskip

The static AND cell performs better also than the dynamic cell, since the latter is implemented in dynamic CMOS logic, while the first one in static CMOS logic. In fact, considering the circuit of the dynamic AND cell in \autoref{fig:Dynamic_AND_gate_and_cell}, it can be noticed that, once the sensing of the AND line is enabled through $\overline{EN}$, it takes a certain amount of time for the dynamic gate to discharge its output, denoted with $AND$, and, hence, disable the pull--down. During this time interval, the pull--down is conducting and prevents the AND line, denoted with $\overline{AND}$, from getting charged by the ANDSA. This leads to an increase in both energy consumption and sensing delay. 

Considering the circuit of the static AND cell in \autoref{fig:Static_AND_gate_and_cell}, one can notice that the output of the AND gate is already at ground voltage before the sensing enabling, for the reasons discussed in \autoref{sec:Arrays_characterization}. At the beginning of the AND operation, the pull--down is already disabled, which means that the line starts immediately to get charged, without having any current flowing to ground. Moreover, at each AND execution, all the AND gates invert their outputs to turn off the pull--down transistor connected to the AND line; this leads to a large increase in the energy consumption, as it can be observed from \autoref{tab:AND_comparison}.

\subsection{Comparison among different operations}
\label{sec:comparison_operations}

In this section, the operations performed are compared and analysed in relation to each other.

\smallskip

From \autoref{fig:energy_delay}, one can notice that write performance worsens more than the read one, as the array size is increased. This is mainly due to the fact that while a read operation does not imply the complete commutation of one of the bitlines (one of the two lines needs to discharge just enough for the sense amplifier to properly read the cell value), a write one does, since a ``strong'' logic `0' has to be put on one of the bitlines to force the desired value to be written to the cell; as a consequence, larger energy consumption for the write operation with respect the read one is measured.

\begin{table}[h]
 \centering
 \begin{tabular}{cccccccc}
    \toprule
    \multicolumn{7}{c}{\textbf{Energy--delay products relative variations: Write v.s. Read}}\\
    \midrule
    & & \multicolumn{5}{c}{\textbf{Read}}\\
   \multirow{7}{*}{\textbf{Write}} &  & \textbf{SRAM} & \textbf{CAM} & \textbf{AND SP} & \textbf{AND ST} & \textbf{AND DYN}\\
    & \textbf{SRAM} & +13.56\% & --- & --- & --- & --- \\
    & \textbf{CAM} & --- & +19.56\% & --- & --- & ---\\
    & \textbf{AND SP} & --- & --- & +18.85\% & --- & ---\\
    & \textbf{AND ST} & --- & --- & --- & +49.92\% & ---\\
    & \textbf{AND DYN} & --- & --- & --- & --- & +32.15\%\\   
    \bottomrule                     
 \end{tabular}
  \smallskip
  \captionsetup{justification=centering}
 \caption{Percentage differences of the write and read energy--delay products in each memory. Each value corresponds to the increase, expressed in percentage, in the energy--delay product of the memory on the corresponding row with respect to the one of the memory on the corresponding column. The data are extracted from \autoref{tab:energy_delay}.}
 \label{tab:write_read_comparison}
\end{table}

In \autoref{tab:write_read_comparison}, the read and write performance, in terms of energy--delay product, are compared in each memory. One can notice that the largest difference between reading and write performance is associated to the static AND memory. This is due to the fact that, as the array size is enlarged, the corresponding bitlines capacitive load increases more than linearly; since the static AND cell is the most complex one, a larger difference in write and read performance is measured for large arrays (e.g. the 256x256 one in \autoref{fig:energy_delay}), while in the other ones a smaller one is obtained. In fact, in \autoref{tab:write_read_comparison}, the write/read discrepancy value follows the cell circuit complexity: the best performing memory is the SRAM, followed by CAM, special--purpose, dynamic and static AND.

\medskip

\begin{table}[h]
 \centering
 \begin{tabular}{cccccccc}
    \toprule
    \multicolumn{7}{c}{\textbf{Energy--delay products relative variations: Search v.s. Write}}\\
    \midrule
    & & \multicolumn{5}{c}{\textbf{Write}}\\
   \multirow{6}{*}{\textbf{Search}} & & \textbf{SRAM} & \textbf{CAM} & \textbf{AND SP} & \textbf{AND DYN} & \textbf{AND ST}\\
    & \textbf{CAM} & +76.12\% & -16.52\% & --- & --- & ---\\
    & \textbf{AND SP} & +435.07\% & +160.72\% & +132.04\% & --- & ---\\
    & \textbf{AND DYN} & +759.7\% & +318.91\% & +272.81\% & +93.29\% & --- \\
    & \textbf{AND ST} & +348.51\% & +118.54\% &  +94.5\% & -91.68\% & -59.9\% \\
    \bottomrule                     
 \end{tabular}
  \smallskip
  \captionsetup{justification=centering}
 \caption{Percentage differences between the read and search energy--delay products among the arrays. Each value corresponds to the increase, expressed in percentage, in the energy--delay product of the memory on the corresponding row with respect to the one of the memory on the corresponding column. The data are extracted from \autoref{tab:energy_delay}.}
 \label{tab:read_search_comparison}
\end{table}

\begin{table}[h]
 \centering
 \begin{tabular}{cccccccc}
    \toprule
    \multicolumn{7}{c}{\textbf{Energy--delay products relative variations: Search v.s. Read}}\\
    \midrule
    & & \multicolumn{5}{c}{\textbf{Read}}\\
   \multirow{6}{*}{\textbf{Search}} & & \textbf{SRAM} & \textbf{CAM} & \textbf{AND SP} & \textbf{AND DYN} & \textbf{AND ST}\\
    & \textbf{CAM} & +100\% & +2.61\% & --- & --- & ---\\
    & \textbf{AND SP} & +507.63\% & +211.74\% & +175.77\% & --- & ---\\
    & \textbf{AND DYN} & +876.27\% & +400.86\% & +343.08\% & +115.43\% & --- \\
    & \textbf{AND ST} & +409.32\% & +161.3\% &  +131.15\% & +33.26\% & --6.65\% \\
    \bottomrule                     
 \end{tabular}
  \smallskip
  \captionsetup{justification=centering}
 \caption{Percentage differences between the write and search energy--delay products among the arrays. Each value corresponds to the increase, expressed in percentage, in the energy--delay product of the memory on the corresponding row with respect to the one of the memory on the corresponding column. The data are extracted from \autoref{tab:energy_delay}.}
 \label{tab:write_search_comparison}
\end{table}

In \autoref{tab:write_search_comparison} and \autoref{tab:read_search_comparison}, the comparisons between the search operation and write and read operations, respectively, energy--delay products are reported. One can notice that in all the cases the search operations perform worse than the read/write one of the SRAM array. However, for the static AND and CAM arrays, the search operation is characterised by 16.52\% and 59.9\%, respectively, lower energy--delay products when compared to the same array write operation; for what concerns the read one, the CAM search operation performs just 2.61\% worse, while the static array performs 6.65\% better.

\medskip

\begin{table}[h]
 \centering
 \begin{tabular}{cccccc}
    \toprule
    \multicolumn{6}{c}{\textbf{Energy--delay products relative variations: AND v.s. Search}}\\
    \midrule
    & & \multicolumn{3}{c}{\textbf{Search}}\\
   \multirow{5}{*}{\textbf{AND}} & & \textbf{CAM} & \textbf{AND SP} & \textbf{AND DYN} & \textbf{AND ST}\\
    & \textbf{AND SP} & -140.81\% & -631.63\% & --- & --- \\
    & \textbf{AND DYN} & +750.85\% & +180.05\% & +74.3\% & --- \\
    & \textbf{AND ST} & -210.52\% & -843.42\% &  -1415.79\% &  -690.79\%\\
    \bottomrule                     
 \end{tabular}
  \smallskip
  \captionsetup{justification=centering}
 \caption{Percentage differences in the AND and Search energy--delay products among the arrays. Each value corresponds to the increase, expressed in percentage, in the energy--delay product of the memory on the corresponding row with respect to the one of the memory on the corresponding column. The data are extracted from \autoref{tab:energy_delay}.}
 \label{tab:AND_search_comparison}
\end{table}

From \autoref{fig:energy_delay}, it can be observed that the AND operation performs better than the search one for the static and special--purpose AND arrays. This is due to the fact that the hardware involved in the AND operation is less complex than the one of the search operation: while in the CAM cell (\autoref{fig:CAM_cell}) there are two pull-down paths of two series transistors connected to the matchline, in the AND cells (\autoref{fig:Static_cell}, \autoref{fig:Special_cell} and \autoref{fig:Dynamic_cell}) there is only one pull--down path. This leads to lower power consumption and latency.

In \autoref{tab:AND_search_comparison}, the AND and search operations energy--delay products values are compared. It can be observed that, apart from the dynamic AND case, the AND operation performs always better than the search one. In the dynamic AND case, this does not hold true due to the dynamic CMOS logic implementation of the gate, which leads to the commutation of all the row cells AND gates every time an AND operation is performed. This leads to a large increase in the energy consumption associated with the AND functionality.

\medskip

\begin{table}[h]
 \begin{tabular}{cccccccc}
    \toprule
    \multicolumn{7}{c}{\textbf{Energy--delay products relative variations: AND v.s. Write}}\\
    \midrule
    & & \multicolumn{5}{c}{\textbf{Write}}\\
   \multirow{4}{*}{\textbf{AND}} & & \textbf{SRAM} & \textbf{CAM} & \textbf{AND SP} & 
  \textbf{AND DYN} & \textbf{AND ST}\\
    & \textbf{AND SP} & -36.7\% & -180.61\% & -215\% & ---\\
    & \textbf{AND DYN} & +1398\% & +630.2\% & +549.84\% & +236.91\% & -- \\
    & \textbf{AND ST} & -76.31\% & -261.84\% &  -306.57\% & -684.21\% & -1164\% \\
    \bottomrule                     
 \end{tabular}
  \smallskip
  \centering
 \captionsetup{justification=centering}
 \caption{Percentage differences in the write and AND energy--delay products among the arrays. Each value corresponds to the increase, expressed in percentage, in the energy--delay product of the memory on the corresponding row with respect to the one of the memory on the corresponding column. The data are extracted from \autoref{tab:energy_delay}.}
 \label{tab:AND_write_comparison}
\end{table}

\begin{table}[h]
 \centering
 \begin{tabular}{cccccccc}
    \toprule
    \multicolumn{7}{c}{\textbf{Energy--delay products relative variations: AND v.s. Read}}\\
    \midrule
    & & \multicolumn{5}{c}{\textbf{Read}}\\
   \multirow{5}{*}{\textbf{AND}} & & \textbf{SRAM} & \textbf{CAM} & \textbf{AND SP} & \textbf{AND DYN} & \textbf{AND ST}\\
    & \textbf{AND SP} & -20.41\% & -134.69\% & -165.31\% & --- & ---\\
    & \textbf{AND DYN} & +1601.7\% & +773.04\% & +672.31\% & +345.23\% & ---\\
    & \textbf{AND ST} & -55.26\% & -202.63\% &  -242.1\% &  -1164.47\% & -743.2\% \\
    \bottomrule                     
 \end{tabular}
  \smallskip
  \captionsetup{justification=centering}
 \caption{Percentage differences in the read and AND energy--delay products among the arrays. Each value corresponds to the increase, expressed in percentage, in the energy--delay product of the memory on the corresponding row with respect to the one of the memory on the corresponding column. The data are extracted from \autoref{tab:energy_delay}.}
 \label{tab:AND_read_comparison}
\end{table}

For what concerns the AND operation and the conventional ones, one can notice from \autoref{fig:energy_delay} that the AND operation, in the static and special--purpose arrays, performs better than both read and write ones in the SRAM array, for an array size equal to 256x256. This is due to the fact to perform the AND operation there is no need to access the cell content, thank to the additional cell circuitry, which allows for lower latency and energy consumption; in fact, the SRAM core circuit is highly inefficient, as observed in the previous discussion.

In \autoref{tab:AND_write_comparison} the comparison between AND and write performance is detailed. One can notice that, apart from the dynamic AND case, the AND operation always outperforms the write one, even when comparing it with the conventional SRAM architecture: for the special--purpose case, a 36.7\% reduction in the AND energy--delay product is measured with respect the SRAM write one, while in the static AND case the reduction is equal to 76.31\%.

In \autoref{tab:AND_read_comparison}, the comparison between AND and read performance is analysed. Also in this case, the AND operation always outperforms the read one, apart from the dynamic AND case, even when compared with the SRAM: for the special--purpose AND, a 20.41\% reduction in the AND energy--delay product with respect to the SRAM read one is measured; for the static AND case, a reduction of 55.26\% is obtained.

\smallskip

This implies that performing an in--memory operation, such as the AND one, is more convenient from both energetic and latency points of view, even when compared with a conventional SRAM memory. It has to be highlighted that in this analysis the overhead associated to the extraction of the data from the array --- i.e. the energy and latency contributions due to the data transfer between the memory and the CPU, and due to the data processing inside the process --- is not taken into account; as a consequence, the advantages resulting from the in--memory approach are being heavily underestimated.

\section{Conclusions}
\label{sec:Conclusions}

In this work, a LiM array with 3 memory cell variants is designed and implemented at physical level in Cadence Virtuoso, by implementing the cells layout and extracting the parasitic netlists from these. The resulting circuit is compared against conventional memory arrays, such as SRAM and CAM ones, by evaluating the overheads associated to the LiM hardware on the standard memory operations.

From the results, an increase in energy consumption and latency is observed for the read and write memory operations in the LiM array (+120.34\% and +13.04\% for the read operation w.r.t. SRAM and CAM, respectively, in the best case). The results also highlight that the in--memory processing cost, represented by the energy--delay product associated with the LiM operation, is 55.26\% lower than the one associated to the read operation of an SRAM memory, in the best case, even without considering the energy and delay contributions due to the out--of--chip transfer of the data to the CPU. This implies that processing the data directly in memory is much more convenient than extracting them from the array and performing the computations in the CPU, despite the previously discussed drawbacks due to the additional hardware complexity.

These results highlight that Logic--In--Memory arrays, in which the memory cell is modified by adding computational elements to it, are best suited for applications with a low number of reading and write operations and a large number of in--memory logic operations. These represent a suitable alternative for the design of algorithm accelerators, that can be also used as secondary low--density conventional memories for data storage.


\begin{thebibliography}{10}

\bibitem{cam_sizes_0}
A.~Fritsch, M.~Kugel, R.~Sautter, D.~Wendel, J.~Pille, O.~Torreiter,
  S.~Kalyanasundaram, and D.~A. Dobson, ``A 4ghz, low latency tcam in 14nm soi
  finfet technology using a high performance current sense amplifier for ac
  current surge reduction,'' in {\em ESSCIRC Conference 2015 - 41st European
  Solid-State Circuits Conference (ESSCIRC)}, pp.~343--346, 2015.

\bibitem{cam_sizes_1}
E.~Seevinck, P.~van Beers, and H.~Ontrop, ``Current-mode techniques for
  high-speed vlsi circuits with application to current sense amplifier for cmos
  sram's,'' {\em IEEE Journal of Solid-State Circuits}, vol.~26, no.~4,
  pp.~525--536, 1991.

\bibitem{cam_sizes_2}
N.~Mohan, W.~Fung, D.~Wright, and M.~Sachdev, ``Design techniques and test
  methodology for low-power tcams,'' {\em IEEE Transactions on Very Large Scale
  Integration (VLSI) Systems}, vol.~14, no.~6, pp.~573--586, 2006.

\bibitem{neuro_0}
B.~Alimkhanuly, J.~Sohn, and I.~e.~a. Chang, ``Graphene-based 3d xnor-vrram
  with ternary precision for neuromorphic computing.,'' {\em npj 2D Mater},
  vol.~Appl 5, no.~55, 2021.

\bibitem{neuro_1}
M.~Le, T.~K.~H. Pham, and S.~N. Truong, ``Noise and memristance variation
  tolerance of single crossbar architectures for neuromorphic image
  recognition,'' {\em Micromachines}, vol.~12, no.~6, 2021.

\bibitem{neuro_2}
H.~Abbas, Y.~Abbas, G.~Hassan, A.~S. Sokolov, Y.-R. Jeon, B.~Ku, C.~J. Kang,
  and C.~Choi, ``The coexistence of threshold and memory switching
  characteristics of ald hfo2 memristor synaptic arrays for energy-efficient
  neuromorphic computing,'' {\em Nanoscale}, vol.~12, pp.~14120--14134, 2020.

\bibitem{nmc_gp_6}
S.~Angizi and D.~Fan, ``Redram: A reconfigurable processing-in-dram platform
  for accelerating bulk bit-wise operations,'' in {\em 2019 IEEE/ACM
  International Conference on Computer-Aided Design (ICCAD)}, pp.~1--8, 2019.

\bibitem{blade_EPFL}
W.~A. {Simon}, Y.~M. {Qureshi}, M.~{Rios}, A.~{Levisse}, M.~{Zapater}, and
  D.~{Atienza}, ``{BLADE: An in-Cache Computing Architecture for Edge
  Devices},'' {\em IEEE Transactions on Computers}, vol.~69, no.~9,
  pp.~1349--1363, 2020.

\bibitem{nmc_ml_0}
H.~{Jiang}, S.~{Huang}, X.~{Peng}, J.~W. {Su}, Y.~C. {Chou}, W.~H. {Huang},
  T.~W. {Liu}, R.~{Liu}, M.~F. {Chang}, and S.~{Yu}, ``{A Two-way SRAM Array
  based Accelerator for Deep Neural Network On-chip Training},'' in {\em 2020
  57th ACM/IEEE Design Automation Conference (DAC)}, pp.~1--6, 2020.

\bibitem{lim_ml_0}
Z.~{Jiang}, S.~{Yin}, J.~{Seo}, and M.~{Seok}, ``{{C3SRAM}: An
  In-Memory-Computing SRAM Macro Base on Robust Capacitive Coupling Computing
  Mechanism},'' {\em IEEE Journal of Solid-State Circuits}, vol.~55, no.~7,
  pp.~1888--1897, 2020.

\bibitem{nmc_gp_0}
K.~{Lee}, J.~{Jeong}, S.~{Cheon}, W.~{Choi}, and J.~{Park}, ``{Bit Parallel 6T
  SRAM In-memory Computing with Reconfigurable Bit-Precision},'' in {\em 2020
  57th ACM/IEEE Design Automation Conference (DAC)}, pp.~1--6, 2020.

\bibitem{nmc_ml_1}
H.~{Jiang}, X.~{Peng}, S.~{Huang}, and S.~{Yu}, ``{CIMAT: A Compute-In-Memory
  Architecture for On-chip Training Based on Transpose SRAM Arrays},'' {\em
  IEEE Transactions on Computers}, vol.~69, no.~7, pp.~944--954, 2020.

\bibitem{nmc_gp_1}
A.~K. {Rajput} and M.~{Pattanaik}, ``{Implementation of Boolean and Arithmetic
  Functions with 8T SRAM Cell for In-Memory Computation},'' in {\em 2020
  International Conference for Emerging Technology (INCET)}, pp.~1--5, 2020.

\bibitem{nmc_ml_2}
S.~{Yin}, Z.~{Jiang}, J.~{Seo}, and M.~{Seok}, ``{XNOR-SRAM: In-Memory
  Computing SRAM Macro for Binary/Ternary Deep Neural Networks},'' {\em IEEE
  Journal of Solid-State Circuits}, vol.~55, no.~6, pp.~1733--1743, 2020.

\bibitem{lim_gp_0}
A.~{Jaiswal}, A.~{Agrawal}, M.~F. {Ali}, S.~{Sharmin}, and K.~{Roy}, ``{i-SRAM:
  Interleaved Wordlines for Vector Boolean Operations Using SRAMs},'' {\em IEEE
  Transactions on Circuits and Systems I: Regular Papers}, vol.~67, no.~12,
  pp.~4651--4659, 2020.

\bibitem{nmc_ml_3}
A.~{Agrawal}, A.~{Kosta}, S.~{Kodge}, D.~E. {Kim}, and K.~{Roy}, ``{CASH-RAM:
  Enabling In-Memory Computations for Edge Inference Using Charge Accumulation
  and Sharing in Standard 8T-SRAM Arrays},'' {\em IEEE Journal on Emerging and
  Selected Topics in Circuits and Systems}, vol.~10, no.~3, pp.~295--305, 2020.

\bibitem{santoro_turvani_graziano_2019}
G.~Santoro, G.~Turvani, and M.~Graziano, ``{New Logic-In-Memory Paradigms: An
  Architectural and Technological Perspective},'' {\em Micromachines}, vol.~10,
  no.~6, p.~368, 2019.

\bibitem{lim_ml_1}
G.~{Saha}, Z.~{Jiang}, S.~{Parihar}, C.~{Xi}, J.~{Higman}, and M.~{Ahosan Ul
  Karim}, ``{An Energy-Efficient and High Throughput in-Memory Computing
  Bit-Cell With Excellent Robustness Under Process Variations for Binary Neural
  Network},'' {\em IEEE Access}, vol.~8, pp.~91405--91414, 2020.

\bibitem{lim_ml_2}
H.~{Kim}, Q.~{Chen}, T.~{Yoo}, T.~T.~H. {Kim}, and B.~{Kim}, ``{A Bit-Precision
  Reconfigurable Digital In-Memory Computing Macro for Energy-Efficient
  Processing of Artificial Neural Networks},'' in {\em 2019 International SoC
  Design Conference (ISOCC)}, pp.~166--167, 2019.

\bibitem{nmc_ml_4}
H.~{Shin}, J.~{Sim}, D.~{Lee}, and L.~S. {Kim}, ``{A PVT-robust Customized 4T
  Embedded DRAM Cell Array for Accelerating Binary Neural Networks},'' in {\em
  2019 IEEE/ACM International Conference on Computer-Aided Design (ICCAD)},
  pp.~1--8, 2019.

\bibitem{lim_ml_3}
A.~{Agrawal}, A.~{Jaiswal}, D.~{Roy}, B.~{Han}, G.~{Srinivasan}, A.~{Ankit},
  and K.~{Roy}, ``{Xcel-RAM: Accelerating Binary Neural Networks in
  High-Throughput SRAM Compute Arrays},'' {\em IEEE Transactions on Circuits
  and Systems I: Regular Papers}, vol.~66, no.~8, pp.~3064--3076, 2019.

\bibitem{lim_ml_4}
J.~{Saikia}, S.~{Yin}, Z.~{Jiang}, M.~{Seok}, and J.~{Seo}, ``{K-Nearest
  Neighbor Hardware Accelerator Using In-Memory Computing SRAM},'' in {\em 2019
  IEEE/ACM International Symposium on Low Power Electronics and Design
  (ISLPED)}, pp.~1--6, 2019.

\bibitem{nmc_ml_5}
H.~Kim, H.~Oh, and J.-J. Kim, ``{Energy-Efficient XNOR-Free in-Memory BNN
  Accelerator with Input Distribution Regularization},'' in {\em Proceedings of
  the 39th International Conference on Computer-Aided Design}, ICCAD '20, (New
  York, NY, USA), Association for Computing Machinery, 2020.

\bibitem{nmc_ml_6}
S.~Huang, H.~Jiang, X.~Peng, W.~Li, and S.~Yu, ``{XOR-CIM: Compute-in-Memory
  SRAM Architecture with Embedded XOR Encryption},'' in {\em Proceedings of the
  39th International Conference on Computer-Aided Design}, ICCAD '20, (New
  York, NY, USA), Association for Computing Machinery, 2020.

\bibitem{nmc_gp_2}
M.~{Ali}, A.~{Jaiswal}, S.~{Kodge}, A.~{Agrawal}, I.~{Chakraborty}, and
  K.~{Roy}, ``{IMAC: In-Memory Multi-Bit Multiplication and ACcumulation in 6T
  SRAM Array},'' {\em IEEE Transactions on Circuits and Systems I: Regular
  Papers}, vol.~67, no.~8, pp.~2521--2531, 2020.

\bibitem{nmc_ml_7}
H.~{Jiang}, R.~{Liu}, and S.~{Yu}, ``{8T XNOR-SRAM based Parallel
  Compute-in-Memory for Deep Neural Network Accelerator},'' in {\em 2020 IEEE
  63rd International Midwest Symposium on Circuits and Systems (MWSCAS)},
  pp.~257--260, 2020.

\bibitem{nmc_ml_8}
A.~{Biswas} and A.~P. {Chandrakasan}, ``{CONV-SRAM: An Energy-Efficient SRAM
  With In-Memory Dot-Product Computation for Low-Power Convolutional Neural
  Networks},'' {\em IEEE Journal of Solid-State Circuits}, vol.~54, no.~1,
  pp.~217--230, 2019.

\bibitem{lim_gp_1}
A.~{Agrawal}, A.~{Jaiswal}, C.~{Lee}, and K.~{Roy}, ``{X-SRAM: Enabling
  In-Memory Boolean Computations in CMOS Static Random Access Memories},'' {\em
  IEEE Transactions on Circuits and Systems I: Regular Papers}, vol.~65,
  no.~12, pp.~4219--4232, 2018.

\bibitem{lim_gp_2}
Q.~{Dong}, S.~{Jeloka}, M.~{Saligane}, Y.~{Kim}, M.~{Kawaminami}, A.~{Harada},
  S.~{Miyoshi}, M.~{Yasuda}, D.~{Blaauw}, and D.~{Sylvester}, ``{A 4 + 2T SRAM
  for Searching and In-Memory Computing With 0.3 V $V_{DDmin}$},'' {\em IEEE
  Journal of Solid-State Circuits}, vol.~53, no.~4, pp.~1006--1015, 2018.

\bibitem{marco}
M.~{Vacca}, Y.~{Tavva}, A.~{Chattopadhyay}, and A.~{Calimera},
  ``{Logic-In-Memory Architecture For Min/Max Search},'' in {\em 2018 25th IEEE
  International Conference on Electronics, Circuits and Systems (ICECS)},
  pp.~853--856, 2018.

\bibitem{nmc_gp_3}
K.~{Yang}, R.~{Karam}, and S.~{Bhunia}, ``{Interleaved Logic-in-Memory
  architecture for energy-efficient fine-grained data processing},'' in {\em
  2017 IEEE 60th International Midwest Symposium on Circuits and Systems
  (MWSCAS)}, pp.~409--412, 2017.

\bibitem{nmc_gp_4}
K.~C. {Akyel}, H.~{Charles}, J.~{Mottin}, B.~{Giraud}, G.~{Suraci},
  S.~{Thuries}, and J.~{Noel}, ``{DRC2: Dynamically Reconfigurable Computing
  Circuit based on memory architecture},'' in {\em 2016 IEEE International
  Conference on Rebooting Computing (ICRC)}, pp.~1--8, 2016.

\bibitem{nmc_gp_5}
S.~{Jeloka}, N.~B. {Akesh}, D.~{Sylvester}, and D.~{Blaauw}, ``{A 28 nm
  Configurable Memory (TCAM/BCAM/SRAM) Using Push-Rule 6T Bit Cell Enabling
  Logic-in-Memory},'' {\em IEEE Journal of Solid-State Circuits}, vol.~51,
  no.~4, pp.~1009--1021, 2016.

\bibitem{cam_literature}
K.~{Pagiamtzis} and A.~{Sheikholeslami}, ``{Content-addressable memory (CAM)
  circuits and architectures: a tutorial and survey},'' {\em IEEE Journal of
  Solid-State Circuits}, vol.~41, pp.~712--727, March 2006.

\bibitem{cam_sa}
I.~{Arsovski} and A.~{Sheikholeslami}, ``{A mismatch-dependent power allocation
  technique for match-line sensing in content-addressable memories},'' {\em
  IEEE Journal of Solid-State Circuits}, vol.~38, no.~11, pp.~1958--1966, 2003.

\bibitem{coluccio}
A.~Coluccio, M.~Vacca, and G.~Turvani, ``Logic-in-memory computation: Is it
  worth it? a binary neural network case study,'' {\em Journal of Low Power
  Electronics and Applications}, vol.~10, no.~1, 2020.

\bibitem{sram_sa}
T.~Kobayashi, K.~Nogami, T.~Shirotori, and Y.~Fujimoto, ``A current-controlled
  latch sense amplifier and a static power-saving input buffer for low-power
  architecture,'' {\em IEEE Journal of Solid-State Circuits}, vol.~28, no.~4,
  pp.~523--527, 1993.

\bibitem{rabaey_book}
J.~M. Rabaey, A.~Chandrakasan, and B.~Nikolic, {\em Digital Integrated
  Circuits}.
\newblock USA: Prentice Hall Press, 3rd~ed., 2008.

\end{thebibliography}
\end{document}